# POLITECNICO DI BARI

RESEARCH DOCTORATE
IN
INFORMATION ENGINEERING

XXV CYCLE
2010-2012
SSD: ING-INF/01

DOCTORAL THESIS

# OPTICAL BIOCHEMICAL PLATFORMS FOR NANOPARTICLES DETECTION

**Supervisor:**

Prof. C. CIMINELLI

**Coordinator of the Ph.D. Course:**

Prof. B. TURCHIANO

**Ph.D. Candidate:**

Clarissa M. CAMPANELLA

*Per aspera sic itur ad astra.*

*Thanks to all the minds and hearts I met in the past three years*

*And to those pure souls I'll never forget.*

LIST OF ABBREVIATIONS:

| | |
|---|---|
| CW | Clockwise |
| CCW | Counter-Clockwise |
| CMT | Coupled Mode Theory |
| ER | Extinction Ratio |
| FDTD | Finite Difference Time Domain |
| FWHM | Full Width at Half Maximum |
| IO | Integrated Optics |
| LOD | Limit Of Detection |
| NA | Numerical Aperture |
| NAIL | Numerical Aperture Increasing Lens |
| NP | Nanoparticle |
| Q | Quality Factor |
| RI | Refractive Index |
| RIU | Refractive Index Unit |
| SIL | Solid Immersion Lens |
| SPR | Surface Plasmon Resonance |
| WGM | Whispering Gallery Mode |

TABLE OF CONTENTS

# Chapter 1.
# INTRODUCTION

Preserving, monitoring, or anyway dealing with the human and environmental health state should be one of the major goals to be pursued in a world always more dominated by a strong correlation between industrial processes and technological advances.

Due to their enormous potential, biosensors are widely used for detecting a wide range of analytes in the health care and food industries and in the environmental monitoring.

Mainly, the widespread employment of these devices is attributed to the enormous demands of diseases diagnosis and control, as well as to the property of biosensors of offering a convenient, hygienic, rapid, and compact method for personal monitoring [1].

In recent years, indeed, the sensors market has been dominated by an increasing demand in the field of medical diagnostics relying on the individuation of disposable, reliable, user-friendly, cost-efficient devices that also demonstrate fast response times and that are suitable for mass production. All these criteria are potentially fulfilled by biosensor technologies that combine interdisciplinary approaches coming from nanotechnology, chemistry and medical science.

The ability to assess health status is thus related to an early disease diagnosis or to the monitoring of its progression. Specifically, the main interest in the diagnostic research field is devoted to the detection of the most relevant mankind diseases in terms of worldwide incidence, prevalence, morbidity and mortality such as diabetes, cardiovascular disease and cancer. Despite their hereditary nature, the increasing diffusion of some of these diseases has been associated to the dominant presence of small-sized particles in air, water, soil and food as follows from the adverse effects noticed in vivo and in vitro [2]. These small-sized particles are introduced into the environment via several sources such as novel industrial processes that are giving rise to the undesired production of contaminant agents concurring to the modification of particulate matter. The cited agents include, as an example, ultrafine dust emitted by industrial plants, reactive particles produced by incinerators for waste disposal, catalyst nanoparticle leaked from automotive catalytic converters and others [3].

Although the wide diffusion of novel industrial processes, natural phenomena (i.e. volcanic eruptions, wind erosion, photochemical reactions) are still considered mostly responsible for nanoparticulate matter production than human activities [4].

Despite the great chemical and morphological diversity exhibited by nanoparticles (NPs), their nanometric-scale size allow them to enter the living organism following inhalation, translocate within it and damage it by acting as some of the most dangerous human body intruders, i.e. viruses.

Even though the exact definition of nanoparticles differs depending upon the materials, fields and applications concerned, they could be anyway defined as ultrafine particles in the size of nanometer order, i.e. one thousandth of 1 m.

In the narrower sense, they are regarded as those particles having the characteristic size smaller than a limit value of $10 - 20$ nm corresponding to that value where the physical properties of solid materials themselves would drastically change. Really often anyway, particles with characteristic size ranging from 1 to 100 nm are called nanoparticles and they are considered as particles smaller than those conventionally called "submicron particles".

The relationship between nanoparticles toxicity and dimensions has been widely investigated [2], [5], [6] and the main toxicity parameters have been identified in their size and chemical composition, in the exposure time to which human body is subjected and thus bioactivity and in-vivo accumulation [3]. This is valid for both natural and engineered particles.

Despite the undesired effects induced by NPs on human body, their small size is attractive from a pharmaceutical point of view due to several features, including NPs high surface to mass ratio, their quantum properties and their ability to adsorbs or carry other compounds (such as drugs, probes and proteins) that make NPs good candidates for drug delivery. Ultimate advances in novel diagnostic and therapeutic applications (theragnostics) [7] rely, indeed, on the production and employment of engineered nanoparticles to be used as carriers.

Both industrial and bio-chemical field have been characterized by a widespread use of engineered nanoparticles including metal NPs for single cancer cell detection, for medical imaging enhancement or for cleaning up $CCl_4$ (Carbon tetrachloride) pollution in groundwater, or dielectric NPs for the delivering of chemotherapy drugs directly to cancer cells or for the increasing of lithium-ion battery power and reduction of recharge time.

Despite the importance of detecting and characterizing potentially toxic NPs, the study of nano-sized natural particles could lead also to the investigation of new phenomena at nano-scale. All the solid particles, indeed, consist of atoms or molecules of a certain material whose behavior affects NPs by conferring them different properties from those of the bulk solid of the same material. This is attributable to the change of the bonding state of the atoms or the molecules constructing the particles [8].

An accurate measurement of nanoparticles size would lead to understand the change in properties of NPs in terms of morphological, structural, thermal, electromagnetic, optical and mechanical characteristics. This change is usually indicated as " size effect". Actually, the most basic method to discriminate NPs in terms of size and size-distribution relies on analyzing images acquired through a really expensive microscope, i.e. the transmission electron microscope (TEM), due to its very high spatial resolution.

In biochemical sensing field, a fervent research activity related to the development of real time, low cost, compact and high throughput devices for the detection and characterization of natural or synthetic nanoparticles NPs actually exist. In this research scenario, different platforms for biosensing purposes have been developed according to the huge amount of physical effects involved in the transduction of the biochemical-signal into a measurable output signal.

In the present work two different optical platforms for NP detection have been investigated, one based on integrated optics and the other based on microscopy.
Both the approaches rely on the study of the interaction of an electromagnetic wave with a small particle in the hypothesis of dealing with a Rayleigh scatterer, i.e. a nanoparticle having a size really smaller than the one of the wavelength of the incident light and scattering light elastically [9].
The study deals with a class of problems known as "the inverse problem" which relies on describing the characteristics of the particle or particles responsible for a light scattering phenomenon once the scattered field is known.

The first examined configuration is based on the employment of a high quality factor microresonator supporting Whispering Gallery Modes (WGMs) for NP detection purposes.
A planar optical resonant cavity has indeed been designed to detect and size a single NP by analyzing the resonator spectrum. First, some analytical formulae modeling the resonator-NP interaction have been investigated in order to find a correlation between the transmission spectrum of a WGM resonator and nanoparticles characteristics (i.e. size and refractive index). Quantum electrodynamic (QED) laws have been taken into account for the modeling process, according to literature.
On the basis of the investigated model, some criteria for designing a planar cavity for single NP detection have been deduced and a cavity with a high Q-factor, a low modal volume $V_c$ and a high extinction ratio ER value has been designed and tested for NPs detection and sizing by employing an FDTD method based commercial software. NPs having a radius in the range 30 : 100 nm have been employed as testing elements and the estimation error in NP sizing has been evaluated to be in the order of 2%.
The sensitivity of the cavity has also been investigated in terms of detection of glucose concentration in solution. A refractometric detection scheme has been employed and an analytical sensitivity of about 550 nm/RIU and a LOD of $10^{-6}$ RIU have been measured.

The second investigated platform relies on the employment of an interferometric microscopy set-up. The starting point for this study is the IRIS (Interferometric Reflectance Imaging Sensor) platform developed at the Boston University by the Ünlü group [10]. The IRIS setup has yet demonstrated detection and size discrimination capability for polystyrene beads ranging from 70 nm to 150 nm in diameter, as well as specific capture and identification of single H1N1 virus particles. An improvement of the existing configuration has been investigated in the present study in order to realize a real time, sub-surface detection of single molecules and NPs. A setup based on confocal illumination aided by the employment of a numerical aperture increasing lens (NAIL) has been modeled in order to confer an ultra-high resolution to the microscopy system. In order to increase the contrast, a multilayer substrate has been assumed so that the light intensity at the photo-detector is maximized. The study has been conducted at the Department of Electronic and Computer Engineering, College of Engineering, Boston University, under the guidance of Prof. M.S. Ünlü.

## 1.1 DISSERTATION OVERVIEW

The manuscript is organized as follows: an introduction to bio-chemical sensing is given in Chapter 2. The main figures of merit of a biosensor, including sensitivity, detection limit, throughput and selectivity are described. First, the attention is focused on the functionalization of the biosensor surface (comprising the opportune choice of the receptors and their immobilization on the surface) which is one of the most critical aspect in the development of a device devoted to biosensing purposes. Then, the main sensing mechanisms and transducing methods employed in this sensing field are described and particular attention is paid to optical devices due to their potentiality. The main configurations of integrated optics biosensor are then introduced, i.e. planar waveguides, photonic crystals based devices, fiber optics biosensors and optical ring resonators. Due to

- 5 -

the theoretical characteristics of high sensitivity, low detection limit, high throughput and small footprint achievable by employing resonant cavities for sensing purposes, in Chapter 3 an overview of optical ring resonators is presented. Properly, the performances of resonators in planar, optofluidic or spherical configuration, i.e. resonators supporting whispering gallery modes WGMs, are investigated and a discussion concerning the design of an optical resonant cavity is presented. Some expedients useful to improve biosensor performances have been also described, including an opportune choice of the resonator geometry and technology in order to overcome issues related to bending loss, thermal noise and other.

After describing resonant cavities performances for generic purposes in the bio-chemical sensing field in Chapter 3, the attention is focused on the different transducing methods and devices employed for nanoparticles detection in the following Chapter (i.e. 4). An overview of the most relevant detection methods based on mechanical, electrical and optical transduction is presented. Single nanoparticle detection schemes realized in integrated optics and discrete optics are mainly described in the same Chapter including WGM based- and microscopy based- nanoparticle detection configurations, and their advantages and limits are pointed out.

In Chapter 5, a method for NPs detection and sizing has been analyzed. It relies on the coupling of CW and CCW travelling wave modes propagating within a WGM based cavity occurring when light is perturbed along its path by the presence of a scattering element, i.e. a NP or surface roughness. On the basis of the analytical formulae modeling the resonator-NP interaction, some criteria for designing a planar cavity for single NP detection have been deduced and a cavity has been designed and tested for NPs detection and sizing.

Microresonant optical cavities anyway do not allow for nanoparticles shape discrimination in a complex solution. Thus, a sensor based on interferometric reflectance imaging (IRIS) has been modeled in Chapter 6, in confocal, back-side illumination scheme. After describing some mathematical instruments useful to model of the setup, the most important results have been presented for the investigated sensor in terms of optimization of the platform contrast through an opportune choice of the substrate.

Finally, a comparison between both the designed and investigated optical platforms and techniques is presented in Chapter 7.

## 1.2 REFERENCES

1. C. Qi, G.F. Gao and G. Jin, "Label-free Biosensors for Health Applications, " in : "Biosensors for Health, Environment and Biosecurity", P.A. Serra Edt., 2011.
2. W.H. De Jong, P.J.A. Borm, "Drud delivery and nanoparticles: Applications and hazards," International Journal of Nanomedicine, Vol. 3, No. 2, pp. 133-149, 2008.
3. C. Ciminelli, C.M. Campanella, R. Pilolli, N. Cioffi, M.N. Armenise, "Optical Sensor for Nanoparticles", XIII International Conference on Transparent Optical Networks (ICTON) 2011, Invited Paper, Stockholm (Sweden), 27-30 June 2011.
4. D.A. Taylor, "Dust in the wind," Environmental Health Perspectives, Vol. 110, No. 2, 2002.
5. C. Buzea, I. Pacheco Blandino, K. Robbie, "Nanomaterials and nanoparticles: sources and toxicity", Biointerphases, vol. 2, no. 4, 2007.
6. M. Osier and G. Oberdorster, "Intratracheal Inhalation vs Intratracheal Instillation: Differences in Particle Effects", Fundamental and Applied Toxicology, Vol. 40, pp. 220-227, 1997.
7. D.A. LaVan, T. McGuire, and R. Langer, "Small-scale systems for in vivo drug delivery," Nature Biotechnology, No. 10. pp. 1184–1191, 2003.
8. M. Hosokawa, K. Nogi, M. Naito and T. Yokoyama, "Nanoparticle Technology Handbook," Elsevier 2007.
9. C.F. Bohren, D.R. Huffman, "Absorption and scattering of light by small particles," Wiley-VCH 2004.

10. M.S Ünlü, "IRIS: Interferometric Reflectance Imaging Sensor - Multiplexed Assays and Single Virus Detection," Laser Science (LS) Conference, (LTh3I), Rochester, NY, October 14, 2012.

# Chapter 2.
# BIOCHEMICAL SENSORS

Due to the wide relevance assumed in ordinary life by bio-chemical sensors, some general concepts on biochemical sensing are here reported. Then the most common mechanisms and transducing methods employed in biosensing field are investigated. The attention is mainly focused on integrated optics biosensors, including planar waveguides, photonics crystal devices, fiber optic biosensors and optical resonators.

## 2.1 DEFINITION

Usually an ambiguity occurs when referring to biosensors. Here some definitions and recommendations given by the Commission on General Aspects of Analytical Chemistry IUPAC are reported in order to clarify the concepts.

"*A chemical sensor is a device that transforms chemical information, ranging from the concentration of a specific sample component to total composition analysis, into an analytically useful signal*".

"*A biosensor is an analytical device incorporating a biological material or a bio-mimic e.g. tissue, microorganisms organelles, cell receptors, enzymes, antibodies, nucleic acids etc. intimately associated with or integrated with a physicochemical transducers or transducing micro-system using optical, electrochemical, thermometric, piezoelectric or magnetic properties etc*".

This definition is recommended by the Chemistry and the Environment Division.

Thus, a biosensor is a self-contained integrated device that should be clearly distinguished from an analytical system, which incorporates additional separation steps, such as high performance liquid chromatography (HPLC), or additional hardware and/or sample processing such as specific reagent introduction, e.g. flow injection analysis (FIA).

Even though a biosensor should be a reagent-less analytical device, the presence of ambient co-substrates, including water for hydrolases or oxygen for oxidoreductases, may be required for the analyte determination [1].

According to [1], biosensors constitute a subgroup of chemical sensors where biological host molecules, such as natural or artificial antibodies, enzymes or receptors or their hybrids, are equivalent to synthetic ligands and are integrated into the chemical recognition process.

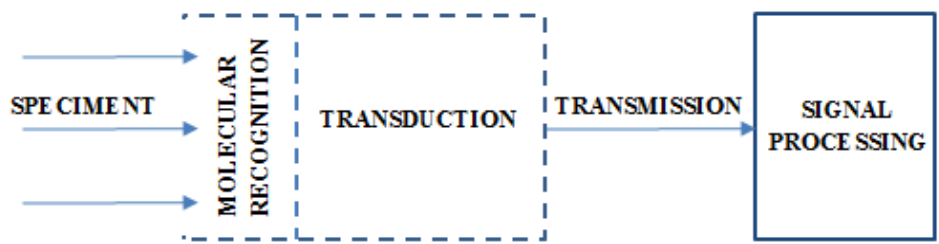

**Figure 1**. Schematic of a bio-chemical sensor.

In order to consider a sensor as bio-chemical, it should be a combination of a physical device, i.e. a transducer, and analyte-selective layers of biochemical or chemical substrates, as emphasized in Fig.1.

Thus, a complete sensor system will contain the transduction principle, the sensitive layer, the signal processing and an output device to collect the generated signal.

The sensitive layer, as explained in the following sections, is mainly responsible of selectivity, sensitivity, stability and reversibility of the sensor which are the fundamental, but not unique, requirements of the latter. An efficient biosensor should thus be subjected to surface chemistry and modification of the sensitive layer in order to provide high selectivity and sensitivity.

## 2.2 BIO-SENSORS PROPERTIES

The major figures of merit characterizing a bio-sensor are here described.

An ideal biochemical sensor should cover several properties including high sensitivity and selectivity, low detection limit, high throughput, good reliability, fast and reversible response, robustness, easy portability, low cost and simple fabrication processes.

- *Selectivity*, i.e. the capability of the sensor to detect a specific analyte in a sample containing different chemicals, can be assured by employing a functionalization layer of the relevant receptors in order to immobilize the desired target molecule upon the resonator surface.

This kind of sensing, i.e. the one employing receptors by assuring high selectivity, is termed surface sensing.

Despite this, the other sensing technique is the one called homogeneous sensing where the bio-sensor is covered by a fluid (liquid or gaseous medium) and biomolecules are inserted inside this solution by altering device properties. This way, all the molecules contributes to a change in the device response and it is not possible to select a specific molecule.

- *Detection limit* (indicated with both the acronyms DL or LOD that stands for limit of detection) is defined as the minimal change in the analyte properties that the sensor is capable to detect.
- *Limit of quantitation* (LOQ) is defined as the lowest concentration at which not only the analyte can be reliably detected, but also at which some predefined goals concerning measurement errors, i.e. bias and imprecision, are met. It is an additive figure to LOD used to describe the smallest concentration of a measurand that can be reliably measured by an analytical procedure.

- *Limit of blank* (LOB) is defined as the highest apparent analyte concentration expected to be found when replicates of a blank sample (containing no analyte) are tested.

A standard method for determining these last three figures has been provided in the guideline EP17, Protocols for Determination of Limits of Detection and Limits of Quantitation [2].

- *Sensitivity* (S) is related to the variation of the output signal generated by the device with respect to the variation of the analyte properties, i.e. a mass or a concentration.
- *Analytical Sensitivity* (AS) is defined as the slope of the calibration curve representing the device sensitivity and having a linear shape. Sometimes the AS figure is confused with LOD. This should be avoided since LOD could reside below the linear range of the assay where the calibration curve has not validity.
- *Throughput* is the rate at which a biosensor system performs separate assays.

## 2.3 SENSITIVE LAYER

Most of bio-chemical sensors rely on a surface-based detection of the molecules composing the interest analyte. That means that the device surface is activated in order to allow a surface-molecule binding by suiting some probes that are immobilized on the device surface. These probes are also indicated as biorecognition elements and their employment depends on the specific target analyte to be detected. Such biorecognition elements are immobilized in the form of a sensitive layer on the surface of the device and they are capable of recognize and capture the target analytes.

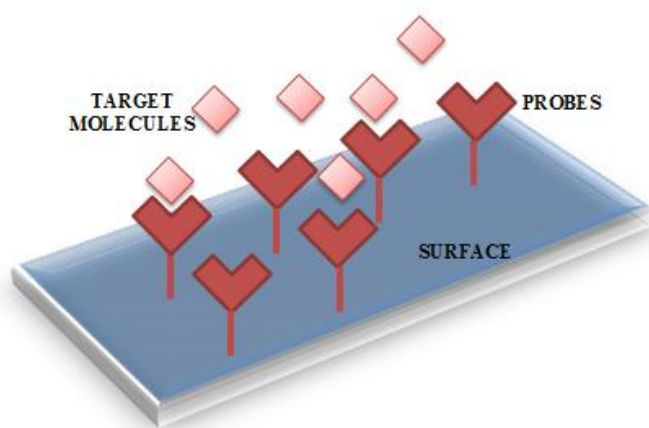

**Figure 2**. Schematic of bio-recognition process.

Antibodies and polymeric molecules are the most commonly used capturing probes.

Properly, antibodies in monoclonal and polyclonal form are employed as binding elements for proteins due to their robustness, large binding affinity and high specificity. Other widely employed probes are polymeric molecules such as aptamers that not only show the same peculiarities of antibodies, but can also be chemically synthesized without using animal or cell cultures [3].

When capturing probes are immobilized on the surface, two crucial issues need to be taken in consideration: one related to a homogeneous coverage of the surface and one related to the orientation of the probe on the surface. A failure in this immobilizing probes procedure gives rise to a lack of selectivity. If the spatial orientation of an antibody on the surface is not correct, indeed, the formation of a biomolecular complex coming from the reaction of the antibody with the associated antigen could not be verified.

There exist several techniques that could be used to achieve the correct orientation. They are essentially based on the use of surface modifying agents that couple to the antibody in such a way that the antibody sites that are reactive to the binding with the antigen are orientated away from the surface of the device. More details on different antibodies immobilization techniques aimed to a correct orientation control could be found in [4].

Here we only resume the most employed surface activation techniques useful to bind the biorecognition elements to the device surface. The main interactions employed for this purpose include the physical adsorption, the formation of a covalent bond, and the use of functional molecules [5].

2.3.1 RECEPTORS IMMOBILIZATION ON THE SUBSTRATE

The simplest technique to be employed for biosensing operations where no directional orientation of biomolecules is necessary is the *physical adsorption* where electrostatic or hydrophobic interactions are suited to immobilize biorecognition elements.

The first cited immobilization strategy relies on the electrostatic attraction between the charged molecules and the oppositely pre-charged surface that makes the technique interesting for the simple and inexpensive activation surface procedure. On the other hand, charged surfaces are prone to nonspecific adsorption of charged molecules and they should not be suited for the analysis of complex samples due to the lack of selectivity of the method.

The hydrophobic interaction is instead based on the interaction between non polar molecules and it is favorite in water where hydrophobic molecules are unable to form hydrogen bonding with polar molecules like water, as the name suggests.

A different approach is based on *covalent immobilization* of biorecognition molecules on the sensor surface. The functional group of the capturing probes is coupled to the functional group of the surface that has been pretreated in order to confer more reactivity to its functional group. The formation of covalent bond could rely on amine, thiol group or aldehyde coupling. Amine coupling involves the activation and stabilization of carboxylic groups in order to make them reactive to nucleophile groups. This method is employed for protein immobilization. The thiol coupling is another widely used immobilization method. It is based on the reaction of thiol groups (-SH) usually located on the receptors and thiol-reactive functional groups located on the device surface. The location could anyway be the opposite. The formed bond may be easily disrupted in solutions containing thiols and this property can be exploited for regeneration of the sensor surface and its repeated use. The aldehyde coupling method is particularly useful for the immobilization of glycoproteins and it is based on the Schiff-base condensation of aldehyde groups to amines and hydrazines. The latters are preferred for Schiff-base formation since they are more stable in aqueous environment than amine groups.

Although the employment of nonspecific interactions or covalent bonds for the immobilization of biorecognition elements on the device surface is rapid, simple and cheap, really often the immobilized molecules are randomly orientated on the surface and this could lead to a partial loss of the receptor activity as yet mentioned.

A strategy that could be employed for the creation of well-defined and oriented biomolecular assemblies is to immobilize the capturing probes via *non-covalent interactions*. The strongest non-covalent biological interaction known is the biotin-avidin (streptavidin) system that is one of the most used affinity pairs in the field of molecular, immunological, and cellular assays due to its strength and specificity of interaction. It consists of the binding between a protein, biotin, with a vitamin, avidin, or with a bacterial homologous protein to avidin, called streptavidin that is anchored to the sensor surface. Streptavidin is more employed than avidin since commercially available in a number of engineered forms. In addition, avidin could increase the undesired nonspecific adsorption. The peculiarity of this complex is that the bond forms very rapidly and is stable in wide ranges of pH and temperature. Since the affinity constant* of streptavidin–biotin interaction is really high, the system is considered one of the strongest and almost irreversible bonds available in bio-interactions.

An additional non-covalent immobilization technique is the one based on the employment of complementary nucleic-acid strand. It suites the site-directed immobilization of biorecognition elements via a sequence-specific hybridization of DNA and nucleic acids. One DNA strand is indeed incorporated into the biorecognition element, usually a protein, that is site-directly immobilized to the complementary DNA strand located on the sensor surface. Another widely employed complex for immobilization is the one called His Tag system. His Tag is a short amino acid sequence consisting of histidine (His) residues in recombinant proteins that is used as biorecognition element. Nickel nitrilotriacetic acid is usually employed for the immobilization of this receptor in a non-covalent bond. An oriented and specific immobilization of antibody could instead be assured by binding protein A/G to the surface. The capturing probe here is the system protein-antibody. Proteins A or G are immobilized on the sensor surface via a covalent immobilization for example, and they bind to antibodies specifically in certain regions containing antigen-binding element. Both Protein A and Protein G are native and recombinant proteins of microbial origin that bind to mammalian immunoglobulins (Ig) composing antibodies. These can come in different varieties known as isotypes or classes. One of the most versatile isotype is the antibody isotype Immunoglobuline G (IgG) which is in the binding domain of both protein A and G.

---

*The affinity constant, or association constant, is a numerical constant used to describe the bonding affinity of two molecules at equilibrium.

In order to activate a surface, a widely employed technique is based on silane chemistry. It is largely used in presence of silicon, silicon oxide and silicon nitride that are the most common materials used for sensitive layer realization in optics. Silane coupling agents have, indeed, the ability to form a durable bond between organic and inorganic materials. They are employed for different biomaterials, including DNA, oligonucleotides, protein, cell-organelle, whole cell and tissue. A typical application in which these coupling agents are suited is the activation of an inorganic surface having hydroxyl groups (–OH) that can be converted into stable oxane bonds by reaction with the silane. Thermal stability up to 350°C for certain silanes and their excellent effectiveness on inorganics materials make them real popular for surface chemical activation.

### 2.3.2 MOLECULAR RECOGNITION ELEMENTS

Different elements could be employed as recognition elements, including proteins, peptides, nucleic acids, carbohydrate structures and small organic or inorganic molecules. These biorecognition elements bind to the functional groups that have been previously immobilized on the activated surface through the techniques described in the previous section. Target molecules then anchor to the biorecognition elements and their detection could be realized through different schemes. The choice of scheme or format depends usually on the size of the target molecule.

The most employed detection formats are the one named *direct-, sandwich-, competition-* and the *inhibition-detection format.* In the first approach, the target molecule is bound to the biorecognition element immobilized on the device surface giving rise to a response signal from the device. Another detection format is the one called sandwich which is often used to enhance the sensor response. It is based on a primary binding of the target molecule to the biorecognition elements and then to the binding of a second antibody to the target molecule in order to create a double ligand with the target molecule that is sandwiched. This way the useful signal generated by the sensing device is related to the second antibody presence. The competitive detection format, instead, is based on the competition of two target molecules for the binding to a specific site on the device surface. One of the target is free, while the other is typically conjugated with a large molecule. Finally, the inhibition detection format relies on the immobilization of the target molecule on the device surface and the flowing of a solution containing a mixture of biorecognition and free target molecules on the device surface. A binding is generated between the unreacted biorecognition molecules and target molecules immobilized on the sensor surface.

A schematic of these detection formats is depicted in Fig. 3.

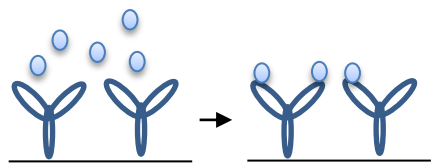

(A)

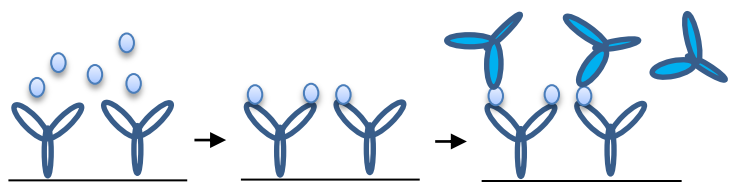

(B)

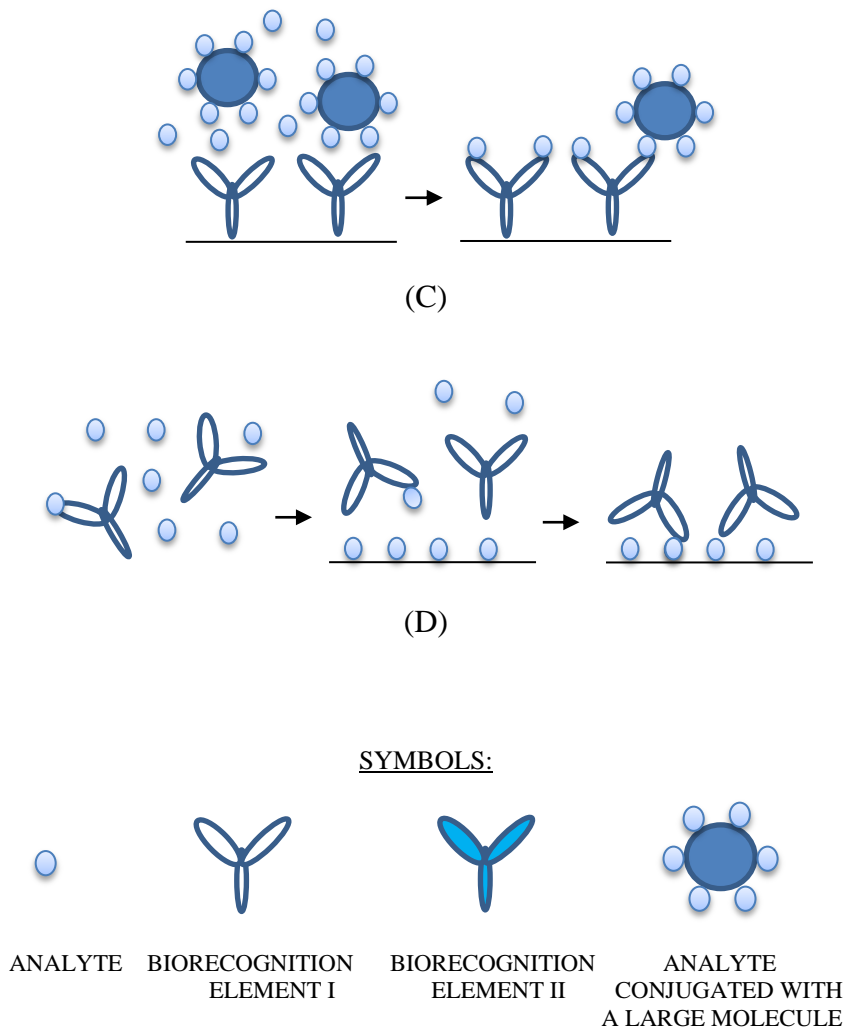

**Figure 3**. Detection formats: (A) direct detection, (B) sandwich detection, (C) competition detection, (D) inhibition detection.

A widely employed classification of modern biological and chemical sensors is the one based on the transduction mechanism supported by the device, i.e. electrical, thermal, mechanical or optical.

2.4 SENSING MECHANISMS AND TRANSDUCING METHODS: OPTICAL DEVICES

Biochemical sensors are developed for numerous kinds of analytes: gases, vapors and humidity, ions, organic chemicals, and macromolecules. As written previously, different chemical approaches and transduction mechanisms can be adopted to produce the final signal.

According to [1], the following Table shows the main classes of biosensors and the relative possible physical principle employed.

**Table 1**. Bio-chemical sensors classification.

| CLASS OF SENSORS | OPERATING PRINCIPLE |
|---|---|
| *Optical devices (optodes)* | Absorbance |
| | Reflectance |
| | Luminescence |
| | Refractive index |
| | Optothermal effect |
| | Light scattering (Raman) |
| *Electrochemical* | Voltammetry |
| | Potentiometry |
| | Chemically sensitized field effect transistor |
| | Potentiometry with solid electrolytes for gas sensing |
| *Electrical* | Metal oxide semiconductivity |
| | Organic semiconductivity |
| | Electrolyting conductivity |
| | Electric permettivity |
| *Mass sensitive* | Piezoelectric |
| | Surface acoustic wave propagation |
| *Magnetic* | Changes of paramagnetic gas properties |
| *Thermometric* | Heat effects of a specific chemical reaction |
| *Others* | Emission of α-, β-, or γ-radiation |

As reported in Table 1, several classes of sensors have been investigated as biochemical sensors including electrochemical, electrical, mass sensitive, magnetic and thermometric sensors.

Despite other transducing methods, optical devices based biosensing seems to be the most promising in nanoscale environment. The immunity to electromagnetic interference, the capability of both performing remote sensing and multiplexed detection within a single device [6] makes indeed photonics devices good candidates for a quick on market penetration. Photonic or optical devices, or optodes, suite various transducing properties for biosensing purposes including light absorption spectroscopy, reflectance, luminescence, optothermal effect, the monitoring of either the effective refractive index RI change or of the Raman scattering effect.

The terms optics and photonics are usually interchanged when referring to biosensing since these two scientific branches are deeply dependent on each other. Properly photonics (from the Greek φωτός that means light) is being hailed as the dominant technology for the 21st century due to the several application fields it has been suited, including information processing, transmission, data storage, and display. Prerogative of this science is to study phenomena related to light–matter interactions which have also been explored in biosensing field since last two decades.

Thus, in order to understand the physical processes concerning optical sensing, some light-matter interactions will be briefly described, following [7]. All the spectroscopic techniques that will be mentioned rely on the measurement of the energy-matter interaction.

## 2.4.1 ELECTRONIC ABSORPTION SPECTROSCOPY

This sensing mechanism relies on the employment of a lamp (e.g., a Xe lamp which provides a continuous distribution of the electromagnetic radiation from UV to near IR) or of any broadband source emitting light that is oriented to impinge on a sample and is then absorbed by a spectrometer. The latter measures the linear electronic absorption defined by the Beer–Lambert's law which states that the attenuation of an incident beam of intensity $I_0$ (at frequency ν) is described by an exponential decay behavior and the output intensity I is given as:

$$I(v) = I_0 e^{-k(v)Lc} = I_0(v) 10^{-\varepsilon(v)Lc} \tag{1}$$

The coefficient $\varepsilon(v)$ is called "*molar extinction*" coefficient at frequency *v* and is expressed in [L/(mol·cm)]; *c* is the molar concentration (mol/L) of absorbing species in the material.

The term *L* in Eqn. (1) (measured in cm) is the optical path length defined by the length of the absorbing medium through which the light travels.

Another figure used to describe absorption or attenuation is the transmittance *T*, defined as the ratio between the incident beam intensity and the output intensity:

$$T(v) = \frac{I(v)}{I_0(v)} \qquad (2)$$

The absorbance A is then defined as:

$$A(v) = \log_{10} \frac{1}{T} \qquad (3)$$

The absorption spectrum is usually exhibited as a plot of *T* versus *v* (or equivalently *λ*) and a dip at the absorption frequency *v* is visible in the spectrum.

Though the absorption spectrum it is possible to detect and identify a *chromophore*, i.e. a molecular unit responsible for the color of a molecule. Indeed, when light hits a molecule, the latter absorbs certain wavelengths and transmits or reflects others, showing a color.

A more accurate definition of chromophore is the one that considers it as a region in a molecule where the energy difference between two different molecular orbitals falls within the range of the visible spectrum.

Visible light that hits the chromophore can thus be absorbed by exciting an electron from its ground state into an excited state. This transition (from ground to excited state) produces absorption at a specific frequency that can be dependent on the environment of the chromophore.

The detection method consists of reading the shift of the absorption band and probing the interaction in which the chromophore or the chromophore containing bioassemblies (biological media) may be involved.

## 2.4.2 ELECTRONIC LUMINESCENCE SPECTROSCOPY

Luminescence is a process where a medium emits radiations after absorbing energy. If the process concerns emission of light due to the relaxation of an electronic excited state generated by light absorption, the term fluorescence is used. By monitoring the fluorescence life-time it is possible to extract information concerning inter-molecular activity in terms of energy transfer.

## 2.4.3 VIBRATIONAL SPECTROSCOPY

Vibrational spectroscopy refers to both infra-red IR and Raman spectroscopy. The first relies on the change of atoms vibrational levels due to the absorption of a photon with frequency in the infrared range.
Raman spectroscopy is instead a spectroscopic technique based on the inelastic scattering of monochromatic light impinging on the biological sample, i.e. the frequency of photons in monochromatic light changes upon interaction with the analyte. Photons are indeed absorbed by the sample and then reemitted at a frequency that is shifted up or down in comparison with the original frequency. This is called the Raman effect. Vibrational, rotational and other low frequency transitions in molecules can be investigated by monitoring this shift.

## 2.4.4 REFRACTIVE INDEX CHANGE

Most integrated optics configurations for biosensing applications suite effective refractive index $n_{eff}$ change induced by the interaction of the analyte with the device.
While refractive index $n$ of a material is an intrinsic property depending on the optical wavelength or frequency*, the effective refractive index (or effective index) is not just a material property, but depends on the whole optical device design.

---

*This dependence is called dispersion.

Formally, the effective index is defined as "a number quantifying the phase delay per unit length in a waveguide, relative to the phase delay in vacuum" [8].

The refractive index can, indeed, be used to quantify the increase of the vacuum wavenumber $k_0$ (phase change per unit length) caused by the presence of the medium, that is:

$$k_m = n_m k_0 \qquad (4)$$

where $k_0 = 2\pi/\lambda$. The subscript $m$ stands for medium.

At the same manner, the effective refractive index $n_{eff}$ is related to the propagation of light in a guiding structure as previously formally defined; the propagation constant $\beta$ of the waveguide is indeed equivalent to the effective index times the vacuum wavenumber $k_0$:

$$\beta = n_{eff} k_0 \qquad (5)$$

As yet indicated, the RI change based detection method accounts for the *effective index change* but usually this method is referred to as refractive index RI change based detection method.

This detection method, widely proposed in literature, suffers from the dependence of the effective index $n_{eff}$ from the temperature, so a $\Delta n_{eff}$ can be induced not only from the variation in the concentration of the analyte to be detected, but also from a temperature instability.

Thus a thermal control consisting in choosing the appropriate material for the sensor fabrication, i.e. a polymer, or an external device for thermal stabilization should be performed when this sensing method based on RI change is applied whatever is the employed configuration, i.e. a planar waveguide or an optical resonator as it will be detailed later (Section 3.2.4).

The detection principle based on effective RI change will be described with more details in subsection 2.6.1.

## 2.5 OPTICAL DETECTION PROTOCOLS: label-free and label-based detection

Optical biosensing can be implemented mainly through two different protocols: label-free and label-based detection. The latter relies on the labeling of either target or biorecognition molecules usually with fluorescent tags or dyes, but also with radioactive molecules or plasmonic and magnetic nanoparticles. It is also named fluorescence-based detection.

This detection method could be classified as an indirect method since molecules composing the analyte are not detected in their natural form, but are structurally and functionally modified. A wide range of labels is commercially available and well-established detection techniques based on optical microscopy are experimentally used.

The label based detection is indeed widely employed when the target molecules are so tiny not to be easily detected with a conventional microscopy technique and it is the intensity of the fluorescence coming from the dyes anchored to the target molecules that is collected by the photodetector.

This technique is extremely sensitive because it can measure down to a single molecule. Some issues are anyway associated to a label-based detection which is not only time consuming, but it also results to be more expensive than label-free or direct detection method. Additional chemical steps are indeed required before transducing the device output signal which accounts for the presence (i.e. concentration) of the target molecules.

Another limitation is given by the complexity that the label adds to the reaction under analysis and, thus, by the greater risk of interference and misinterpretation of the data. Finally, since the labels are typically added in a further processing step, most of the methods label-based can provide data only at the end of the reaction and do not provide any real-time information on the reaction kinetics.

Since the research on the biosensors is driven by the need to reduce assay costs and complexity by ensuring at the same time quantitative and kinetic measurement of molecular interactions, much of the research effort in biological sensing is focused on label-free detection procedure. It relies on the employment of a sensitive layer "reactive" to some physical properties of the analyte ,

including mass, volume, electric permittivity or others. A change in one of these properties is then transduced into a signal that can be detected and measured by an appropriate instrument.

This detection protocol is easy and cheap to be performed.

## 2.6 GEOMETRIES OF INTEGRATED OPTICS BIOSENSORS

Previously, the main goals that a biosensor should reach have been pointed out including robustness and easy portability.

These properties could be assured in the optics field by employing compact platforms monolithically integrated on a chip. Here the basic configurations adopted in the design of devices for biosensing applications and realized in integrated optics are introduced. These could be merely employed in their native structure or designed in order to create more complex configurations.

These basic configurations are:

- Planar waveguides;
- Photonic crystal devices;
- Optical fiber;
- Resonant cavity devices (ring, disk, toroid, sphere configurations).

And a brief description of these is given in the followings.

### 2.6.1 PLANAR WAVEGUIDES

Planar waveguides are the simplest optical devices employed for biosensing purposes.

Light confinement in the waveguide and propagation within it can be physically explained by employing the Snell's law (known also as refraction law) or equivalently by describing the total internal reflection (TIR) physical phenomenon.

Snell's law states that when a wave impinges on an interface between two media with different indices of refraction ($n_1$ and $n_2$), the phase of the wave has to be constant on any given plane since it is valid the relation of continuity of the wave across a boundary, i.e.

$$n_1 \sin\theta_1 = n_2 \sin\theta_2 \tag{6}$$

The relation described in Eqn. (6) is known as Snell's law and it is applied to every interface that light encounters along its path. Here $n_1$ is assumed to be the refractive index of the incidence medium, $\theta_1$ the incidence angle and $n_2$ and $\theta_2$ the refractive index of the refraction medium and the refraction angle, respectively. For certain values of the incidence angle $\theta_1$, most of the light is refracted at the interfaces, a portion is instead back reflected.

The critical angle is defined as the angle for which light is not refracted at the interface between the media $n_1$ and $n_2$, with $n_1 > n_2$, but propagates along the interface (i.e. $\theta_2 = \pi/2$):

$$\theta_c = \sin^{-1}\left(\frac{n_2}{n_1}\right) \tag{7}$$

When light strikes at this interface with an angle of incidence larger than the critical angle, the total internal reflection (TIR) phenomenon occurs, i.e. light is totally back reflected at the interface. The situation of an incidence medium sandwiched between two media having lower refractive index is depicted in Fig. 4. Here we assume $n_1 > n_2$ and $n_1 > n_3$.

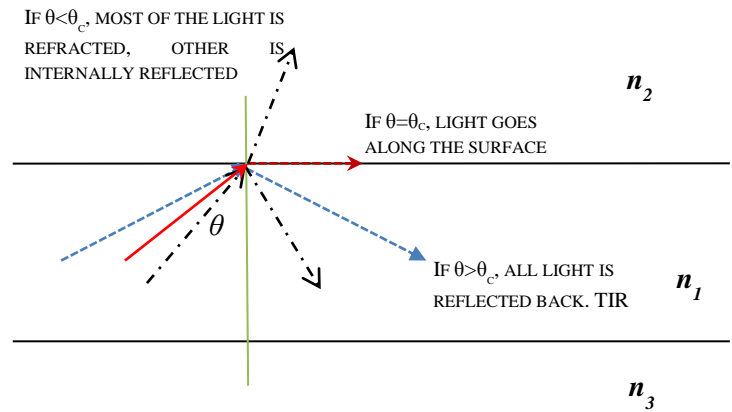

**Figure 4.** Schematic of a ray of light striking at an interface.

In Figure 4, the physical principle of TIR suited for light propagation, according to the Snell's law, is depicted. As an example, assuming $n_1 = n_{Si}$, the relations $n_{Si} > n_{air}$ and $n_{Si} > n_{SiO2}$ are satisfied in SOI technology since $n_{air}$= 1 RIU, $n_{Si}$= 3.48 RIU and $n_{SiO2}$=1.44 RIU at the operative wavelength of 1550 nm and light could propagate by TIR in the waveguide core (TIR occurs at both the interfaces).

The operative principle of planar waveguide-based biosensors depends on the detection of effective index change induced by the immobilization of bio-recognition molecules upon waveguide surface and the subsequent binding of target molecules.

Figure 5 shows the generic detection protocol (named *surface sensing*) of a RI-based optical label-free biosensor. As previously indicated, here we use both refractive index change and effective index change as synonyms.

Generally, light propagating in the sensor is confined near its surface showing an evanescent field exponentially decaying into the bulk solution. Its characteristic decay length (ranging from few tens to a few hundreds of nanometers) allows to detect the RI change induced by the analyte binding within the decay length. Those analytes that are far away from the sensing surface are indeed not captured by the biorecognition molecules and they do not interfere with the sensor response.

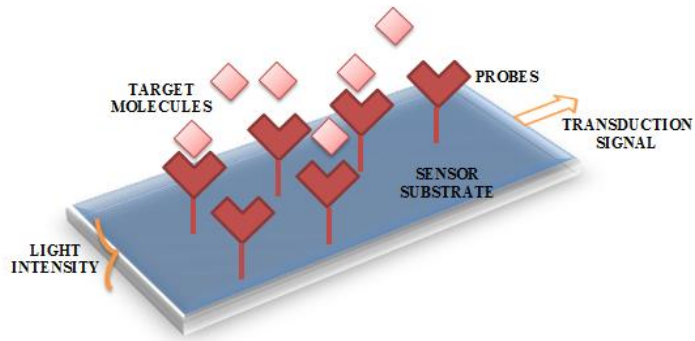

**Figure 5.** Surface sensing detection method.

The same principle of RI change detection is suited by other optical geometries employed for biosensing and is here theoretically described.

In order to analytically explore refractive index change concept, a modeling of an integrated optics (IO) sensor based on this detection scheme is plotted in Fig. 6.

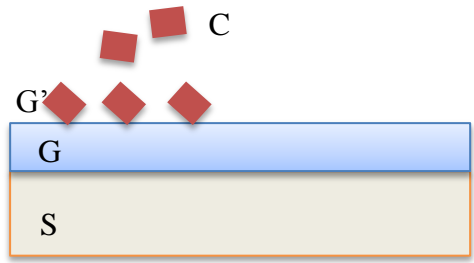

**Figure 6.** Sensing guide section.

We define C as the covering medium (a fluid), G' as the add-layer of molecules absorbed by or bounded to the waveguide surface when this latter is exposed to the medium C, G as the guiding layer and S as the substrate.

When a source of light (input wave) propagates along the guiding layer, it always shows an evanescent tail that interacts with the surrounding medium.

The penetration depth of the input wave is strictly related to the index contrast between the effective refractive index and the cover refractive index:

$$\Delta z_c = \frac{\lambda}{2\pi}\left(n_{eff}^2 - n_C^2\right) \qquad (8)$$

If the cover refractive index is increased, a change in the waveguide effective refractive index is induced and the resonance wavelength is red-shifted (i.e. shifted toward higher wavelengths).

Properly the effective refractive index change $\Delta n_{eff}$ can be induced by three different effects [10]:

a) the presence of the new layer G';
b) the variation in the covering medium refractive index $\Delta n_c$;
c) the eventual porose nature of the guiding layer G that induces a change $\Delta n_G$ in the refractive index of the same layer.

According to the above mentioned effects, it is possible to define three optical sensitivity constants strictly related to the effective refractive index change $\Delta n_{eff}$ through the formula:

$$\Delta n_{eff} = \left(\frac{\partial n_{eff}}{\partial d_{G'}}\right) d_{G'} + \left(\frac{\partial n_{eff}}{\partial n_C}\right) \Delta n_C + \left(\frac{\partial n_{eff}}{\partial n_G}\right) \Delta n_G \qquad (9)$$

where $d_{G'}$ is the add-layer thickness.

If we consider the definition of cut-off thickness of the guiding layer for TE polarized light:

$$\hat{d}_G = \frac{1}{2} \lambda \sqrt{n_G^2 - n_S^2} \left[\frac{1}{\pi} arctg\left(\frac{\sqrt{n_S^2 - n_C^2}}{\sqrt{n_G^2 - n_S^2}}\right)\right] \qquad (10)$$

it is possible to realize waveguides having a cut-off thickness (thickness that establishes the condition for which just a single mode propagates down the active region/guiding layer) really smaller than the incident wavelength only if the contrast index between the guiding layer and the substrate is big enough (> 0.3). For example, if $n_G$-$n_S$ = 0.2 RIU, the mode supported by the guide is $TE_0$ and the operative wavelength is 1550 nm, then the cut-off thickness is comparable (1376 nm) to the operative wavelength.

That is one of the reasons why Silicon on Insulator (SOI) chips (having an index contrast of more than 2 RIU at the wavelength of 1550 nm) are widely employed in optical devices design in order to realize sub-micrometric chemical and biochemical sensors.

A simultaneous detection of different analytes can be realized by creating an array of planar waveguides. This operative method is thus called "simultaneous multichannel multianalyte detection", and it is realizable also through other optical geometries [11], [12].

2.6.2 PHOTONIC CRYSTAL DEVICES

A photonic crystal device is constituted by a dielectric substrate characterized by the presence of periodic structures that, if excited by a source of light, allows for the propagation of a range of wavelengths (photonic bandgap). If the periodicity of the structure is changed (due to the introduction of a defect) and/or the refractive index contrast is modified (due to biomolecules absorption on photonic crystal surface), then the photonic bandgap is shifted. The shift is due to the effective index change experienced by the light source when propagating through it, as in planar waveguide biosensors [13].

2.6.3 FIBER OPTIC BIOSENSORS

Fiber optic biosensors are devices relying on the employment of a fiber optic device as transduction element. Light propagates along the optical fiber by suiting the principle of total internal reflection reaching the site of analysis. A signal proportional to the concentration of the target molecules bound to the receptors is then produced. If the biosensor is employed in an "extrinsic" configuration, then the optic fiber acts only as a channel useful to transmit light to the sensing elements and to receive it back. Spectroscopic techniques, including absorption, fluorescence, phosphorescence or surface plasmon resonance (SPR) are then employed to process the signal. Otherwise, the fiber could act as a sensing element since one or more of its properties change in response to the presence of the analyte. This is called intrinsic sensor configuration of the fiber optics biosensor [13].

2.6.4 OPTICAL RESONATORS

An additional class of "light-matter interaction" based biosensors is the one composed by devices suiting optical resonances. Different geometries consisting of a guiding structure (i.e. a bus waveguide or a fiber) and a circularly shaped element (ring, disk, toroid, sphere) composing both an optical cavity can be employed to realize these devices known also as optical ring resonator based biosensors or Whispering Gallery Mode (WGM) biosensors.

Their basic working principle is described in the followings. A more detailed theoretical investigation is presented in Appendix 1.

A resonant mode, excited by an input signal, is guided to propagate within the cavity due to the TIR principle. The light pulse propagating through the bus (a waveguide or fiber) couples to the circularly shaped element only when the wavelength of the exciting optical signal equals the optical length of the cavity:

$$\lambda = \frac{L n_{eff}}{m} \quad (11)$$

where $L$ is the geometrical length of the cavity and it is equal to $2\pi R$ with R the cavity outer radius. Also, $m$ is an integer number called modal number and $n_{eff}$ is the mode effective index.

The observable variable in Eqn. (11) is the resonant wavelength $\lambda$. For a specific cavity, i.e. $L$ is fixed, the $\lambda$ value is modified when an effective index change occurs. According to this, one of the mainly employed method for optical resonators based bio-chemical sensing relies on Refractive Index (RI) change. When the operative conditions of a resonator are modified in terms of concentration of molecules introduced into the bulk solution under test and interacting with the resonator, the effective index of light propagating through the medium changes as yet explained in section 2.6.1.

By evaluating the resonator output signal, i.e. the shift that the resonance wavelength experiences, it is possible to derive the effective index change according to Equations (9) and (11), i.e.

$$\Delta\lambda \propto \left(\frac{\partial n_{eff}}{\partial d_{G'}}\right)d_{G'} + \left(\frac{\partial n_{eff}}{\partial n_C}\right)\Delta n_C + \left(\frac{\partial n_{eff}}{\partial n_G}\right)\Delta n_G \quad (12)$$

This detection method leads to a sensitivity definition in terms of [length/RIU] where RIU is the Refractive Index Unit, i.e. $S = \Delta\lambda/\Delta n_{eff}$.

The above mentioned approach is also known as refractometric sensing scheme or resonant wavelength shift scheme since it relies on monitoring the $\Delta\lambda$ between the reference resonance and the resonance under sensing, as depicted in Fig. 7(a). An additional transducing method relies on monitoring the intensity variation $\Delta I$ of the reference and sensing resonance at a fixed wavelength (Fig. 7(b)).

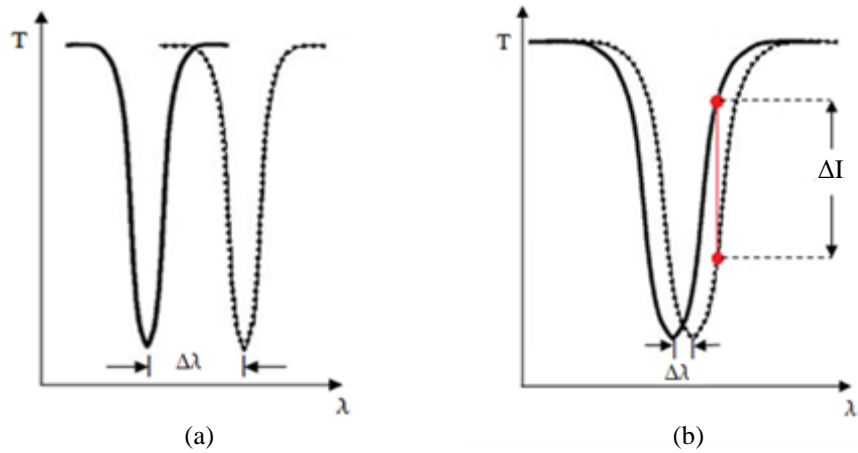

**Figure 7**. (a) Resonant wavelength shift-scheme. (b) Intensity variation-scheme.

Peculiarity of optical resonators is to enhance light-matter interaction due to resonances that basically means that light passes through the sensing area where the analyte is placed more than once. The interaction time between light and matter is thus increased and, in turn, the sensitivity is higher than the one of a simple planar waveguide where a single light-analyte interaction occurs.

As mentioned at the beginning of the present subsection, optical ring resonator based biosensors support WGMs. Briefly, these modes consists of waves propagating along the resonator outer surface and they are characterized by low reflection losses showing a very high quality factor Q value.

This figure of merit can be defined for all the resonant devices (i.e. mechanical, optical, electrical) as the ratio of the energy stored in the resonator to the energy dissipated per cycle.

In optics, the Q-factor is practically calculated as the ratio between the resonant wavelength and the 3dB resonance line width (also known as Full Width at Half Maximum, FWHM):

$$Q = \frac{\lambda}{\Delta\lambda_{3dB}} \qquad (13)$$

The Q-factor affects the device sensitivity and detection limit in both the above mentioned detection schemes, i.e. the resonant wavelength shift- and the intensity variation-scheme.

From a theoretical point of view, indeed, $\Delta\lambda$ (Eqn. (12)) should be determined only by the instrument resolution, independently from the resonance shape or bandwidth. If anyway the resonance curve is broad and some noise perturbs resonance spectra, it results really difficult to determinate the resonance wavelength shift $\Delta\lambda$. A narrower resonance (i.e. a higher Q-factor), instead, enhances the accuracy in detecting the $\Delta\lambda$ value. In the intensity variation scheme, instead, a reduction of the resonant bandwidth (i.e. a higher Q-factor) improves the spectrum slope used to define the device sensitivity which is $S = \Delta I/\Delta n_{eff}$.

The achievement of a low detection limit and a high sensitivity is one of the main goals characterizing devices to be employed for biosensing application, as yet described in Section 2.2. Besides the Q-factor, other figures have to be taken into account to achieve these features by employing an optical resonator. These are the extinction ratio (or contrast ratio) ER and the free spectral range FSR that both account for an unambiguous detection of the resonances.

To get an high contrast ratio between ON and OFF resonance allows indeed to mitigate the noise disturbance that means to increase the signal to noise ratio SNR and thus to make the resonance distinguishable. The free spectral range FSR is instead a figure defined as the distance in frequency (or wavelength) units between two adjacent resonant wavelengths of a mode propagating within the resonator. Since a real time tracking of the sensing resonance could fail if resonance spectra are too adjacent, a wide FSR is desirable for sensing applications.

If compared to waveguides, optical ring resonators offer the advantage of reducing the sensor footprint since the effective length of interaction between the light and the analyte is determined by the number of revolutions that light experiences inside the resonator (which is related to Q) and not to the sensor physical size. Despite their sub-millimeter dimensions, optical micro-resonators indeed show a Q-factor value resulting in an effective light/matter interaction of the order of centimeters.

2.7 CONCLUSIONS

An introduction to biosensing field has been here presented starting with the definition of bio-chemical sensor and the description of the principal characteristics and components a biosensor should rely on. Due to their potentialities, the attention has been focused on optical devices with a brief description of the main sensing mechanisms and platforms so far proposed in the biosensing research field. According to the properties of robustness and easy portability that a biosensor should possess, the basic configurations adopted in the design of devices for biosensing applications and realized in integrated optics have been introduced, including planar waveguides, photonic crystal devices, fiber optics and optical resonators. The mainly investigated transducing method associated to these configuration has been also described. It relies on the change of the effective index of the guided mode propagating within the optical device due to the interaction of guided light with the analyte. This method is referred to as RI change method.

In the integrated optics field, optical resonators seem to be the most promising configuration for biosensing purposes due to the high performances in terms of sensitivity and detection limit achievable with a reduced sensor footprint due to resonances.

## 2.8 REFERENCES

1. M. Trojanowicz, "Main concepts of chemical and biological sensing," in: R.A. Potyrailo, V.M. Mirsky Ed.s, "Chemical sensors and Biosensors for medical and biological applications," Springer 2009.
2. U.E. Spichiger-Kelle, "Chemical Sensors and Biosensors for Medical and Biological Applications," John Wiley & Sons, 2008.
3. A.B. Iliuk, L. Hu, W.A. Tao, "Aptamers in bioanalytical applications," Anal Chem., 83(12), pp. 4440-4452, 2011.
4. M. Campbell, "Systems for environmental monitoring. Volume One: Sensors Technologies," Blackie Academic and Professional, London, 1997.
5. P. Adam, M. Piliarik, H. Sipova, T. Springer, M. Vala and J. homola, "Surface Plasmons for Biodetections," in: Photonic Sensing: Principles and Applications for Safety and Security Monitoring, First Edition. G. Xiao and W.J. Bock Editors, John Wiley & Sons, 2012.
6. X. Fan, I.M. White, S.I. Shopova, H. Zhu, J.D. Suter and Y. Sun "Sensitive optical biosensors for unlabeled targets: a review," Analytica Chimica Acta 620, pp. 8-26, 2008.
7. P.N. Prasad, "Introduction to Biophotonics", WILEY- Interscience, 2003.
8. R. Paschotta, "Encyclopedia of Laser Physics and Technology", WILEY-VCH, 2008.
9. A. Potisatityuenyong, R. Rojanathanes, G. Tumcharern, and M. Sukwattanasinitt, "Electronic Absorption Spectroscopy Probed Side-Chain Movement in Chromic Transitions of Polydiacetylene Vesicles", Langmuir, 24 (9), pp. 4461–4463, 2008.
10. W. Lukosz, "Integrated optical chemical and direct biochemical sensors," Sensors and Actuators B, Vol. 29, pp. 37-50, 1995.

11. K.B. Gylfason, C.F. Carlborg, A. Kaźmierczak, F. Dortu, H. Sohlström, L. Vivien, C.A. Barrios, W. van der Wijngaart, and G. Stemme "On chip temperature compensation in an integrated slot-waveguide ring resonator refractive index sensor array," Optics Express, Vol. 18, No. 4, pp. 3226-3237, 2010.
12. A.L. Washburn, L.C. Gunn, and R.C. Bailey, "Label-free quantization of a cancer biomarker in complex media using silicon photonic microring resonator," Analytical Chemistry, Vol. 81, No. 22, pp. 9499–9506, 2009.
13. C. Ciminelli, C.M. Campanella, F. Dell'Olio, C.E. Campanella, and M.N. Armenise, "Label-free optical resonant sensors for biochemical applications," (in press) Progress in Quantum Electronics, 2013.

# Chapter 3.
# WHISPERING GALLERY MODE BASED BIO-CHEMICAL SENSORS

Different resonant cavity configurations have been investigated in the design of chemical and biochemical optical sensors for quantitative detection of a specific analyte or for nanoparticles detection, including ring, disk, cylinder, toroid or sphere configurations.

These are also known as whispering gallery mode (WGMs) sensors because they involve light that is guided in a way similar to the whisper heard by Lord Rayleigh in an ancient gallery located under the dome of St. Paul's Cathedral in London [1].

This is an acoustic phenomenon relying on the propagation of sound along the perimeter of a round gallery without a significant reduction of sound intensity.

At the same manner, waves composing electromagnetic modes propagating within an optical cavity are pushed to travel along the cavity outer periphery, i.e. at the interface between the resonator and the surrounding medium, due to the total internal reflection phenomenon which also produces an evanescent field in the medium outside the cavity (see Chapter 2).

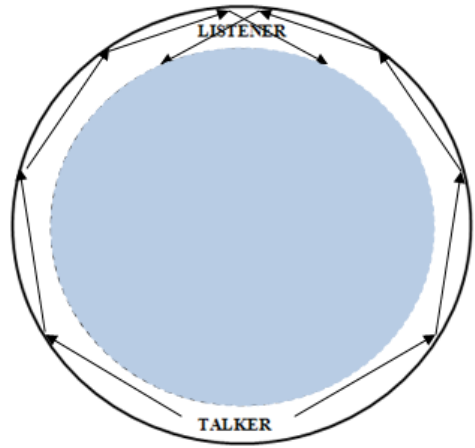

**Figure 1**. A possible path taken by the whisper.

These circular modes are excited when the light introduced into the resonator constructively interferes with itself by completing an integer number of optical cycles in the time required to make one revolution around the cavity [2], i.e. at the cavity resonant wavelength $\lambda_R$.

The physical binding of a cloud of particles (composing the fluid solution under test) to the device implies the interaction of the particles with the evanescent tail of the electromagnetic field propagating within the resonator as WGM. This gives rise to a change of the refractive index immediately around the device (i.e. a background index change) that affects the cavity resonant wavelength in terms of shift.

According to this phenomenon, WGMs based resonators are widely investigated as biosensing platforms since the analyte concentration is solution can be quantified by measuring the wavelength shift. In addition, the device sensitivity and limit of detection are improved by the high quality factor characterizing WGMs due to the low reflection losses within the cavity.

The Chapter is organized as follows: first an overview of the main results achieved in the bio-chemical sensing field by suiting WGM based devices is presented. Different configurations (i.e. planar ring resonators, liquid core ring resonators and microsphere and microtoroids) are examined in terms of sensing capabilities, such as sensitivity and limit of detection. Advantages and disadvantages of the different analyzed configuration are pointed out. Then issues concerning the design of a cavity to be employed for biosensing purposes are discussed, including the choice of the technology and the geometrical configuration in accordance with cavity loss and potential sources of noise.

## 3.1 OPTICAL RING RESONATOR CONFIGURATIONS AND PERFORMANCES OVERVIEW

### 3.1.1 PLANAR RING RESONATORS

Planar ring resonators are optical microresonant structures characterized by a structure well fixed to the chip via monolithic integration, thus assuring the properties of compactness and reproducibility and their intrinsic advantages (i.e. mass production, lowering of cost production, and reliability).

Different material systems, such as polymers [3], [4], [5], [6], glass [7], silicon nitride [8], [9], [10], silicon on insulator (SOI) [11], [12], [13], [14], have been utilized for planar ring resonator fabrication. Polymers seem to be promising for realizing disposable passive optical resonator based biosensors due to the low cost, simple fabrication techniques, wide commercial availability and biocompatibility. Respect to silicon equivalent, a reduced sensitivity has been demonstrated by employing polymers. This issue could be anyway overcome by suiting porous polymers because the internal surface area of the material provides a host medium to immobilize specific biorecognition elements [15].

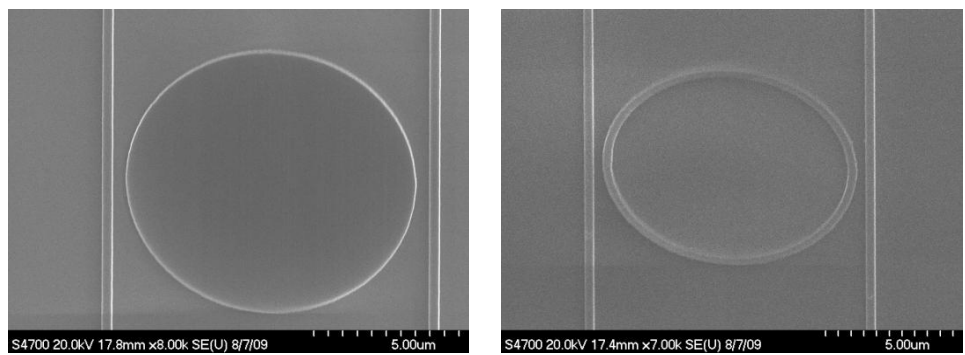

**Figure 2.** SEM images of a disk (left) and a ring (right) optical microresonator.

A widely used material system for fabricating on-chip biosensors is SOI whose main advantages are the light confinement, the high refractive index contrast, and the low propagation losses [16]. In particular, the most common silicon nanophotonic waveguides are made by silicon wires surrounded by a silicon oxide bottom cladding and by a low index top cladding (air or silicon oxide). The high index contrast between the core and the cladding allows the light to propagate in very small radius bends with negligible radiation loss.

Single bus waveguide (called two ports configuration) or two distinct bus waveguides (called four ports configuration), parallel or crossing, configurations have been employed for exciting the resonator and extract the signal from that. The coupling path between the waveguides and the resonator could be vertical [5], [7] or lateral [3].

Different planar ring resonator configurations have been proposed for sensing applications. The most common are disk [17], circular ring [7], and racetrack geometries [12], [18], but also slot-waveguides based resonators [19], [20], [21], [22], spiral [23] and SNOW [24] have been employed as biosensors

Several analytes have been investigated in order to test devices sensitivity and detection limit in terms of minimum detectable value of analyte concentration in solution (expressed either as effective index change [RIU] or as mass adsorbed per unit area [pg/mm$^2$] on sensor surface).

Due to the strength and specificity of interaction, the biotin-avidin system is not only employed as capturing probe (see Chapter 2) but it serves also as specimen useful to test a device sensitivity. Several planar ring resonators have indeed been employed as optical platforms for proteins detection, including the most common biotin-avidin (streptavidin system) [9], [18], [22], [25], [26], and bovine serum albumin (BSA) [12], [13]. Planar ring resonator configurations have also been employed as gas sensors. Ethanol [19] and acetylene [23] detection have been demonstrated with a LOD of the order of 10$^{-6}$ RIU and a sensitivity of about 500 nm/RIU, respectively. Also cytokine [27], 5-TAMRA cadaverine molecule [28], DNA hybridization [11] and toxic organic compounds such as N-methylaniline [29] have been detected by employing planar resonators. Due to the relevance of some metabolic diseases, such as diabetes, and the reduced cost associated to the sample preparation, sucrose solutions have been widely employed as analytes to be investigated [4], [30], [31], [32]. The strong and specific reaction between the enzymes glucose oxidase (GOD) and glucose (β–D-glucose) with the consequent conversion of β–D-glucose to gluconic acid is usually suited for glucose molecules immobilization. A LOD of the order of $10^{-6}$ RIU [30] and a sensitivity of few hundreds of nm/RIU [4] have been demonstrated in sugars detection.

**Table 1.**

| Analyte | Technology | Radius | Q-factor | Sensitivity | LOD | Ref. |
|---|---|---|---|---|---|---|
| Glucose/sucrose solution | $Si_3N_4$ | 1 mm | - | - | $5 \times 10^{-6}$ RIU | [30] |
| | PS | 30 $\mu$m | $5 \times 10^3$ | - | $5 \times 10^{-5}$ RIU | [32] |
| | Polymer | - | - | 200 nm/RIU | $10^{-5}$ RIU | [4] |
| | $Si_3N_4$ | 15 $\mu$m | - | 23 nm/RIU | $10^{-4}$ RIU | [8] |
| Streptavidin-biotin/avidin-biotin | PS | 45 $\mu$m | $2 \times 10^4$ | - | 250 pg/mm$^2$ | [3] |
| | $Si_xN_y$ | 2 mm | - | - | 6.8 ng/mL | [9] |
| | $SiN_x$ | - | - | 5.7 nm/nm | 30 ng/mL | [22] |
| | SOI | - | - | - | 3 pg/mm$^2$ | [25] |
| | Tb: SiON | 5 $\mu$m | 500 | 0.26 nm/nm | $10^{-7}$ RIU | [26] |
| 5-TAMRA cadaverine molecule | polymer | 180 $\mu$m | $3.5 \times 10^4$ | 23 nm/fg | 0.22 af | [5] |
| BSA | SOI | 15 $\mu$m | $4.3 \times 10^4$ | 163 nm/RIU | - | [13] |
| | SOI | 5 $\mu$m | - | - | 17 pg/mm$^2$ | [12] |
| DNA hybridization | SOI | - | - | 683 nm/RIU | - | [11] |
| N-methylaniline | - | 100 $\mu$m | $10^5$ | - | 2 nL | [29] |
| - | Si | 5 $\mu$m | - | 240 nm/RIU | 1 Å | [24] |
| - | $Si_3N_4$ | 70 $\mu$m | - | 200 nm/RIU | $2 \times 10^{-4}$ RIU | [20] |
| NaCl | Si | 5 $\mu$m | 330 | 298 nm/RIU | $4 \times 10^{-5}$ RIU | [21] |

## 3.1.2 OPTO-FLUIDIC RING RESONATORS

An additive configuration for the optical sensing is represented by liquid core optical ring resonators (LCORRs) also indicated as optofluidic ring resonators (OFRRs). The device has the unique property of acting as a resonator and as a microfluidic channel, at the same time. When an optical fiber is used to couple light to a micro-capillary, WGMs can be excited to travel around the capillary ring-section which is in contact with the optical fiber. It has been demonstrated that this micro-capillary annulus should have a size lower than 4 μm in order to excite the WGM whose evanescent tail will interact with the fluid solution flowing within the capillary [33].

In a multiplexed configuration, light is coupled to the LCORR through the employment of different optical fibers or waveguides placed in proximity of the capillary and the resonance conditions can be satisfied at each bus line.

The sensing scheme relies on the to the occurring of a light-matter interaction on the capillary inner surface enhanced by the reduced thickness of the capillary wall. An opportune surface functionalization leads to the selective detection of the target molecules contained in the solution flowing within the capillary in terms of effective index change which is transduced into a resonant wavelength shift.

This way, a compact device is realized where the capillary acts as fluidic guide and it is possible to obtain an array of micro-resonators for a multi-analyte detection by employing a single glass capillary.

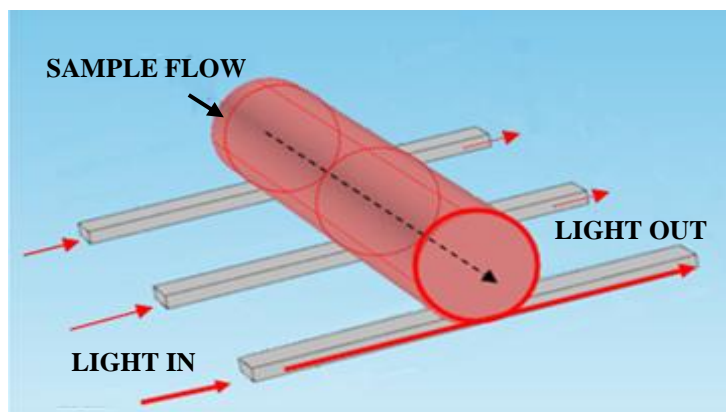

**Figure 3**. LCORR schematic view. Multiplexed configuration: a ring resonator is created in proximity of every bus waveguide

LCORRs have been employed for DNA or proteins detection showing a LOD of the order of few pg/mm$^2$ [33], [34]. Zhu et al. [35] demonstrated an opto-fluidic ring resonator based on a capillary structure for bovine-serum-albumin (BSA) detection showing a LOD of about 3 pM, corresponding to a surface density of 0.5 pg/mm$^2$ and a total mass of few femto-grams. The device response to various DNA samples was also theoretically and experimentally analyzed. The analysis demonstrated that the LCORR can detect DNA concentration down to 10 pM and a mass density detection limit of 4 pg/mm$^2$ on the surface ($10^{10}$ molecules/cm$^2$). The same group demonstrated also virus detection employing LCORRs [36]. A filamentous bacteriophage 900 nm long and 10 nm wide was used to model human pathogens. In a detection dynamic range of seven orders of magnitude ($2 \times 10^3 - 2 \times 10^{10}$ PFU/mL), a LOD of about $2 \times 10^3$ PFU/mL corresponding to a mode shift of a few picometers has been detected. Even with a very low virus concentration, the circular nature of the capillary enables an efficient capture and the detection of the analyte.

Suter et al. [34] also employed LCORRs to detect DNA. They experimentally analyzed the response of a LCORR to a variety of DNA samples having different strand lengths (25-100 bases), number of base- mismatches (1-5), and concentrations (10 pM to 10 μM) in order to evaluate the LCORR sequence detection capability. The device has been demonstrated to be sensitive enough to differentiate DNA with only a few base-mismatches based on the raw sensing signal and kinetic analysis. A detection of 10 pM bulk DNA concentration was demonstrated with a sensitivity of about 37 nm/RIU.

Detection of the HER2 breast cancer biomarker in human serum samples has also been demonstrated by employing an opto-fluidic ring resonator [37]. Results have shown that the OFRR is able to detect HER2 at medically relevant concentrations in serum ranging from 13 to 100 ng/mL in 30 min.

Recently, a quantum dot-based "fluorescent-core microcapillary" (FCM) sensor has been proposed by Manchee et al. [38]. A layer of fluorescent silicon quantum dots (QDs) is indeed employed to coat the channel inner surface. Due to the high effective index of the QD layer, the electric field is confined near the capillary channel and WGM resonances appears in the fluorescence spectrum as peaks, rather than dips in the laser transmission spectrum. The detection method is then based on reading the resonances shift in the fluorescence mode. The device showed an experimental refractometric sensitivity close to 10 nm/RIU and a LOD of the order of $10^{-3}$ RIU when a sucrose solution is employed as testing fluid. A proposed potential way to increase both this figures is related to the optimization of the QD film thickness.

Besides the classical configuration, known as optofluidic cylindrical configuration, recently bottle and bubble-shaped optofluidic ring resonators have been investigated. Li et al. [39] investigated silica micro-bubble resonators for single particle detection. The theoretical LOD has been estimated to be less than 20 nm in terms of smallest detectable particle radius. Scaling the size of detectable particle implies the possibility of detecting different kind of small viruses otherwise not detectable by employing photonic devices. Experimental results are related to gas sensing, i.e. ethanol have been demonstrated in [40] showing a LOD of $10^{-6}$ RIU. Promising results are also presented in [41] where single and double ports micro-bubble are reported.

The main advantage coming from the employment of optofluidic geometries is that they directly incorporate the fluidic function into the sensor itself, avoiding this way all the issues related to microchannel-optical device on chip integration. Their sensing capability is really high, while their LOD is comparable to the one obtained by employing planar microresonator as sensing platforms.

**Table 2.**

| Analyte | Technology | Radius | Sensitivity | LOD | Ref. |
|---|---|---|---|---|---|
| BSA | $SiO_2$ | 50 $\mu$m | 50nm/RIU | $10^{-7}$ RIU | [35] |
| virus | $SiO_2$ | - | - | $2 \times 10^3$ pfu/mL | [36] |
| DNA | $SiO_2$ | - | 37 nm/RIU | 10 pM | [34] |
| | $SiO_2$ | - | - | 4 pg/mm$^2$ | [33] |
| HER2 breast cancer biomarker | $SiO_2$ | 72 $\mu$m | 30 nm/RIU | 1 u/mL | [37] |
| Sucrose Solution | $SiO_2$+ Si QDs | - | 10 nm/RIU | $10^{-3}$ RIU | [38] |
| - | $SiO_2$ | - | - | R=20 nm | [39] |
| Testing liquids | $SiO_2$ | 106 $\mu$m | 800 nm/RIU | - | [42] |

### 3.1.3 MICROSPHERE RING RESONATORS

Also microsphere and microtoroidal ring resonators have been investigated as sensors to be employed in the field of biomedical and environmental monitoring. Since these devices show a very high Q-factor which has been experimentally demonstrated to be >$10^9$ in air and >$10^6$ in water [43], their sensitivity and LOD are widely improved with respect to planar ring resonator configurations. Due to these features, the detection of single virus and air pollutants [44], [45], [46] has been demonstrated experimentally in a label-free and label-based detection scheme. The detection method is based on the transduction of the effective index change (and thus of a resonant wavelength shift), or on other observable phenomena including the resonance splitting due to a modal coupling caused by the scattering of light induced by a local RI change.

Microtoroidal resonators have been demonstrated to be able to detect the presence of single metal atoms (i.e. Cesium atoms) [47] or to detect and size a single nanoparticle or virus [48], [49]. An exhaustive analysis of their optical response has been presented by Spillane et al. [50] who investigated microtoroid potentiality and advantages with respect to spherical resonators. Their smaller modal volume has been demonstrated to enhance device sensitivity.

Lu et al. [48] proposed a microtoroid resonator in aqueous solution (Q = $10^8$) for single NPs detection. In their work, they have been able to detect a single polystyrene bead with a radius of 12.5 nm (a size approaching the single protein molecule) and InfA virion by employing a thermal-stabilized reference interferometer in conjunction with an ultrahigh-Q microcavity.

Armani and Vahala [51] demonstrated the detection of a single species in a mixture of chemically similar molecules by employing a toroidal microcavity. $D_2O$ in $H_2O$ mixtures has been detected by employing a high Q resonator (Q = $10^7$) and by monitoring the change of the Q factor value when the resonators is immersed in $D_2O$ rather than $H_2O$. This way, concentrations of 1 part in $10^6$ per volume of $D_2O$ in $H_2O$ have been detected. By employing a silicon toroidal microcavity, the same Group [52] realized a highly specific and sensitive optical sensor based on an ultrahigh quality factor (Q > $10^8$) cavity. Label-free, single-molecule detection of interleukin-2 was demonstrated in serum, with a dynamic range of about $10^{12}$ in concentration. A concentration of interleukin-2 ranging from $10^{-19}$ M to $10^{-6}$ M in solution has been indeed tested in order to evaluate the microtoroid dose response.

The first experimental demonstration of a microsphere biosensor is reported by Vollmer et al. [53]. A 300 μm diameter silica glass spheroidal cavity with a Q factor of $2 \times 10^6$ was indeed used to detect both non-specific binding of BSA and specific binding of streptavidin. The device response saturates for BSA concentrations as low as 20 nM, representing the upper limit of the dynamic range. By sensitizing the surface of a quartz microsphere by chemical modification, the adsorption of BSA on the surface has been demonstrated also by Arnold et al. [54].

Sensors based on microspheres were also proposed for detecting protease enzyme [55], [56]. Hanumegowda et al. [55], proposed a biosensor based on a 100 μm radius sphere for the BSA/trypsin pair detection. After a primary attachment of BSA molecules to the sphere surface and the subsequent WGM spectral position shift, the interaction of BSA with trypsin gives rise to a cleaving operation of amino acids from the BSA, i.e. to a BSA etching, and to a shift of WGM resonance in the opposite direction. From the measurement of this shift, the existence of proteolytic activity has been detected to be at the level of 10 pg/ml or $10^{-4}$ Units/ml within 15 minutes which corresponds to the device LOD.

Zhu et al. [56] proposed a 125 μm radius microsphere for thrombin detection. The sphere surface have been modified with anti-thrombin aptamer, which has excellent binding affinity and selectivity in order to detect thrombin levels in a buffered solution. A LOD of about 1 NIH Unit/mL*, corresponding roughly to 6 nM has been experimentally demonstrated.

Another important application of microsphere sensors is the detection of single beads such as NPs, proteins or viruses.

Vollmer et al. [57] demonstrated label-free DNA quantification with a high sensitive device composed by two silica microspheres evanescently coupled to a common optical fiber. Measurements are realized by employing first a single sphere and then both. The binding on the sphere surface of polarizable DNA material gave rise to a detectable spectroscopic shift of the resonance with a limit mass loading of only 6 pg/mm$^2$ of DNA material which corresponds to the device LOD. The signals from the two spheres allowed for a single nucleotide mismatch detection with a signal-to-noise ratio of 54 with the advantage of removal of common mode noise in experiments due to a differential measurements with the two spheres.

Single-base-mismatched DNA was also detected by using 7.5 μm silica microspheres coated with dye labeled DNA probes [58]. The DNA targets were 5 μM using lengths between 10 and 40 nucleotides. The same type of experiment was proposed also with polystyrene spheres with 1.5-20 μm diameters for testing BSA solutions, achieving a LOD of about 213 pg/mm$^2$.

Low detection limits of a few femtograms have also been demonstrated by employing silica microspheres in low-Q limit [59].

As microtoroidal resonators, also microspheres have been employed to detect large bio-particles, like bacteria and viruses, through simple adsorption effect [60], [61], [62]. Ren et al. [60] theoretically and experimentally investigated the detection of bacteria through a resonant microsphere. They model the bacterial cell of E.coli as a cylindrical cell with its axis aligned parallel to the surface of the cavity upon binding and they experimentally measured the shift of resonance

---

wavelength and broadening of line-width of the WGM propagating within the cavity induced by the adsorption. The detection limit of the setup has been estimated to be ~$1.2 \times 10^2$ E.coli/mm$^2$, which corresponds to ~44 bacteria bound to the ~0.36 mm$^2$ total area of the sphere. Also MS2 bacteriophage has been detected by employing a resonant cavity sensor based on a micro-sphere [61]. Keng et al. [61] indeed in their work propose an oblate spheroid to detect infectious agents as MS2 bacteriophage, i.e. a positive-sense single-stranded RNA virus infecting the Escherichia coli bacterium. A concentration of $10^9$ PFU/mL was sensed with a detection limit of 2 pg/mm$^2$ for surface density.

Single influenza A virus particle has been detected by using 50 μm silica microsphere [62]. The discrete change in frequency of whispering gallery modes (WGMs) propagating within the microsphere cavity when Influenza A virions bind to it has been observed. Both the virus size and mass have been identified. The proposed detection scheme is a classical one and it relies on the employment of a tunable laser whose output signal is guided in tapered optical fiber to excite a WGM along the microsphere equator by evanescent coupling. The transmitted light T is then collected by a photo-detector where dips are visible at the resonant wavelengths.

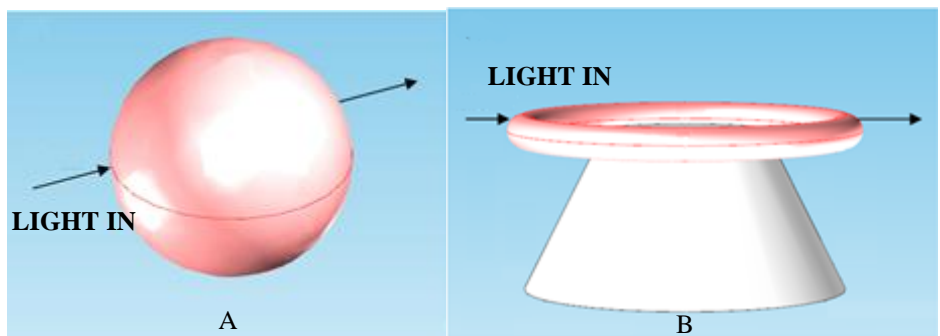

**Figure 4**. (A) Microsphere and (B) microtoroid optical resonator.

A silica microsphere biosensor functionalized with alkylsilane self-assembled monolayers (SAMs) and integrated in a microfluidic flow cell under laminar flow conditions has been also proposed to detected the adsorption of protein onto alkysilane [63]. Recognition elements are indeed needed for a specific detection goal when a complex sample such as blood plasma is employed as buffer solution.

---

* PFU = plaque forming units. It is a measure of concentration used in virology.     

Recently, a 200 μm diameter silica microsphere has been proposed [64] to study the behavior of fluorescent molecules embedded in biomaterials, in terms of photobleaching.

Silica has been replaced by other materials as germanium [65], bismuth-silicate [66] or chalcogenide glass [67] to study physical phenomena useful to the environmental sensing. Ahmad et al. [67] investigated indeed a chalcogenide microsphere resonator for temperature sensing.

Environmental sensing has been performed also by Lin et al. [68] that proposed a theoretical study concerning a polymer-coated silica microsphere resonator showed a refractive index detection limit of $2 \times 10^{-8}$ in chemical vapor sensing. An advantage coming from this configuration is that the employed polymer (PMMA) has a thermo-optic coefficient which is opposite in sign to the one of silica, reducing this way temperature induced resonance drifts.

Several mechanisms for sensitivity enhancement in WGM biosensing have been investigated recently. The Vollmer group [69], [70] demonstrated that sensitivity for single-molecule detection can be greatly improved by boosting the field intensity at the analyte binding site without reducing significantly the resonator Q-factor value. The proposed method consists in suiting a hybrid photonic–plasmonic mode produced by immobilizing gold nanoparticles within the evanescent field of a microsphere cavity. The WGM propagating along the equator of microsphere interact evanescently with a layer of Au NPs, thus exciting plasmon resonances at several sites in the layer. This way an array of intensity hotspots is generated. Its sensitivity is really high at the hotspots locations where proteins binding induces a resonance shift related to the Plasmon-coupled WGM wavelength.

Despite their impressive sensing properties, microspheres suffers from reproducibility issues due to the procedure followed for their fabrication. They are indeed fabricated by melting the tip of a standard single mode fiber with a $CO_2$ laser and the controllability of their final dimensions is not well established. Thus a mass production of biosensors based on microresonant spheres is a challenge by now.

The same considerations are still valid for micro-torus structure where the torus realized with a $CO_2$ laser reflow process is supported by a silicon pedestal. Technological steps to be followed for the fabrication of a toroidal microresonator are more complex than the one associated with a micro-sphere. The preliminary creation of a silica disk on the substrate (Si) by employing standard lithographic and etching techniques is followed by the employment of an isotropic $XeF_2$ etch in order to undercut the silica pads. This way, a silica disk is suspended on a silicon pillar. A $CO_2$ laser reflow process is then used to create the final toroidal structure having a high-Q value ($>10^8$ in water and air) [48]. Recently an improvement in fabrication procedure has been achieved. A microtoroid has indeed been fabricated by deforming an optical microsphere in order to fabricate a high performance optical microdisk resonators [71]. The improved surface quality, the easy handling and alignment due to the flattened shape, makes this resonator a realistic candidate for the developing of a high performance integrated bio-photonic device.

**Table 3.**

| Analyte | Technology | Radius | Q-factor | LOD | Ref. |
|---|---|---|---|---|---|
| Cs | $SiO_2$ | 22 $\mu$m | $10^8$ | Single atom | [47] |
| Interleukin-2 | $SiO_2$ | 40 $\mu$m | $10^8$ | Single molecule | [52] |
| PS | $SiO_2$ | - | $>10^8$ | *R=30 nm | [49] |
| PS | $SiO_2$ | - | $>10^8$ | *R=12.5 nm | [48] |
| PS | $SiO_2$ | 50 $\mu$m | $10^6$ | *R=140 nm | [44] |
| Inf A virion | $SiO_2$ | - | $>10^8$ | *R=50 nm | [48] |
| Inf A virion | $SiO_2$ | 50 $\mu$m | - | Single binding | [62] |
| BSA | $SiO_2$ | 300 $\mu$m | $2\times10^6$ | - | [53] |
| BSA | PS | 10 $\mu$m | - | 213pg/mm$^2$ | [58] |
| $D_2O$ | $SiO_2$ | - | $10^7$ | - | [51] |
| Thrombin | $SiO_2$ | 125 $\mu$m | - | 1 NIH unit/mL | [56] |
| Trypsin | $SiO_2$ | 100 $\mu$m | - | 10 pg/mL | [46] |
| DNA | $SiO_2$ | 200 $\mu$m | - | 6 pg/mm$^2$ | [57] |
| Ammonia | Polymer coated $SiO_2$ | 50 $\mu$m | - | $2\times10^{-8}$ | [68] |

*R = radius of the single detected element.

A comparative analysis of label free optical resonant sensors highlights their capability to detect small amount of analyte in solution (of the order of few pg/mm$^2$ in terms of mass LOD or $10^{-8}$ RIU in terms of RI LOD). Appreciable results have been obtained by employing planar configurations realized in silicon technology, despite their Q-factor value lower than the one of other WGMs based resonators. In order to bypass issues related to the integration of the resonant device with the fluidic platform designed to drive the testing solution toward the resonator, LCORRs have been proposed as more compact devices showing appreciable sensitivity values. Very high quality factor resonators such as microspheres or microtoroids have been demonstrated for the detection of the event of binding of a single molecule (a virus or a nanoparticle) to the resonator surface in label-free detection scheme with a LOD of few tens of nanometers in terms of radius of the detected particle (or virus, molecule or other).

Despite their high potentialities, the market penetration of ultra-high-Q resonators is limited by several factors, including reproducibility, possibility of on-chip integration and thus portability.

The obtained experimental results based on whispering gallery mode sensing provide, anyway, addressable challenges toward the employment of label free resonant devices for bio-chemical sensing.

## 3.2 CAVITY DESIGN: DISCUSSION

### 3.2.1 GUIDING STRUCTURE

Silicon on insulator (SOI) technology is one of the most employed configuration for planar cavity realization.

In SOI structures a strong light confinement can be realized and a single mode operation, i.e. only one guided mode for each polarization is supported by the guide, can be reached by employing a waveguide cross-section dimension in the nanometer range.

A multimode operation is indeed undesirable since it can cause effects such as intermodal dispersion or optical signal distortion that degrade waveguide performances.

Other commonly employed materials for the realization of guiding structures in the field of biochemical sensing are polymers and Silicon compounds such as Silicon Nitride ($Si_xN_y$) as mentioned in Section 3.1.1.

Here we anyway will empathize SOI characteristics since this is the technology employed in the present study.

Single-mode operation in waveguides depends on waveguide dimensions, polarization state of light (TE-like or TM-like mode), and refractive index of materials composing the waveguide. If we focus the attention on an SOI asymmetric waveguide composed by a Silicon wire located on a Silicon Oxide substrate, the dimensions of the wire (its width and height) are usually chosen by considering SOI wafers commercially available in terms of Si layer thickness, whose typical values are h = 200 nm, h = 220 nm, and h = 250 nm, and then by considering the achievement of a single-mode operation.

Fixing the wire height, the mono-modality condition can be easily verified by employing a finite element method (FEM) based software or a beam propagation software.

The mono-modality condition in this study has been verified for waveguide dimensions (w,h) = (450, 220), (480, 220), (500, 220)nm by assuming a refractive index value of 3.48 for the guiding layer (Si) and 1.44 for the substrate ($SiO_2$) at the incident wavelength of 1550 nm for both TE and TM polarized light.

Beyond the mono-modality condition, an important figure of merit to be considered in integrated optics biological sensing is the waveguide molecular sensitivity, defined as [72]

$$S_{mol} = \frac{\partial n_{eff}}{\partial D_{opt}} \quad (1)$$

This is properly a figure characterizing *guided mode molecular sensors* and accounts for the dependence of the sensor response on the refractive index and thickness of the add-layer bound to the waveguide surface (see Chapter 2). $D_{opt}$ is a term known as optical thickness and it is defined as the product of the index of refraction and the thickness of the molecular layer adsorbed onto the device surface. As an example, for a quasi-TE mode propagating within a waveguide, molecular sensitivity data are reported in Table 4.

**Table 4.**

| WAVEGUIDE WIDTH [nm] | MOLECULAR SENSITIVITY [pm$^{-1}$] |
|---|---|
| 450 | 1.739 |
| 480 | 1.652 |
| 500 | 1.481 |

According to the specific application, the cross-section of the Si wire could be selected by considering the molecular sensitivity of the waveguide. Data of Table 4 have been collected for a molecular layer thickness of 10 nm and refractive index of 1.45 RIU and for an SOI waveguide having a 220 nm height.

3.2.2 COUPLER SECTION

One of the most important step in the design of a cavity is related to the study of the coupler section in order to control the amount of energy transferred from the input guiding structure (a fiber or a waveguide) to the ring structure considered in its general acceptation.
An appropriate choice of the coupling parameters enables indeed the achievement of the desired Q-factor and ER within the cavity.
As reported in Section 3.1.1, coupling schemes employed so far are based on a vertical or lateral coupling. Peculiarities of both the approaches are related to fabrication steps. In vertical coupling, the exposure of the waveguide to surface treatments is prevented and the coupling gap is optimized by easily control the waveguide-ring gap. On the other side, anyway, multiple and complicated fabrication steps are required. Laterally side-coupled approach instead suffers from low fabrication tolerance errors in defining fine waveguide-resonator gap with the advantage, anyway, of employing a reduced number of fabrication steps.

When a cavity based on microsphere configuration needs to be excited, instead, a moving stage allows to place a tapered fiber close to the equator of the sphere where fundamental modes are confined [73].

Despite the different approaches, evanescent coupling techniques (i.e. fiber based and waveguide based) should satisfy three fundamental conditions:

- a spatial overlap between the evanescent field decaying outside the waveguide core and the mode in the resonator should exist;
- the wavelength (frequency) of the exciting mode should be resonant with a target WGM of the resonator;
- a phase matching between the resonator mode and the waveguide mode should exist.

The strength of coupling is also determined by the interaction length between the resonator and the waveguide.

This can be easily controlled by employing a resonator in racetrack configuration. The racetrack is an extension of ring structure and it is composed by two directional couplers interconnected via two half portions of ring, as depicted in Fig. 5.

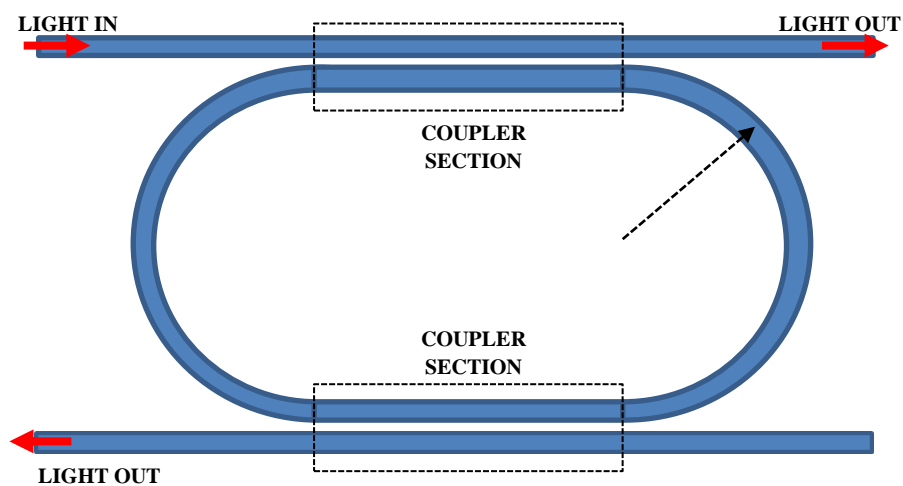

**Figure 5.** Racetrack configuration

The length of cavity $L_{cav}$ is given by the sum of the length of the two rectilinear tracts and the ring circumference, while the amount of tolerable phase mismatch is inversely related to the coupling (or interaction) length $L_c$. The phase mismatch is indeed zero when the coupling length equals the beat length $L_{beat}$. The beat length or cross over length expression is:

$$L_{beat} = \frac{\pi}{2\Delta\beta} \tag{2}$$

and it corresponds to that length z travelled by light that allows power to be totally transferred from one to the other waveguide of the directional coupler. $\beta$ is the propagation constant and its gradient accounts for $\beta$ of the two waveguides composing the directional coupler and could be calculated by employing a FEM based software.

The $E_y$ mode profile for a quasi- TE mode propagating in an SOI waveguide in air in depicted in Fig. 6 (A), while in Fig. 6 (B) the modal profile at the coupler section.

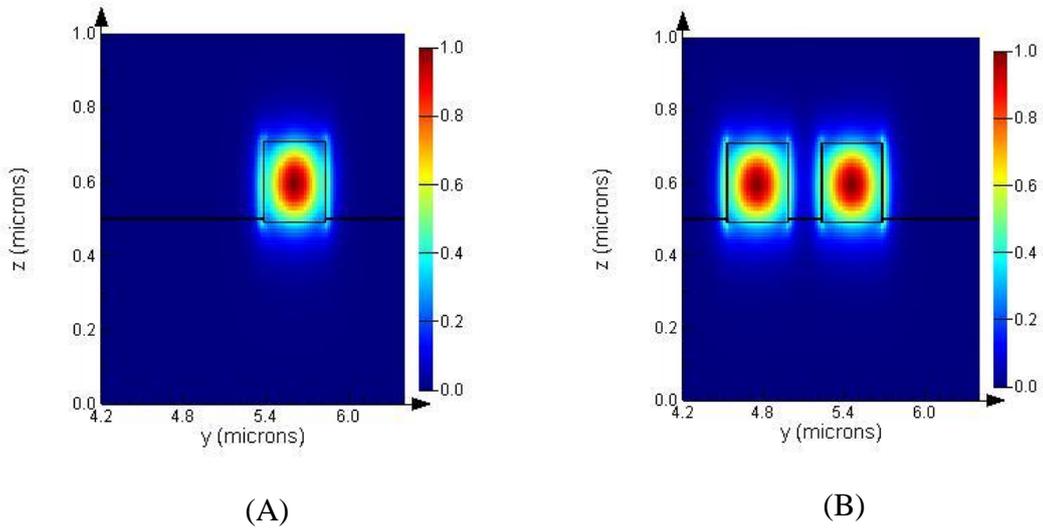

(A)                          (B)

**Figure 6.** Mode profile at the input waveguide (A) and at the coupler section (B).

Analytically, the total transfer of power corresponds to:

$$\left.\frac{P_2}{P_1}\right|_{z=L_{beat}} = \sin^2(k^* L_{beat}) = 1 \tag{3}$$

where $k^*$ is the coupling per length and it is expressed in [μm$^{-1}$]. This condition of total transfer of power is satisfied if k*$L_{beat}$ = π/2 or equivalently:

$$k^* = \frac{\pi}{2L_{beat}} \tag{4}$$

For a generic amount of power transfer, another figure can be introduced. It is the coupling efficiency $K$:

$$0 \leq K = \left.\frac{P_2}{P_1}\right|_{z=L_c} = \sin^2(k^* L_c) \leq 1 \tag{5}$$

The term $L_c$ is the coupling length associated with a fixed efficiency $K$, that means:

$$L_c = \frac{\sin^{-1}(\sqrt{K})}{k^*} \tag{6}$$

In Fig. 7, the coupling efficiency trend versus the coupling length is plotted for different coupling gap values for a Si wire cross-section of (450, 220) nm.

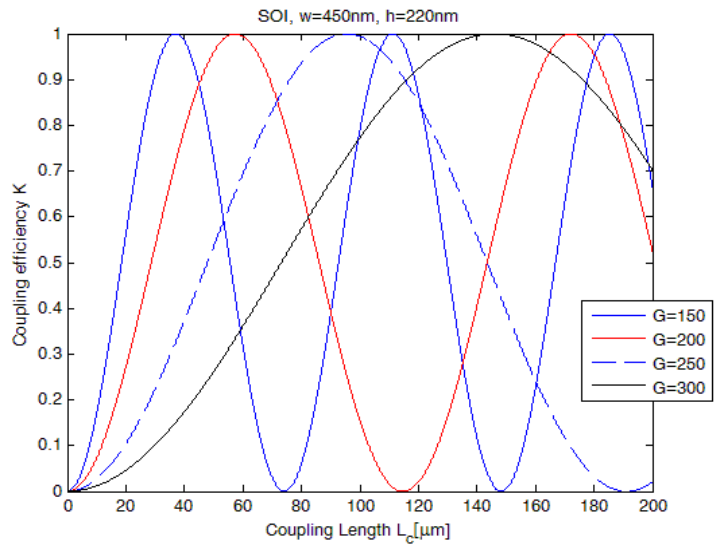

**Figure 7.** Coupling efficiency curves as function of coupling length and coupler gap. Data have been plotted by assuming an SOI technology in Si wire configuration, with wire dimensions (w, h) = (450, 220) nm.

A precise control of the coupling efficiency can be obtained by employing a racetrack configuration through the manipulation of the length of the straight section. Although, to increase the straight section length implies a reduction of cavity FSR that could be detrimental for biosensing application. Since FSR is dependent on $L_{cav}$, a possible way to overcome this issue is to increase the curvature of bend section, i.e. to reduce the ring radius. This way bending loss increases and the resonator could experience a strong reduction of its quality factor due to radiation loss.

Also, an excessive reduction of racetrack radius gives rise to a modal mismatch between the straight and the bent portions of the cavity that could anyway be controlled by introducing a geometrical off-set at the connection side.

The description of coupling so far presented is related to the application of the Coupled Mode Theory (CMT) in the space domain. The same approach can be employed for classical ring resonators [74]. For more details, see Appendix 1.
An additive coupling regime analysis can be developed in the time domain [75], [76].

Here we briefly present the fundamentals of the Coupled Mode Theory in the time domain since a similar analysis will be performed later in Chapter 5 by taking in consideration this classical theory.

*3.2.2.1 MODAL COUPLING IN TIME DOMAIN*

Usually the theoretical investigation of the coupling between a mode and a travelling wave resonator is described by employing the model of a RLC circuit or a spring-mass –damper system, i.e. through a second order differential equation. By following a rigorous formalism, indeed, this second order differential equation can be solved by obtaining two first order uncoupled differential equations associated with the positive and negative frequency component of the mode amplitude, that are conjugate. Due to this relation, only the positive frequency component $a_+$ of the mode amplitude can be employed to fully describe the system. It will be indicated as *a* in the following and it is normalized so that its squared magnitude coincides with the energy W in the system (i.e. the resonator):

$$|a|^2 = W \qquad (7)$$

The temporal response of the isolated system could be written as:

$$\frac{da}{dt} = \left(i\omega_0 - \frac{1}{\tau_0}\right)a \qquad (8)$$

In Eqn. (8), *i* is the imaginary unit, $\omega_0$ is the mode frequency according to the dependence $exp(i\omega_0 t)$ and $1/\tau_0$ is the mode decay rate due to loss with dependence $exp(-t/\tau_0)$. This loss is usually indicated as internal loss and defines the resonator inverse unloaded or internal quality factor $Q_0$ though the formula

$$\frac{1}{Q_0} = \frac{2}{\tau_0 \omega_0} \qquad (9)$$

*3.2.2.1.1 RESONATOR-WAVEGUIDE COUPLING*

When the resonator is coupled to an external waveguide, some phenomena has to be considered. Two degenerate traveling wave modes, i.e. propagating at the same frequency but in opposite direction, i.e. a clockwise (CW) $a_{CW}$ and counterclockwise (CCW) $a_{CCW}$ mode, coexist within the resonator. If a source of light is propagating within the waveguide in forward direction that we assume to be the same of the CW mode of the resonator, the coupling between the forward and the CW mode can occur with a rate *k*.

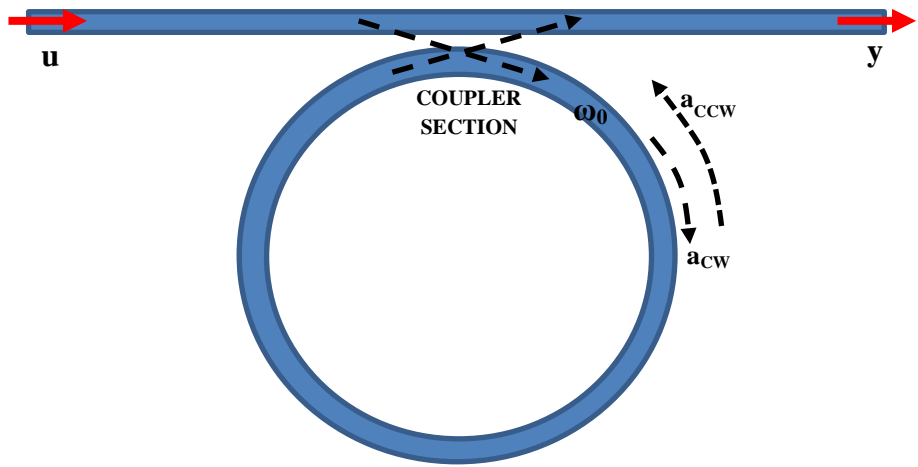

**Figure 8.** Schematic of the coupling between a single waveguide and a WGM based resonator.

A strong phase mismatch exists between the forward mode of the waveguide indicated as *a* and the CCW mode of the resonator and their coupling rate is so weak to be considered negligible. In the coupling region, indeed, they propagates in opposite direction.

Thus, the time-domain coupling (in the 1$^{st}$ order approximation) can be expressed through the following set of equations:

$$\frac{da}{dt} = \left( i\omega_0 - \frac{1}{\tau_0} - \frac{1}{\tau_e} \right) a + ku \tag{10}$$

$$y = u - k^* a \tag{11}$$

$$\frac{1}{Q_e} = \frac{2}{\tau_e \omega_0} \tag{12}$$

where $y$ is the system output; $1/\tau_e$ is the mode decay rate due to the external loss associated with the resonator-waveguide coupling and defining the external quality factor $Q_e$; $k$ is the resonator-source coupling rate which is dependent on the time constant $\tau_e$:

$$k = \sqrt{\frac{2}{\tau_e}} \tag{13}$$

Solving the first order differential equation in frequency domain in steady state regime, assuming an *exp(iωt)* dependence for the input signal *u*, the mode amplitude is:

$$a(\omega) = \frac{ku(\omega)}{i\Delta\omega + \frac{1}{\tau_0} + \frac{1}{\tau_e}} \tag{14}$$

Where

$$\Delta\omega = \omega - \omega_0 \tag{15}$$

Thus, the system transfer function $T=|y/u|^2$ is:

$$T = \left| 1 - \frac{k^2}{i\Delta\omega + \frac{1}{\tau_l}} \right|^2 \tag{16}$$

That could be written as function of the cavity quality factor $Q_l$ (*l* stands for loaded) which is usually indicated simply as $Q$:

$$T = \left|1 - \frac{k^2}{i\Delta\omega + \frac{\omega_0}{2Q_l}}\right|^2 \qquad (17)$$

$$\frac{1}{Q_l} = \frac{1}{Q_0} + \frac{1}{Q_e}$$
$$\frac{1}{\tau_l} = \frac{1}{\tau_0} + \frac{1}{\tau_e} \qquad (18)$$

*3.2.2.1.2 INTERNAL MODAL COUPLING*

When a perturbation occurs along light path, the two degenerate CW and CCW modes propagating within the resonator could couple each-other. The orthogonality condition, indeed, inherits the possibility for CW and CCW modes to exchange energy each-other. The perturbation could be due to the presence of roughness on the resonator sidewall or to the presence of any kind of defect inducing scattering of light within the cavity. If this cavity modes inter-coupling happens, the CW and CCW mode amplitude is described by the following set of equations:

$$\frac{da_{CW}}{dt} = \left(i\omega_0 - \frac{1}{\tau_0} - \frac{1}{\tau_e}\right)a_{CW} + iga_{CCW} + ku \qquad (19.a)$$

$$\frac{da_{CCW}}{dt} = \left(i\omega_0 - \frac{1}{\tau_0} - \frac{1}{\tau_e}\right)a_{CCW} + ig^*a_{CW} \qquad (19.b)$$

By solving Equations (19) in steady state regime, the following expressions are found:

$$a_{CW}(\omega) = \frac{ku(i\Delta\omega + \frac{1}{\tau_l})}{(i\Delta\omega + \frac{1}{\tau_l})^2 + |g|^2};$$

$$a_{CCW}(\omega) = \frac{ig^* a_{CW}}{(i\Delta\omega + \frac{1}{\tau_l})}$$

(20)

Consequently, the system output $y$ and the system transfer function $T$ expressions are:

$$y = u\left(1 - |k|^2 \left(\frac{\frac{1}{2}}{\left((i\Delta\omega + \frac{1}{\tau_l}) + |g|\right)} + \frac{\frac{1}{2}}{\left((i\Delta\omega + \frac{1}{\tau_l}) - |g|\right)}\right)\right)$$

(21)

$$T = \left|1 - \frac{1}{2}|k|^2 \left(\frac{1}{\left((i\Delta\omega + \frac{1}{\tau_l}) + |g|\right)} + \frac{1}{\left((i\Delta\omega + \frac{1}{\tau_l}) - |g|\right)}\right)\right|^2$$

(22)

If compared with Eqn. (16), the system transfer function identified by Eqn. (22) clearly demonstrates that two distinct dips now exist instead of a single resonance. This phenomenon is known as resonance mode splitting and it visible in the resonator spectrum as a doublet centered around the $\omega_0$ frequency (characterizing the resonator in absence of inter-modal coupling), and shifted of a quantity $\pm|g|$ with respect to $\omega_0$. That means that the splitting doublet should appear as symmetric with respect to the $\omega_0$ frequency (see Fig. 9).

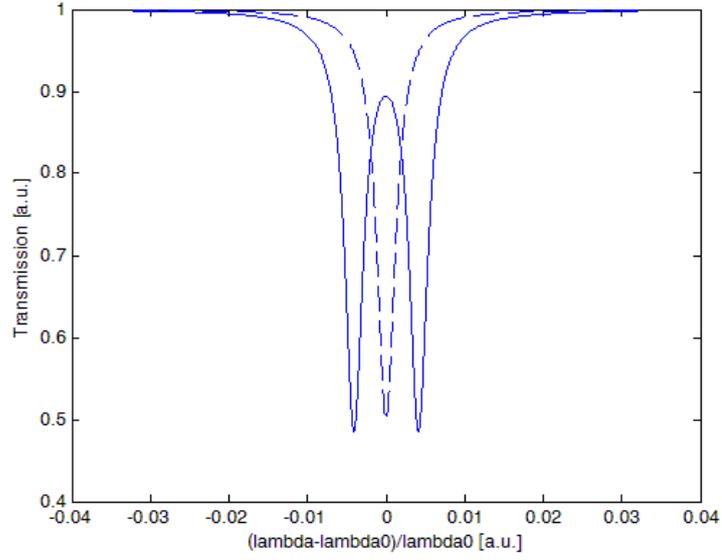

**Figure 9**. (Dotted line) Resonance dip in absence of inter-modal coupling. (Continuous line) Splitting doublet symmetrically shifted of |g| from the central wavelength $\lambda_0$.

To solve the state equation of Equations (19), a matrix form has been employed:

$$\dot{A} = FA + GU$$

$$\begin{bmatrix} \dot{a}_{CW} \\ \dot{a}_{CCW} \end{bmatrix} = \begin{bmatrix} f_{11} & f_{12} \\ f_{21} & f_{22} \end{bmatrix} \begin{bmatrix} a_{CW} \\ a_{CCW} \end{bmatrix} + \begin{bmatrix} k \\ 0 \end{bmatrix} u$$

$$f_{11} = f_{22} = i\omega_0 - \frac{1}{\tau_l}$$

$$f_{12} = ig$$

$$f_{21} = ig^*$$

(23)

Where $g$ and $g^*$ are the inter-mode coupling rate and its conjugate, respectively; $\tau_l$ is the total mode decay rate within the cavity as indicated in Eqn. (18), and $F$ is the characteristic matrix of Eqn. (21). Its eigenvalues are calculated from the characteristic equation $\det(F-\lambda I) = 0$ where $I$ is the identity matrix:

$$\lambda_{1/2} = i\omega_0 - \frac{1}{\tau_l} \pm i|g| \qquad (24)$$

While the eigenvectors are:

$$V_{1/2} = \frac{1}{\sqrt{2}} \begin{bmatrix} \frac{g}{|g|} \\ \pm 1 \end{bmatrix} \tag{25}$$

Equation (23) can now be written by introducing two new variables, $A_+$ and $A_-$:

$$A_+ = \frac{\frac{|g|}{g} a_{CW} + a_{CCW}}{\sqrt{2}}$$

$$A_- = \frac{\frac{|g|}{g} a_{CW} - a_{CCW}}{\sqrt{2}} \tag{26}$$

in order to obtain a set of uncoupled equations:

$$\dot{A} = FA + GU$$

$$\begin{bmatrix} \dot{A}_+ \\ \dot{A}_- \end{bmatrix} = \begin{bmatrix} \lambda_1 & 0 \\ 0 & \lambda_2 \end{bmatrix} \begin{bmatrix} A_+ \\ A_- \end{bmatrix} + \frac{1}{\sqrt{2}} \begin{bmatrix} \frac{gk}{|g|} \\ \frac{gk}{|g|} \end{bmatrix} u \tag{27}$$

$$\lambda_1 = i\omega_0 - \frac{1}{\tau_l} + i|g|$$

$$\lambda_2 = i\omega_0 - \frac{1}{\tau_l} - i|g|$$

Where the $A_+$ and $A_-$ represent the amplitude of two modes not coupled each-other as resulting from the value (null) of the off-diagonal elements.

Their phase variation is no more depending on an exponential term, but on a cosine and sine term for the $A_+$ and $A_-$, respectively, conferring them a standing wave (SW) nature in contrast to the travelling wave (TW) nature of the native uncoupled $a_{CW}$ and $a_{CCW}$ modes.

These generated standing wave modes could anyway be characterized by a different intrinsic lifetime, i.e. they could have different quality factors and so a different line-width.

In Chapter 5 we will investigate this splitting doublet arising from the CW and CCW cavity modes coupling and we will show that a more realistic prediction of cavity transmission spectrum in presence of a perturbative source can be given by employing quantum electrodynamics (QED) theory. The modal splitting also will have not a symmetric behavior with respect to the unperturbed resonance. This concept is depicted in Fig. 10 where a fixed splitting strength has been assumed to characterize different cavities (everyone identified by a Q-factor value).

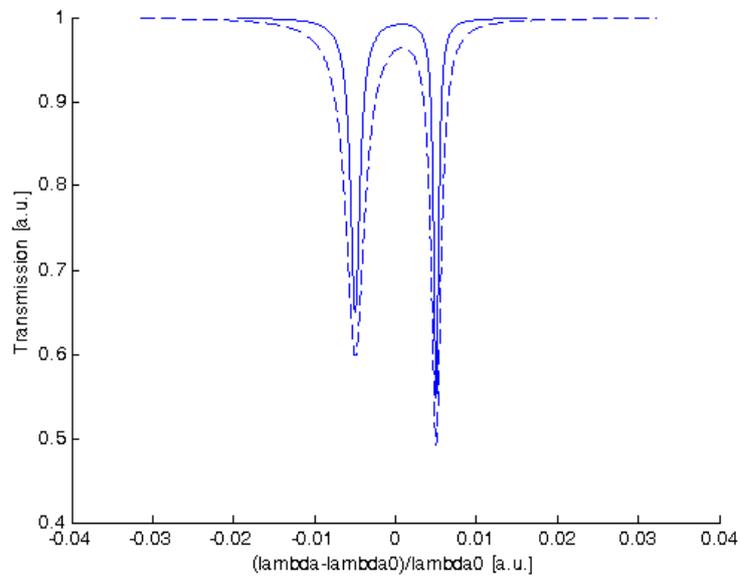

**Figure 10**. Asymmetric splitting of resonance wavelength. Every resonance is characterized by a different life-time.

3.2.3 LOSS SOURCES

*3.2.3.1 INTERNAL LOSS*

Internal losses are also indicated as propagation losses and they account for the loss experienced by the resonator in absence of an external load (i.e. not in a coupling regime). The internal Q-factor defined in Eqn. (9) accounts for these losses.

*3.2.3.2 BENDING OR RADIATION LOSS*

As the name suggest, these losses affect bent waveguides since light propagating within them experiences a shift of the mode toward the outer edge of the waveguide. This could induce either a mode radiation, or an enhancement of losses due to increased interaction of the mode with the sidewall surface roughness, or reflections due to phase mismatch between the mode in the bend waveguide and the one in the straight section of the coupler.

Bending losses $\alpha_r$ are usually described as having a simple exponential dependence on the bend radius R [77]:

$$\alpha_r = Ce^{-kR} \tag{28}$$

Where *C* depends on optical and geometrical waveguide characteristics, i.e. the thickness and refractive index of the waveguide core and the cladding, while *k* depends on the modal effective index $n_{eff}$ and propagation constant *β*. The exponential decay behavior of bending loss leads to the calculation of the critical radius, i.e. the bent radius for which the loss trend reaches its linear regime value. For a TE-like polarized mode and an SOI strip waveguide with a (500, 220) nm cross-section, the critical radius value has been calculated to be 3μm with bending loss corresponding to 0.16 dB/cm as reported in Table 5. Obtained data are in good correspondence with experimentally measured values reported in literature.

**Table 5.**

| R [μm] | Bending Losses [dB/cm] | Bending losses per round trip [dB] |
|---|---|---|
| 3 | 0.16391 | 3.0896e-4 |
| 5 | 0.046958 | 1.4752e-4 |
| 10 | 0.004598 | 2.889e-5 |

Data have been calculated by employing a commercial mode solver and do not account for surface sidewall roughness.

*3.2.3.3 MODAL MISMATCH LOSS*

This loss is due to the mode mismatch at the transition between the straight and curved section of a racetrack resonator and could be minimized by employing a junction offset. It is function of the bending radius of the curved section. When indeed the bending radius decreases below a certain value corresponding usually to the resonator critical radius, the modal mismatch at the junctions between the straight and curved waveguide becomes the dominant loss mechanism in the cavity. To avoid this loss, classical circular-shaped rings with no straight sections should be used if the goal is to realize resonators having a really small footprint.

3.2.4 NOISE SOURCES

Several noise sources are encountered when the designed cavity is experimentally tested, including photodetector noise and light source intensity fluctuation in the bandwidth range of interest. Usually these phenomena are well described by assuming white noise acting on the system.
An additional noise source for optical resonator based biosensor is the temperature fluctuation since it leads to a random variation of the resonant wavelength. This phenomenon can be alleviated by taking into account the thermo-optic properties of the sensor. It is possible to estimate the device capability of stabilizing or compensating the temperature fluctuation according to the formula [78]:

$$\Delta n_T \propto \left(\frac{dn}{dT}\right)_{dev} \cdot \Delta T_{sys} \qquad (29)$$

Where $\Delta n_T$ expresses the refractive index resolution limited by temperature noise, *dn/dT* is the thermo-optic coefficient of the material constituent the cavity and $\Delta T_{sys}$ is the average value of temperature fluctuation when a compensation/stabilization scheme is employed. In order to reduce the $\Delta n_T$ value, different approaches could be adopted.

It is possible to provide a compensation for the temperature variations of the system $\Delta T_{sys}$ by placing a reference channel in close proximity to the sensor device. As emphasized by Hu et al. [78], the main requirement and also lack of this technique is that the reference channel and the sensor device should have at the same time identical thermo-optic characteristics and a distinctive response to the analyte solution, which is quite difficult to be realized. An additional way to reduce the sensor temperature dependence relies on reducing the thermo-optic coefficient *dn/dT*. That means to choose a technology for the resonator fabrication relying on a dielectric material with low thermo-optic coefficients placed over semiconductors for compensation. Most dielectric materials have a positive thermo-optic coefficient ranging from of $10^{-6}$ to $10^{-4}$ $K^{-1}$ that could be compensated by embedding the dielectric in a negative thermo-optic coefficient material such as a polymer [79]. Also water has a negative thermo-optic coefficient of $-10^{-4}$ $K^{-1}$ and thus sensors working in an aqueous environment result to be insensitive to temperature fluctuations. Otherwise, a combination of materials having thermo-optics coefficients with opposite signs should be employed to realize an athermal design of the sensor. When operating in water, anyway, the dielectric material chosen for the sensor design should keep high optical confinement in the interrogation volume, i.e. a high RI sensitivity.

## 3.3 CONCLUSIONS

An overview of WGMs based bio-chemical sensors has been presented. Planar configurations, optofluidic resonators, microspheres and microtoroids resonant cavities have been investigated in terms of performances. Advantages and disadvantages of the different analyzed configuration have been pointed out. Then issues concerning the design of a cavity to be employed for biosensing purposes have been also discussed. Some expedients useful to improve biosensor performances have been described, including an opportune choice of the resonator geometry and technology in order to overcome issues related to bending loss, thermal noise and other.

3.4 REFERENCES

1. Lord Rayleigh, "The problem of the whispering gallery", Scientific Papers, Cambridge University Press, 1912.
2. X. Lopez-Yglesias, J.M. Gamba, and R.C. Flagan, "The physics of extreme sensitivity in whispering gallery mode optical biosensors," J. Appl. Phys., Vol. 111, Issue 8, 2012.
3. C.Y. Chao, W. Fung, and L.J. Guo, "Polymer microring resonators for biochemical sensing applications", IEEE J. Sel. Topics Quantum Electron., 12(1), pp. 134-142, 2006.
4. G-D. Kim, G-S. Son, H-S. Lee, K-D. Kim, S-S. Lee, "Integrated photonic glucose biosensor using a vertically coupled microring resonator in polymers", Opt. Commun., 281(18), pp. 4644-4647, 2008.
5. C. Delezoide, M. Salsac, J. Lautru, H. Leh, C. Nogues, J. Zyss, M. Buckle, I. Ledoux-Rak and C. T. Nguyen, "Vertically Coupled Polymer Microracetrack Resonators for Label-Free Biochemical Sensors", IEEE Photonics Technol. Lett., 24(4), pp. 270-272, 2012.
6. L. Wang, V. Kodeck, S. Van Vlierberghe, lun Ren, lie Teng, X. Han, X. lian, R. Baets, G. Morthier and M. Zhao, "A Low Cost Photonic Biosensor Built on a Polymer Platform," Proc. of SPIE-OSA-IEEE Asia Communications and Photonics, 8311(22), pp. 1-6, 2011.
7. A. Yalcin, K.C. Popat, J.C. Aldridge, T.A. Desai, J. Hryniewicz, N. Chbouki, B.E. Little, O. King, V. Van, S. Chu, D. Gill, M. Anthes-Washburn., M.S. Ünlü, and B.B. Goldberg, "Optical sensing of biomolecules using microring resonators", IEEE J. Sel. Topics Quantum Electron., 12(1), pp. 148-155, 2006.
8. E. Krioukov, D.J.W. Klunder, A. Driessen, J. Greve, and C. Otto, "Sensor based on an integrated optical microcavity", Opt. Lett., 27(7), pp. 512–514, 2002
9. A. Ksendzov and Y. Lin, "Integrated optics ring-resonator sensors for protein detection", Opt. Lett., 30(24), pp. 3344-3346, 2005.

10. L. Stern, I. Goykhman, B. Desiatov, and U. Levy, "Frequency locked micro disk resonator for real time and precise monitoring of refractive index", Optics Letters, 37(8), pp. 1313-1315, 2012.

11. X. Li, Z. Zhang, S. Qin, T. Wang, F. Liu, M. Qiu and Y. Su, "Sensitive label-free and compact biosensor based on concentric silicon-on-insulator microring resonators", Applied Optics, 48(25), F90-F94, 2009.

12. K. De Vos, J. Girones, S. Popelka, E. Schacht, R. Baets, and P. Bienstman, "SOI optical microring resonator with poly(ethylene glycol) polymer brush for label-free biosensor applications", Biosens. Bioelectron., 24(8), pp. 2528-2533, 2009.

13. M. Iqbal, M.A. Gleeson, B. Spaugh, F. Tybor, W.G. Gunn, M. Hochberg, T. Bachr-Jones, R.C. Bailey, and L. Cary Gunn, "Label-free Biosensor Arrays Based on Silicon Ring Resonators and High-Speed Optical Scanning Instrumentation", IEEE J. of Selected Topics in Quantum Electronics, 16(3), pp. 654-661, 2010.

14. M.S. Luchansky, A.L. Washburn, M.S. McClellan, and R.C. Bailey, "Sensitive on-chip detection of a protein biomarker in human serum and plasma over an extended dynamic range using silicon photonic microring resonators and sub-micron beads", Lab Chip, 11(12), pp. 2042–2044, 2011.

15. M. Mancuso, J. M. Goddard, and D. Erickson, "Nanoporous polymer ring resonators for biosensing", Optics Express, 20(1), pp. 245-255, 2012.

16. X. Fan, I.M. White, S.I. Shopova, H. Zhu, J.D. Suter, and Y. Sun, "Sensitive optical biosensors for unlabeled targets: A review", Analytica Chimica Acta, 620(1-2), pp. 8-26, 2008.

17. R.W. Boyd and J.E. Heebner, "Sensitive disk resonator photonic biosensor", Appl. Optics, 40(31), pp. 5742–5747, 2001.

18. K. De Vos, I. Bartolozzi, E. Schacht, P. Bienstman, and R. Baets, "Silicon-on-Insulator microring resonator for sensitive and label-free biosensing," Optics Express, 15(12), 2007.

19. K.B. Gylfason, C.F. Carlborg, A. Kazmierczak, F. Dortu, H. Sohlström, L. Vivien, C.A. Barrios, W. Van der Wijngaart, and G. Stemme, "On-chip temperature compensation in an integrated slot-waveguide ring resonator refractive index sensor array," Optics Express, 18(4), 2010.

20. A. Barrios, K.B. Gylfason, B. Sánchez, A. Griol, H. Sohlström, M. Holgado, and R. Casquel, "Slot-waveguide biochemical sensor", Opt. Lett., 32(21), pp. 3080-3082, 2007.

21. T. Claes, J.G. Molera, K. De Vos, E. Schacht, R. Baets, and P. Bienstman, "Label-free biosensing with a slot-waveguide-based ring resonator in silicon on insulator", IEEE J. Photonics, 1(3), pp. 197-204, 2009.

22. S. Lee, S.C. Eom, J.S. Chang, C. Huh, G. Yong Sung, and J.H. Shin, "Label-free optical biosensing using a horizontal air-slot SiNx microdisk resonator", Optics Express, 18(20), pp. 20638-20644, 2010.

23. T.J. Robinson, L. Chen and M. Lipson, "On-chip gas detection in silicon optical microcavities," Optics Express, 16(6), 2008.

24. M. Khorasaninejad, N. Clarke, M. P. Anantram, and S. Singh Saini, "Optical bio-chemical sensors on SNOW ring resonators", Optics Express, 19(18), pp. 17575-17584, 2011.

25. D.X. Xu, M. Vachon, A. Densmore, R. Ma, S. Janz, A. Delge, J. Lapointe, P. Cheben, J. Schmid, E. Post, S. Messaoudène, and J.M. Fédéli, "Real-time cancellation of temperature induced resonance shifts in SOI wire waveguide ring resonator label-free biosensor arrays", Optics Express, 18(22), pp. 22867–22879, 2010.

26. H. Jeong, S. Lee, G.Y. Sung, and J.H. Shin, "Design and Fabrication of Tb3+-Doped Silicon Oxy-Nitride Microdisk for Biosensor Applications", IEEE Photonic Technology Letters, 23(2), pp. 88-90, 2011.

27. M.S. Luchansky and R.C. Bailey, "Silicon photonic microring resonators for quantitative cytokine detection and T-cell secretion analysis," Analytical Chemistry, 82(5), 2010.

28. C. Delezoide, M. Salsac, J. Lautru, H. Leh, C. Nogues, J. Zyss, M. Buckle, I. Ledoux-Rak and C. T. Nguyen, "Vertically Coupled Polymer Microracetrack Resonators for Label-Free Biochemical Sensors", IEEE Photonics Technol. Lett., 24(4), pp. 270-272, 2012.

29. A. Nitkowski, L. Chen, and M. Lipson, "Cavity-enhanced on-chip absorption spectroscopy using microring resonators", Optics Express, 16(16), pp. 11930-11936, 2008.
30. H. Sohlström and M. Öberg, "Refractive index measurement using integrated ring resonators", in Proc. of 8th European Conference on Integrated Optics, pp. 322-325, 1997.
31. E. Krioukov, D.J.W. Klunder, A. Driessen, J. Greve, and C. Otto, "Sensor based on an integrated optical microcavity", Optics Letters, 27(7), pp. 512–514, 2002
32. C.Y. Chao and L.J. Guo, "Biochemical sensors based on polymer microrings with sharp asymmetrical resonance", Appl. Phys. Lett., 83(8), pp. 1527-1529, 2003.
33. X. Fan, I.M. White, H. Zhu, J.D. Suter and H. Oveys, "Towards lab-on-a-chip sensors with liquid-core optical ring resonators," 2007, SPIE Newsroom. DOI: 10.1117/2.1200612.0530.
34. J.D. Suter, I.M. White, H. Zhu, H. Shi, C.W. Caldwell and X. Fan, "Label-Free Quantitative DNA Detection using the Liquid Core Optical Ring Resonator," Biosensors and Bioelectronics, 23(27), 2008.
35. H. Zhu, I.M. White, J.D. Suter, P.S. Dale and X. Fan, "Analysis of biomolecule detection with optofluidic ring resonator sensors," Optics Express, 15(15), 2007.
36. H. Zhu, I.M. White, J.D. Suter, M. Zourob, and X. Fan, "Opto-fluidic micro-ring resonator for sensitive label-free detection", Analyst, 132(1), pp. 356-360, 2008.
37. J.T. Gohring, P. S. Dale and X. D. Fan, "Detection of HER2 breast cancer biomarker using the opto-fluidic ring resonator biosensor", Sens. Actuator B-Chem., 146(1), pp. 226–230, 2010.
38. C.P.K. Manchee, V. Zamora, J.W. Silverstone, J.G.C. Veinot and A. Meldrum, "Refractometric sensing with fluorescent-core microcapillaries", Optics Express, 19(22), pp. 21540-21551, 2011.

39. H. Li, Y. Guo, Y. Sun, K. Reddy and X. Fan, "Analysis of single nanoparticle detection by using 3-dimensionally confined optofluidic ring resonators", Optics Express, 18(24), pp. 25081-25088, 2010.
40. S. Berneschi, D. Farnesi, F. Cosi, G. Nunzi Conti, S. Pelli, G. C. Righini, and S. Soria, "High Q silica microbubble resonators fabricated by arc discharge", Optics Letters, 36(17), 2011.
41. R. Henze, T. Seifert, J. Ward and O. Benson, "Tuning whispering gallery modes using internal aerostatic pressure", Optics Letters, 36(23), 2011.
42. M. Sumetsky, R.S. Windeler, Y. Dulashko, and X. Fan, "Optical liquid ring resonator sensor", Optics Express, 15(22), pp. 14376-14381, 2007.
43. H.K. Hunt and A.M. Armani, "Label-free biological and chemical sensors," Nanoscale, Vol. 2, pp. 1544–1559, 2010.
44. S. Arnold, D. Keng, S.I. Shopova, S. Holler, W. Zurawsky and F. Vollmer, "Whispering gallery mode carousel – a photonic mechanism for enhanced nanoparticle detection in biosensing," Optics Express, 17(8), 2009.
45. F. Vollmer, "Optical microresonators: label-free detection down to single viral pathogens," SPIE Newsroom, DOI: 10.1117/2.1201002.002619, 2010
46. N.M. Hanumegowda, I.M. White and X. Fan, "Aqueous mercuric ion detection with microsphere optical ring resonator sensors," Sensors and Actuators B, 120(1), 2006.
47. T. Aoki, B. Dayan, E. Wilcut, W.P. Bowen, A.S. Parkins, T.J. Kippenberg, K.J. Vahala and H.J. Kimble, "Observation of strong coupling between one atom and a monolithic microresonator," Nature, 443, 2006.
48. T. Lu, H. Lee, T. Chen, S. Herchak, J.H. Kim, S.E. Fraser, R.C. Flagand and K.J. Vahala, "High sensitivity nanoparticle detection using optical microcavities," Proc. Natl. Acad. Sci. U S A. 108(15):5976-9, 2011.
49. J. Zhu, S.K. Ozdemir, Y.F. Xiao, L. Li, L. He, D.R. Chen, L. Yang, "On-chip single nanoparticle detection and sizing by mode splitting in an ultrahigh-Q microresonator," Nature Photonics, Vol.4, 2010.

50. S.M. Spillane, T.J. Kippenberg, K.J. Vahala, K.W. Goh, E. Wilcut, H.J. Kimble, "Ultrahigh-Q toroidal microresonators for cavity quantum electrodynamics," Physical Review A 71, 013817, 2005.

51. A.M. Armani and K.J. Vahala, "Heavy water detection using ultra-high-Q microcavities", Optics Letters, 31(12), pp. 1896-1898, 2006.

52. A.M Armani, R.P. Kulkarni, S.E. Fraser, R.C. Flagan, and K.J. Vahala, "Label-free, single-molecule detection with optical microcavities", Science, 317, pp. 783-787, 2007.

53. F. Vollmer, D. Braun, A. Libchaber, M. Khoshsima, I. Teraoka, and S. Arnold, "Protein detection by optical shift of a resonant microcavity", Appl. Phys. Lett., 80(21), pp. 4057-4059, 2002.

54. S. Arnold, M. Khoshsima, I. Teraoka, S. Holler, and F. Vollmer, "Shift of whispering-gallery modes in microspheres by protein adsorption", Optics Lett., 28(4), pp. 272-274, 2003.

55. N.M. Hanumegowda, I.M. White, H. Oveys, and X. Fan, "Label-free protease sensors based on optical microsphere resonators", Sensor Lett., 3(4), pp. 315-319, 2005.

56. H. Zhu, J.D. Suter, I.M. White, and X. Fan, "Aptamer Based Microsphere Biosensor for Thrombin Detection", Sensors, 6(8), pp. 785-795, 2006.

57. F. Vollmer, S. Arnold, D. Braun, I. Teraoka, and A. Libchaber, "Multiplexed DNA quantification by spectroscopic shift of two microsphere cavities", J. Biophys., 85(3), pp. 1974-1979, 2003.

58. E. Nuhiji and P. Mulvaney, "Detection of unlabeled oligonucleotide targets using whispering gallery modes in single, fluorescent microspheres", Small, 3(8), pp. 1408-1414, 2007.

59. A. Weller, F.C. Liu, R. Dahint, and M. Himmelhaus, "Whispering gallery mode biosensors in the low-Q limit", Appl. Phys. B, 90(3-4), pp. 561-567, 2008.

60. H.-C. Ren, F. Vollmer, S. Arnold, and A. Libchaber, "High-Q microsphere biosensor - analysis for adsorption of rodlike bacteria", Opt. Express, 15(25), pp. 17410-17423, 2007.

61. D. Keng, S.R. Mc Ananama, I. Teraoka, and S. Arnold, "Resonance fluctuations of a whispering gallery mode biosensor by particles undergoing Brownian motion", Appl. Phys. Letters, 91(10), 103902(1-3), 2007.

62. F. Vollmer, S. Arnold and D. Keng, "Single virus detection from the reactive shift of a whispering-gallery mode," P. Natl. Acad. Sci. USA, 105(52), pp. 20701-20704, 2008.

63. K.A. Wilson, C.A. Finch, P. Anderson, F. Vollmer, and J.J. Hickman, "Whispering Gallery Mode Biosensor Quantification of Fibronectin Adsorption Kinetics onto Alkylsilane Monolayers and Interpretation of Resultant Cellular Response" Biomaterials, 33(1), pp. 225-236, 2012.

64. L.M. Freeman and A.M. Armani, "Photobleaching of Cy5 Conjugated Lipid Bilayers Determined With Optical Microresonators", IEEE Journal Of Selected Topics In Quantum Electronics, Vol. 18, No. 3, 2012.

65. P. Wang, T. Lee, M. Ding, A. Dhar, T. Hawkins, P. Foy, Y. Semenova, Q. Wu, J. Sahu, G. Farrell, J. Ballato, and G. Brambilla, "Germanium microsphere high-Q resonator", Optics Letters, Vol. 37, No. 4, 2012.

66. P. Wang, G. Senthil Murugan, T. Lee, M. Ding, G. Brambilla, Y. Semenova, Q. Wu, F. Koizumi and G. Farrell, "High-Q Bismuth-Silicate Nonlinear Glass Microsphere Resonators", IEEE Photonics Journal, Vol. 4, No. 3, 2012.

67. H Ahmad, I Aryanfar, K S Lim, W Y Chong and S W Harun, "Thermal response of chalcogenide microsphere resonators", Quantum Electronics, Vol. 42, No. 5, 2012.

68. N. Lin, L. Jiang, S. Wang, Q. Chen, H. Xiao, Y. Lu, and H. Tsai, "Simulation and optimization of polymer-coated microsphere resonators in chemical vapour sensing", Applied Optics, 50(28), 5465-5472, 2011.

69. M. Baaske and F. Vollmer, "Optical Resonator Biosensors: Molecular Diagnostic and Nanoparticle Detection on an Integrated Platform," Chem. Phys. Chem., 13(2), pp. 427 – 436, 2012.

70. M.A. Santiago-Cordoba, S.V. Boriskina, F.Vollmer, and M.C. Demirel, "Nanoparticle-based protein detection by optical shift of a resonant microcavity," Appl. Phys. Lett., 99(7), 073701(1-3), 2011.

71. M.N. Zervas, G. Senthil Murugan, J.S. Wilkinson, "Optical Microdiscus Resonators", Conference on Lasers and Electro-Optics (CLEO), San Jose, CA, USA, 06-11 May 2012

72. S. Janz, A. Densmore, D.-X. Xu, P. Waldron, J. Lapointe, J. H. Schmid, T. Mischki, G. Lopinski, A. Delâge, R. McKinnon, P. Cheben, and B. Lamontagne, "Silicon photonic wire waveguide sensors" in: Advanced Photonic Structures For Biological And Chemical Detection, X. Fan Editor, Integrated Analytical Systems, pp. 229-264, Springer 2009.

73. T.J. Kippenberg, S.M. Spillane, D.K. Armani, B. Man, L. Yang and K.J. Vahala, "Coupling and Nonlinear Optics of Ultra-High-Q Micro-Sphere and Chip-Based Toroid Microcavities," in: "Optical Microcavities," K. Vahala Editor, Advanced Series In Applied Physics Vol. 5, World Scientific Publishing Co., Singapore, 2004.

74. A. Yariv and P. Yeh, "Optical electronics in modern communications," The Oxford Series in Electrical and Computer Engineering, 6$^{th}$ edition.

75. H.A. Haus, "Waves and fields in optoelectronics," Prentice-Hall, 1984.

76. S.L. Chuang, "Physics of optoelectronics devices," Wiley Series in Pure and Applied Optics, USA, 1995.

77. Y.A. Vlasov and S.J. McNab, "Losses in single-mode silicon-on-insulator strip waveguides and bends," Optics Express, Vol. 12, No. 8, 2004.

78. J. Hu, X. Sun, A. Agarwal, and L.C. Kimerling, "Design guidelines for optical resonator biochemical sensors," J. Opt. Soc. Am. B 26, pp. 1032-1041, 2009.

79. N. Lin, L. Jiang, S. Wang, H. Xiao, Y. Lu, and H. Tsai, "Thermostable refractive index sensors based on whispering gallery modes in a microsphere coated with poly(methyl methacrylate)," Applied Optics, Vol. 50, Issue 7, pp. 992-998, 2011.

# Chapter 4.
# NANOPARTICLES DETECTION

Bio-chemical sensors, according to the IUPAC definition, are a class of sensors suited not only for detection of biological analytes, but also for chemical sensing purposes including *nanoparticles detection*. There particles are so called since they have at least one dimension equal or even smaller than 0.1 micrometer.

Due to several processes, natural and industrial, fine and ultrafine particles production is increasing more and more and the research area concerning these particles detection and sizing is assuming a big relevance in bio-sensing field.

Several sectors of civil industry, indeed, ranging from pharmaceutical to environmental to alimentary one, are conditioned by nanoparticles (NPs) in a positive and negative meaning.

NPs results to be really attractive in pharmaceutical industry since their relatively large (functional) surface can be suited to bind, adsorb and carry other compounds such as drugs, probes and proteins [1].

Despite their employment as delivery systems, they also can act as dangerous factors for human health. Thus, a figure called *therapeutic ratio (or index)* expressing the margin between the dose needed for clinical purpose and the dose inducing adverse side effects on human health should be considered when NPs are forced to enter human body for therapeutic purposes.

Majority of data concerning the biological behavior and toxicity of NPs are related to the unintended release of NPs by combustion and the subsequent inhalation by humans.

In these studies, it has been demonstrated that NPs show a higher ability to enter, translocate within and damage living organisms than microparticles; their small size allows them to penetrate physiological barriers and travel within the circulatory and lymphatic system of a host, reaching tissues and organs and potentially disrupting cellular processes and causing disease [2].

Their nature, i.e. chemical composition, is different and it is modifying with technological advancement giving rise also to a change of character of the particulate pollution.

Some studies showed a strong dependence between particulate air pollution levels and diseases interesting not only the respiratory apparatus, but also the cardiovascular one [1], [3].

Nanoparticles potential toxicity depends on different factors that should be considered not separately, including NPs nature, size, exposure time the body is subjected to, in-vivo accumulation and bioactivity [4].

When referring to particulate air pollution, usually nanoparticles are termed fine and ultrafine particles or dusts.

In next figure, most common tiny nanoparticles are reported with their characteristic size.

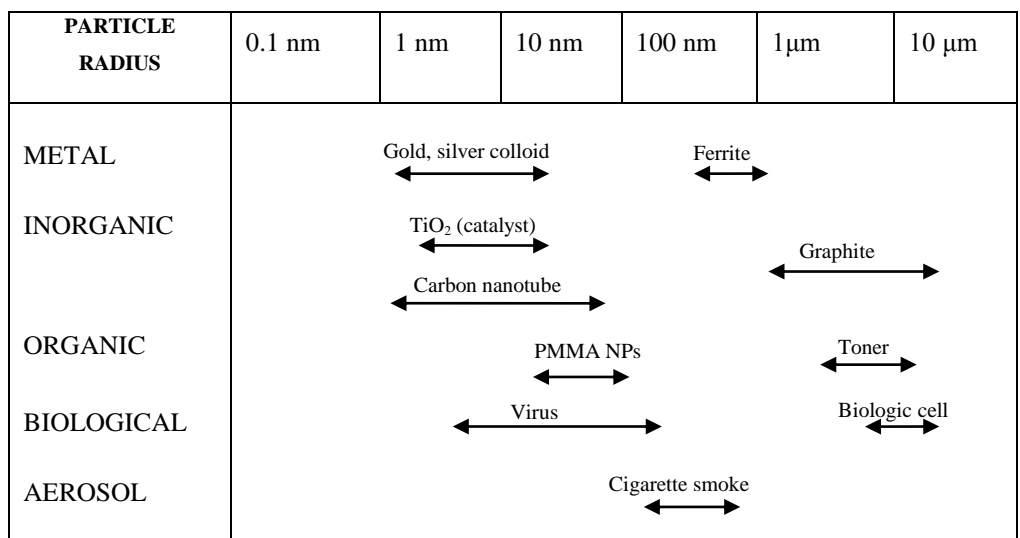

**Figure 1**. Common NPs classified by nature and size.

As previously mentioned, several studies have been carried out to detect and size nanoparticles in order to reduce the risk associated with their toxicity. The development of NPs sensing techniques and sensors could also give an additional benefit related to the environmental monitoring in terms not only of pollutant agents, but also hazardous agents used in bioterrorism, such as chemical compounds or virus [5].

Thus, when referring to NP sensing, all elements having a size comparable to the one of NPs are included in the detection problem and the investigation field extends to virus, molecules, fine and ultrafine dusts and so on.

For these different reasons, several techniques aimed to detect a single nanoparticle (in the explained acceptation) have been experimented in terms of mass, shape, refractive index and size identification.

A brief description of the most relevant detection methods is here introduced. They are based on the electrical, mechanical or optical transduction method.

## 4.1 MECHANICAL DETECTION

Since the mechanical response of a micro-cantilever or a string [6] can be altered by the absorption of a nanoparticle on its surface, these configurations have been employed as single nanoparticle detectors in terms of *mass sensors*.

Micro-cantilevers sensing principle is based on the measurement of the change of deflection or on the resonant frequency shifts that occur when an additional mass, i.e. a nanoparticle absorbed or binding to the surface, perturbs the static equilibrium of the system [7].

One of the earliest work concerning the use of a cantilever as mass sensors was developed about two decades ago [8], when it was demonstrated the detection of a 10 $\mu$m diameter latex bead in terms of mass by employing a silicon dioxide cantilever that exhibited a minimum detectable mass of approximately 0.2 ng. Detection of pathogens, such as bacteria and viruses [9], [10], has also been reported by employing cantilevers. All the experiments were conducted in air since the large damping in liquid associated with viscous effects, and so the resulting decrease in the quality factor value, gives rise to significant errors in the prediction of the resonant frequency shift, whose value is related to the mass sensitivity according to the formula $\Delta m \propto (1/(\Delta f))$. A comparison between cantilever mass sensor performances in air and water is presented in [11] where a minimum of 2 Bacillus anthracis Sterne spores have been detected with an added mass on cantilever of few hundreds of femtograms. As demonstrated in the same work, the magnitude of the minimum detectable mass degrades of about three orders if measurements are conducted in water.

## 4.2 ELECTRICAL DETECTION

Electrochemical sensors operating in potentiometric or amperometric way are one of the mostly employed platforms for NPs and influenza virus detection [12]. Their working principle depends on the nature of the suited electrodes which act themselves as sensing layers. In general, electrochemical sensing relies on the monitoring of the change of conductance or capacitance on a local sensing element induced by the presence of NPs.

Nanowire based field effect transistor devices are one of the common platforms employed for single NP detection. By suiting this configuration, Patolsky et al. demonstrated [13] the detection of Influenza A virus and paramyxo-virus in a buffer solution by opportunely functionalizing the nanowire surface with specific antibodies.

The working principle relies on a variation of conductance affecting the device: if a virus particle binds to the antibody receptor on the surface of the nanowire, then the conductance of that device should change from its baseline value and it should return to the baseline value when the virus unbinds. Also a second nanowire serving as internal control has been employed in the configuration proposed in [13]. The "operative" and the "reference" nanowires have been then embedded in a microfluidic channel for sample delivery.

An alternative approach to the nanowire based field effect devices relies on probing the change of impedance occurring across a nanoscale aperture, i.e. a channel. When a NP having different electrical properties than the solution it is flowing within passes through a channel, the impedance along the channel is perturbed. By measuring this impedance change across the channel, single polystyrene NPs having a diameter down to 50 nm and T7 bacteriophage viruses in different solutions have been detected and sized in a high-throughput microfluidic device [14]. The advantage of very rapid and high-throughput detection characterizing this technique is anyway obscured by the technical impossibility of suiting an affinity-specific detection protocol and this could inhibit the technique from being a candidate for clinical biosensing studies.

4.3 OPTICAL DETECTION

4.3.1 WGM BASED NP DETECTION

Nanoparticles optical detection protocols proposed so far are based on both integrated and discrete optics. The employment of resonant cavities supporting Whispering Gallery Modes is one of the promising techniques for the realization of compact, reliable and portable devices to be employed in civil and military environment for NP detection [14], [15], [16], [17], [18], but also sizing [16], [17], [18], [19], [20], [21], sorting [23], [24] and optical [25].

The pioneering works on WGM based biosensors are the ones of Armani et. al [15] and Vollmer et. al. [16] that demonstrated label free single NP detection. In [16] the discrete change in frequency of whispering gallery modes (WGMs) propagating within a microsphere cavity when Influenza A virions bind to it has been observed. Measurements have been conducted by employing polystyrene nanoparticles having a size similar to the one of viruses (500nm), and a ''reactive'' perturbation of the resonant photon state has been demonstrated. Label-free detection of individual viral-sized nanoparticles in aqueous solution has also been experimentally demonstrated [17] for the monitoring in real-time of binding events between 36 nm radius NPs and a WGM biosensor.

A lower LOD value has been reached in [18] where an ultrahigh-Q microcavity in conjunction with a thermal-stabilized reference interferometer has been employed for nanoparticle and virus detection. A LOD of 12.5 nm in terms of minimum detectable radius has been reached by resolving the shift caused by the binding of individual nano-beads in solution.

Brownian motion of particles in the evanescent field near the surface of a microsphere has also been investigated and the Stokes-Einstein relation has been employed to describe NPs diffusion coefficient and determine their diameter [19]. Brownian motion, indeed, has a big relevance in the resonance wavelength fluctuations of a WGM and could be suited for studying the dynamics of nanoparticles near interfaces as demonstrated in the same work.

The event of binding of a NP to a high Q-factor cavity has not only been examined in terms of shift of the resonant wavelength due to an increase in the optical path followed by light within the resonator, but also in terms of modal splitting [20-21] affecting WGMs at resonance. When indeed a NP binds to the resonator, it perturbs the light propagating within the cavity leading to the coupling between the two degenerate counter-propagating modes existing within the cavity. This phenomenon is visible in resonator transmission spectrum as a splitting doublet. The separation distance between the dips of the doublet and their line width concur to the determination of NP size. An a priori knowledge of NP refractive index is anyway required in order to calculate NP size in terms of radius. The theoretical model concerning NP detection and sizing employed in [20-21] is discussed in Chapter 5.

Dielectric NPs or biological elements such as virus have been investigated in the studies described so far where a resonant wavelength redshift has been observed. Studies concerning also metal NPs detection have been proposed.

Nanoparticles detection has also been investigated by employing a "hurricane" geometry, i.e. a disk resonator with input and output waveguides penetrating its walls tangentially to the surface [22]. A reflection mode-sensing has been suited in this study according to the nature (gold) of the testing NPs. This sensing mode relies on the buid-up of the intensity of backward propagating modes due to the reflection of the incident light from gold NP. An atomic force microscope has been employed to emulate the gold NP which ideally corresponds to the taggant attached to the analyta molecule to be investigated.

In [23] the binding of a gold NP to a microring in air gave rise to a blue shift in the resonance wavelength. This behavior has been explained to be the result of the value assumed by the real and imaginary part of the refractive index of native gold. It has indeed a value of the RI real component lower than unity , while a bigger value of the imaginary component. This way, information concerning the nature of the NP can be extrapolated by observing the direction of the resonant wavelength shift. A differentiation between dielectric NPs and gold NPs is thus possible. On the basis of this result, in [24] a notched ring resonator has been proposed for NPs sorting. A structural defect within the cavity, i.e. a 100 nm notch, has been fabricated by e-beam lithography and then etching. The NPs binding event has been simulated by placing the tip of an atomic force microscope (AFM) within the notch and it has been observed that the pre-existing split resonance (due to the presence of the notch) has been red or blue shifted as function of NP composition, but also broadened.

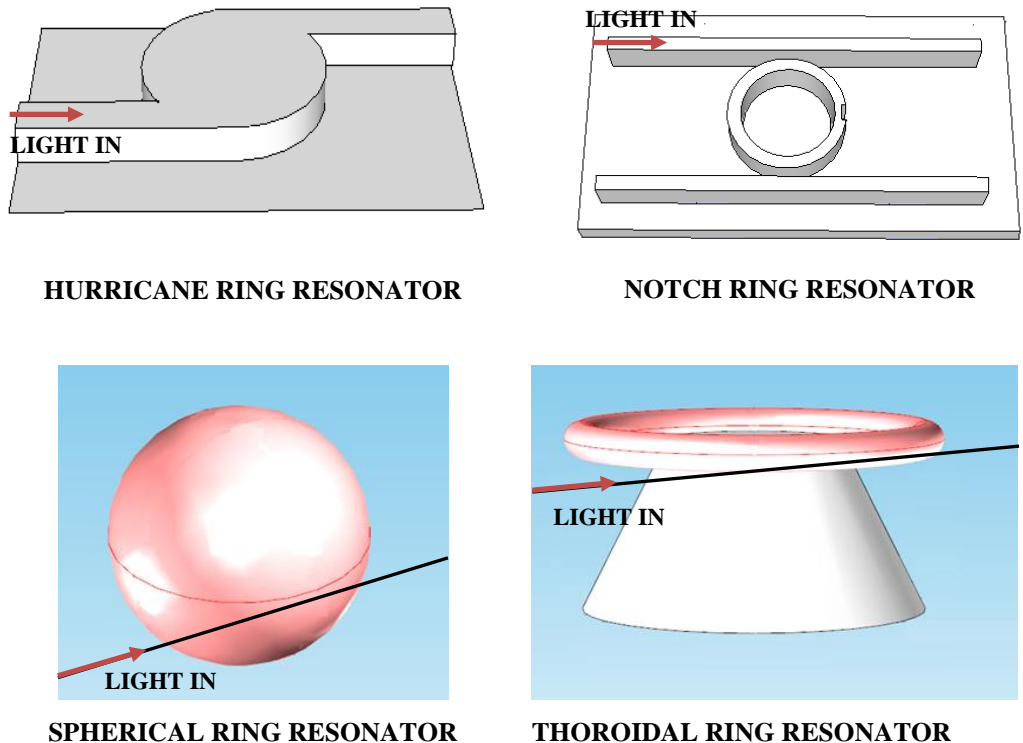

**Figure 2**. Schematics of optical cavities employed for NPs detection.

## 4.3.2 MICROSCOPY BASED NP DETECTION

Several microscopic techniques for NPs detection have been investigated, including interferometric, absorption and surface plasmon resonance microscopy (SPRM).

The latter, also known as *SPR imaging (SPRI)*, couples the capabilities of sensing related to the SPR with the spatial capabilities of an optical detection system. The employment indeed of a glass substrate having the top side covered by a thin metal layer and the excitation of plasmonic resonances on the metal layer surface through a light beam impinging on it enables the reflection of light which is collected by a CCD camera, and whose intensity is dependent on the resonant condition which is in turn dependent on the local refractive index change induced by the binding of a NP (or a molecule) to the surface. Most commonly used metals are gold and silver, but also aluminum and copper may be suitable in SPRM.

Recently, real time detection of nanoparticles and viruses has been demonstrated [25], [27] by employing SPRM. By studying the locally altered reflections from a SPR surface due to the binding of NPs, Zybin et al. [25] demonstrated the detection of polystyrene NPs with sizes between 3μm and 40nm and single virus detection, obtaining a signal to noise ratio of 4 for 40 nm PS nanoparticles. In this study, HIV-VLPs (HIV-virus like particles) have been employed in order to emulate the presence of viruses since they are similar in size, morphology and conformation to the intact viruses, but they are not virulent. Also in [27], single dielectric NPs (silica) and viruses (H1N1) in solution have been detected. Properly, it has been demonstrated that the intensity of the particle image is proportional to the mass of the particle and thus it has been determined the mass and the mass distribution of influenza viral particles. A detection limit of 0.2 fg/mm$^2$ has been achieved in terms of mass per unit area or equivalently the detection of the binding event of a single NP having a mass as small as 1 ag (which is the mass detection limit) on the sensing area. In terms of size, the detection limit is close to 100 nm (diameter) for silica NPs and viruses, with $\sigma = 0.1$. The sensitivity of SPRM could be enhanced by employing nano-antennas rather than a uniform metal layer to excite SPR. This way a stronger interaction light-matter, i.e. between light and NPs, can be achieved since plasmonic waves are localized in highly sensitive regions termed "hot spots" as demonstrated in [28] where a single gold NP with a 5 nm radius has been trapped via a plasmonic dipole antenna.

The above mentioned studies thus demonstrate the suitability of SPRM for single nano-sized object detection , such as a virus or a NP.

A different approach employed for microscopy based NPs detection relies on *absorption microscopy*. This technique is widely employed for very small particles detection and suites existing relation between the efficiency of absorption of a NP (identified through its absorption cross section $\sigma_{abs}$) and its size. Although the simplest way to optically detect metal NPs is via the light they scatter, this approach is less sensitive for very small particles than absorption-based detection methods due to the dependences [29]:

$$\sigma_{sca} = \frac{k^4}{6\pi}|\alpha|^2 \propto R^6 \quad (1)$$

$$\sigma_{abs} = k \, \text{Im}\{\alpha\} \propto R^3 \quad (2)$$

that are valid for spherical particles in the limit of "particles small compared to the wavelength". Here $\alpha$ is the NP polarizability, $R$ the NP radius and $k$ the wave-vector. The dependence of scattered light intensity from the sixth-power of NP size, indeed, hinders the detection of very small NPs.

Two different schemes could be suited for absorption detection, i.e. the direct absorption and the transient absorption. For really small NPs, the direct detection of the light they absorb is evaluated as variation in the transmitted light intensity, which is very small and which can be detected only if this variation is contemporary larger than the laser intensity fluctuation and the photon noise. Arbouet et al. [30] overcame this issues by employing a far-field optical technique based on modulation of the position of a gold NP in order to detect a single gold nanoparticle with average diameter down to 5 nm.

In transient absorption microscopy, a change of absorbance is monitored. The object to be imaged (a metal NP or a biological element) is resonantly excited through a pump laser and then a probe detects this absorbance change. Properly, the output beam of a laser is split in two components in order to get a pump and a probe beam. Through an optical setup composed by translation stages and mirrors, both the beams are focused on the object space via a high NA objective. By spatially and temporally overlapping the pump and probe beams to the sample, it is detected a change in the transmission of the probe induced by the pump. Usually, the beam with higher frequency is used as pump and the one with lower frequency as probe in the imaging of molecules or semiconductors. Near-infrared light is instead used to excite metals [31].

Photothermal detection is an additive method employed for the detection of absorbing nano-objects. The increasing of temperature induced in objects by the light absorption phenomenon leads to a local variation of refractive index that can be monitored in order to detect NPs. By employing a really sensitive photothermal method, Berciaud et al. [32] detected individual gold nanoparticles as small as 1.4 nm in diameter. They indeed introduced the photothermal heterodine imaging (PHI) which actually is one of the most widely used absorption based techniques in nano-science. Small heat capacities and large temperature dependent refractive index are the requisites that the environment surrounding the NP to be detected should satisfy in order to obtain a strong signal. In PHI experiments, for example, water is not a good candidate due to its thermal properties. Individual nanocrystals [32] and DNA micro-arrays based on Au NPs [33] have been also detected via PHI. With this technique, a background-free image can be obtained since no signal is produced by background if non-absorbing. Another widely employed microscopy technique is the *interferometric microscopy* which relies on the interferometric detection of the field scattered by the interest object (NP). The substrate, usually a glass slide, on which the specimen is located is usually illuminated with a coherent or incoherent light source (i.e. a laser or an LED, respectively ); the interference of the scattered light from NPs and the reference light originated from the substrate is then collected and measured via a detector (i.e. a photodiode or CCD). This way, the intensity of light at the detector is not dependent on NP scattering cross section, i.e. on the 6$^{th}$ power of NP radius, but on the interference signal which is in turn dependent on the 3$^{rd}$ power of NP radius.

With no interference, the intensity of light at the detector is:

$$I_{\text{det}} \propto |E_{sca}|^2 \propto \sigma_{sca}^2 \tag{3}$$

While, in an interferometric based detection microscopy:

$$I_{\text{det}} \propto |E_{ref} + E_{sca}|^2 \propto |E_{ref}|^2 + |E_{sca}|^2 + 2|E_{ref}||E_{sca}|\cos\theta \tag{4}$$

where the first term on the right of Eqn. (4) is a background term, the second is a pure scattering term and the last one is the interference contribution. $\theta$ is the phase difference between the reference and the scattered field.

According to NP size, the second or the third term could be dominant in Eqn. (4). For small size NPs, the interference term is more important than the scattering one. As the absorption microscopy, indeed, the interference based microscopy circumvents the problem of detecting small size NPs by relaxing the dependence of the intensity at the detector $I_{det}$ on NP size of 3 orders of magnitude.

Jacobsen et al. [34] demonstrated both confocal and wide-field detection of nanoparticles in aqueous environment in an interferometric detection scheme. The imaging of single gold nanoparticles with a diameter as small as 5 nm has been indeed realized with a detection speed up to 2 µs. Another important goal achieved in the above mentioned work related to particles tracking is the capability to distinguish the signal of the gold nanoparticle from that of other scatterers contained in the sample. The demonstration of wide-field imaging allows the simultaneous observation of several objects within large areas and is really useful in biological investigations. It anyway suffers from some noise sources, i.e. laser noise and spatial irregularities on the interface that both lead to artefacts in the final image. In their work, Jacobsen et al. [34] circumvented this issue through numerical post-processing of the image and they also worked at the plasmonic resonance to enhance NPs contrast.

A way to overcome the issue of laser noise has been proposed by Ignatovich et al. [35] that included a Michelson-like interferometer in the beam path in order to separately pilot the reference field and the one scattered from the NP and , at the same time, keeping the first one larger than the second. The method is not very sensitive to laser intensity noise and result to be background-free due to the self-referencing scheme introduced by the Michelson-like interferometer. The detection of single gold NP with size down to 14 nm and PS beads with size down to 30 nm have been demonstrated.

An additional microscopy technique based on the interferometric detection of scattered light has been proposed by Özkumur et al. [36] in a configuration termed spectral reflectance imaging biosensor (SRIB) actually modified as interferometric reflectance imaging sensor (IRIS) [37], [38]. This technique evolved from the study of optical phase difference due to the accumulation of biological material on a layered solid substrate to the detection of single NP and virus. In [36], indeed, a Si/SiO2 substrate has been designed to optimize phase imaging in a widefield, common path interferometer for the real-time measurement of accumulated biomass in a microarray format. The study have been then extended to the employment of a layered substrate in order to achieve a high spatial resolution and to maintain a relatively constant phase of the reflected incident light [38]. The enhancement of the interference term of Eqn. (4) has indeed been realized by employing a layered substrate opportunely designed to calibrate the phase difference between the reference and the scattered field and so to increase the contrast. The widefield interferometric imaging method demonstrated the detection and the accurate sizing of a range of nanoparticles, including H1N1 influenza virus with an average diameter of 120 nm and PS beads with average diameter up to 70 nm. The high-throughput sensor capability is confirmed by the parallel sensing of $10^5$-$10^6$ individual particle bound anywhere on the entire sensor surface.

The IRIS technique has been also proposed to distinguish metal NPs i.e. gold and silver NPs, in a mixture. The idea relies on adjusting the thickness of the spacer layer between the particle and substrate in order to get a unique optical response for each material type as function of the defocus [39]. An improvement of the IRIS technique is discussed in Chapter 6 where the model of a subsurface imaging setup based on the IRIS is presented [40].

4.4 CONCLUSIONS

Mechanical sensors, mainly in the micro-cantilever configuration, are mainly employed as mass sensors. They are good candidates for applications demanding high-throughput and real-time measurement due to the consolidate scalability fabrication techniques associated to micro-devices. Although these characteristics, sensitivity is enhanced if the cantilever is placed in vacuum. This aspect could highly affects the cost and introduce some handling issue in the employment of cantilever based NPs sensors in point-of-care applications.

The so far proposed electrical methods for NPs detection show the main advantage of compatibility with well-established microelectronic manufacturing technology. An easy integration of the sensing electronics and microfluidic structures on the same chip, indeed, will provide a robust and cost-effective platform for numerous lab-on-chip applications.

Optical techniques based either on microscopy or the perturbation of the resonant condition of a photonic cavity have been widely investigated for single nanoparticle detection showing good response in terms of detection limit and sizing sensitivity. Despite the capability of detection of really small NPs and the high sensitivity, the practical implementation of WGMs based resonators for NPs detection is limited by their low throughput capacity. Also, it has not been demonstrated so far a resonant cavity capable to discern NPs material type in a buffer solution. The high selectivity induced by the employed surface chemistry do not implies specificity of the binding and molecules of different nature could concur to perturb the cavity resonant condition in terms of modal splitting of resonant wavelength shift. The analysis of the modal splitting anyway, leads to overcome the critical issue of the dependence of resonator optical response on the NP-surface binding location. While WGMs based optical resonator offer the advantages of compact size and so potential portability of the sensor inserted in a reduced volume package, microscopy based techniques do not satisfy these requirements. Although these inconveniences, they anyway lead to shape recognition material/affinity specific detection of single nanoparticles and high throughput capacity.

The study of engineered NPs size and shape and their recognition could indeed be useful to understand the way these characteristics affect their functionality when employed as auxiliary media (for drugs delivery for example). This investigation seems to be possible, so far, only through microscopy based techniques.

4.5 REFERENCES

1. W.H. De Jong and P.J.A. Borm, "Drug delivery and nanoparticles: Applications and Hazards," International Journal of Medicine, 3(2), pp. 133-149, 2008.
2. G. Oberdörster, Z. Sharp, V. Atudorei, A. Elder, R. Gelein, W. Kreyling and C. Cox, "Translocation of Inhaled Ultrafine Particles to the Brain," Inhalation Toxicology, 2004, Vol. 16, No. 6-7 , pp. 437-445
3. C. Buzea, I.I. Pacheco Blandino and K. Robbie, "Nanomaterials and nanoparticles: Sources and toxicity," Biointerphases, vol. 2, issue 4, pp. MR17 - MR172, 2007.
4. C. Ciminelli, C.M. Campanella, R. Pilolli, N. Cioffi and M.N. Armenise, "Optical sensor for nanoparticles," 13th International Conference on Transparent Optical Networks (ICTON), 2011.
5. M.R. Hilleman, "Realities and enigmas of human viral influenza: pathogenesis, epidemiology and control," Vaccine, 20, pp. 3068–3087, 2002.
6. S. Schmid, S. Dohn and A. Boisen, "Real-Time Particle Mass Spectrometry Based on Resonant Micro Strings," Sensors, 2010.
7. M. Arroyo-Hernàndez, P.M. Kosaka,, J. Mertens, M. Calleja and J. Tamayo, "Micro- and Nanomechanical Biosensors," in: Handbook of

Nanophysics: Nanomedicine and Nanorobotics, K.d. Sattler (Edt.), CRC Press, Taylor and Francis Group, 2011.

8.  S. Prescesky, M. Parameswaran, A. Rawicz, R.F.B. Turner and U. Reichl, "Silicon micromachining technology for subnanogram discrete mass resonant biosensors," Canadian Journal of Physics, Vol. 70, Issue 10-11, pp. 1178–1183, 1992.

9.  A. Gupta, D. Akin and R. Bashir, "Single virus particle mass detection using microresonators with nanoscale thickness," Applied Physics Letters, Vol. 84, No. 11, pp. 1976–1978, 2004.

10. B. Ilic, D. Czaplewski, M. Zalalutdinov, H.G. Craighead, P. Neuzil, C. Campagnolo and C. Batt, J. Vac. Sci. Technol. B 19, pp. 2825–2828, 2001.

11. A.P. Davila, J. Jang, A.K. Gupta, T. Walter, A. Aronson and R. Bashir, "Microresonator mass sensors for detection of Bacillus anthracis Sterne spores in air and water," Biosensors and Bioelectronics, Vol. 22, pp. 3028–3035, 2007.

12. L. Krejcova, D. Hynek, V. Adam, J. Hubalek and R. Kizek, "Electrochemical Sensors and Biosensors for Influenza Detection," Int. J. Electrochem. Sci., Vol. 7, 2012.

13. F. Patolsky, G. Zheng, O. Hayden, M. Lakadamyali, X. Zhuang and C. M. Lieber, "Electrical detection of single viruses," Proc Natl Acad Sci USA. 101(39), pp. 14017-22, 2004.

14. J.L. Fraikin, T. Teesalu, C.M. McKenney, E. Ruoslahti, and A.N. Cleland, "A high-throughput label-free nanoparticle analyser," Nature Nanotechnology, 6, pp. 308–313, 2011.

15. A. Armani, R.P. Kulkarni, S.E. Fraser, R.C. Flagan and K.J. Vahala, "Label-free, single-molecule detection with optical microcavities," Science, 1145002v1, 2007.

16. F. Vollmer, S. Arnold and D. Keng, "Single virus detection from the reactive shift of a whispering-gallery mode," P. Natl. Acad. Sci. USA 105(52) (2008) 20701-20704.

17. S.I. Shopova, R. Rajmangal, Y. Nishida and S. Arnold, "Ultrasensitive nanoparticle detection using a portable whispering gallery mode biosensor driven by a periodically poled lithium-niobate frequency doubled distributed feedback laser," Review of Scientific Instruments, Volume 81, Issue 10, pp. 103110-103110-4, 2010.
18. T. Lu, H. Lee, T. Chen, S. Herchak, Ji-Hun Kim, S.E. Fraser, R.C. Flagan and K. Vahala, "High sensitivity nanoparticle detection using optical microcavities," P. Natl. Acad. Sci. USA, 108, pp. 5976-5979, 2011.
19. D. Keng, S. R. McAnanama, I. Teraoka, and S. Arnold, "Resonance fluctuations of a whispering gallery mode biosensor by particles undergoing Brownian motion," Appl. Phys. Lett. 91, 103902 (2007).
20. J. Zhu, S.K. Ozdemir, Y.F. Xiao, L. Li, L. He, Da-Ren Chen and L. Yang, "On-chip single nanoparticle detection and sizing by mode splitting in an ultrahigh-Q microresonator", Nature Photonics, Vol.4, 2010.
21. X. Yi, Y-F. Xiao; Y-C. Liu; Bei-Bei Li, You-Ling Chen; Y. Li and Q. Gong, "Multiple-Rayleigh-scatterer-induced mode splitting in a high-Q whispering-gallery-mode microresonator", Physical Review A, vol. 83, Issue 2, 2011.
22. B. Koch, L. Carson, C.-M. Guo, C.-Y. Lee, Y. Yi, J.-Y. Zhang, M. Zin, S. Znameroski and T. Smith, "Hurricane: A simplified optical resonator for optical-power-based sensing with nano-particle taggants," Sensors and actuators B, Vol. 147, Issue 2, pp. 573-580, 2010.
23. A. Haddadpour and Y. Yi, "Metallic nanoparticle on micro ring resonator for bio optical detection and sensing," Biomedical Optics Express, Vol. 1, No. 2, pp. 378-384, 2010.
24. S. Wang, K. Broderick, H. Smith, and Y. Yi, "Strong coupling between on chip notched ring resonator and nanoparticle," Appl. Phys. Lett. 97, 051102, 2010.
25. H. Cai and A.W. Poon, "Optical manipulation of microparticles using whispering-gallery modes in a silicon nitride microdisk resonator," Opt. Lett. 36, pp. 4257-4259, 2011.

26. A. Zybin, Y.A. Kuritsyn, E.L. Gurevich, V.V. Temchura, K. Überla and K. Niemax "Real-time Detection of Single Immobilized Nanoparticles by Surface Plasmon Resonance Imaging," Plasmonics , Vol.5, pp. 31–35, 2010

27. S. Wang, X. Shan, U. Patel, X. Huang, J. Lu, J. Li and N. Tao, "Label-free imaging, detection, and mass measurement of single viruses by surface plasmon resonance," PNAS, Vol. 107, No. 37, pp. 16028-16032, 2010.

28. W. Zhang, L. Huang, C. Santschi and O.J.F. Martin, "Trapping and Sensing 10 nm Metal Nanoparticles Using Plasmonic Dipole Antennas," Nano Lett. 2010, 10, pp. 1006–1011.

29. C. F. Bohren and D. R. Huffman, "Absorption and Scattering of Light by Small Particles," Wiley Interscience, New York, 1983.

30. A. Arbouet, D. Christofilos, N. Del Fatti, F. Vallée, J.R.Huntzinger, L. Arnaud, P. Billaud and M. Broyer, "Direct Measurement of the Single-Metal-Cluster Optical Absorption," Phys. Rev. Lett., Vol. 93, Issue 12, 2004.

31. S. Shang Lo , M.S. Devadas , T.A. Major and G.V. Hartland, "Optical detection of single nano-objects by transient absorption microscopy," Analyst, Vol. 138, Issue 1, pp. 25-31, 2013.

32. S. Berciaud, L. Cognet, G.A. Blab, and B. Lounis, "Photothermal Heterodyne Imaging of Individual Nonfluorescent Nanoclusters and Nanocrystals," Phys. Rev. Lett. Vol. 93, Issue 25, 2004.

33. G.A. Blab, L. Cognet, S. Berciaud, I. Alexandre, D. Husar, J.Remacle, and B. Lounis, "Optical Readout of Gold Nanoparticle-Based DNA Microarrays without Silver Enhancement," Biophysical Journal: Biophysical Letters, 90, L13, 2006.

34. V. Jacobsen, P. Stoller, C. Brunner, V. Vogel, and V. Sandoghdar, "Interferometric optical detection and tracking of very small gold nanoparticles at a water-glass interface," Optics Express, Vol. 14, No. 1, pp. 405-414, 2006.

35. F.V. Ignatovich and L. Novotny, "Real-Time and Background-Free Detection of Nanoscale Particles," Physical Review Letters, vol. 36, 013901, pp. 1-4, 2006.
36. E. Özkumur, J.W. Needham, D.A. Bergstein, R. Gonzalez, M. Cabodi, J.M. Gershoni, B.B. Goldberg, and M.S Ünlü, "Label-free and dynamic detection of biomolecular interactions for high-throughput microarray applications," PNAS, pp. 7988–7992, vol. 105, no. 23, 2008.
37. M.S Ünlü, "IRIS: Interferometric Reflectance Imaging Sensor - Multiplexed Assays and Single Virus Detection," Laser Science (LS) Conference, (LTh3I), Rochester, NY, October 14, 2012.
38. G.G. Daaboul, A. Yurt, X. Zhang, G.M. Hwang, B.B. Goldberg, and M.S. Ünlü, "High-Throughput Detection and Sizing of Individual Low-Index Nanoparticles and Viruses for Pathogen Identification," Nano Letters, pp. 4727-473, Vol. 11, 2010.
39. A. Yurt, G.G. Daaboul, J.H. Connor, B.B. Goldberg and M.S Ünlü, "Single nanoparticle detectors for biological applications," Nanoscale, pp. 715-726, 2012.
40. http://ultra.bu.edu/projects.asp?project=iris

# Chapter 5.
# WISPERING GALLERY MODEs BASED SINGLE NANOPARTICLE DETECTION

Different analytical and theoretical studies concerning whispering-gallery mode based nanoparticle (NP) detection have been proposed. Since one of the major goal of scientific research is to preserve human health, the same investigation concerning NPs detection for generic sensing purposes has been extended to the hazardous biological elements, including virus and other molecules, as reported in Chapter 4.

Really often a polystyrene (PS) NP is employed to emulate the presence of a virus [1], [2] since its refractive index (1.59 RIU) is really similar to the one of proteins composing the virus capsid which is the layer surrounding the virus genome, i.e. the nucleic acid portion of the virus. Due to the spheroidal shape of the capsid or anyway the spheroidal shape assumed by natural and engineered NPs[**], the "model" NP is usually assumed to be spherical.

Here we discuss an analytical method for NPs sizing and propose the employment of a planar microring cavity supporting WGMs as sensing device, despite the spherical and microtoroidal structures adopted in literature so far for NPs detection.

---

[**]NPs are indeed frequently synthetized in presence of an organic stabilizer called capping agent that confers them a spherical shape.

This chapter is organized as follows: we first discuss the behavior of a cavity in presence of a Rayleigh scatterer by employing quantum electrodynamics (QED) principles. Classical electromagnetic theory, indeed, results to be inadequate to study light scattering in a cavity [3]. Then we develop the theoretical foundations of NPs sizing through the analysis of modal splitting occurring when light propagation within a cavity is perturbed by the presence of a scatterer, according to [4], [5], [6].

After, we analyze the limiting factors of the model in terms of detection limit, reliability and applicability. We also present some results obtained by employing a cavity opportunely designed in order to satisfy some requisites of the analyzed model.

## 5.1 THE HARMONIC OSCILLATOR

The theory of the quantum field is based on an equality concerning the representation of the electromagnetic field. Properly, this equality states that a single mode of an electromagnetic field in a one-dimensional cavity acts as a one-harmonic oscillator.

This concept is briefly introduce in the following in order to explain later the analyzed model for NPs detection.

The classical representation of an electromagnetic field that has to satisfy the following Maxwell's equations in free space is:

$$\nabla \cdot E = 0 \tag{1}$$

$$\nabla \times E = -\frac{\partial B}{\partial t} \tag{2}$$

$$\nabla \cdot B = 0 \tag{3}$$

$$\nabla \times B = \varepsilon_0 \mu_0 \frac{\partial E}{\partial t} \tag{4}$$

To quantize the electromagnetic field, we consider a closed one dimensional cavity of volume *V* surrounded by perfectly reflecting mirrors, according to [7]. Then we consider a monochromatic, single-mode electromagnetic field, linearly polarized whose electric and magnetic components have the form:

$$E(z,t) = \bar{x} q(t) E_0 \sin(kz) \tag{5}$$

$$B(z,t) = \bar{y} \dot{q}(t) \frac{E_0}{c^2 k} \cos(kz) \tag{6}$$

where *k* is the wavenumber (=ω/c), *c* is the speed of light, *q(t)* is a measure of the field amplitude, while the expression of $E_0$ is:

$$E_0 = \sqrt{\frac{2\omega^2}{\varepsilon_0 V}} \tag{7}$$

The classical electromagnetic energy density of the electromagnetic field is:

$$U = \frac{1}{2}\left( \varepsilon_0 |E|^2 + \frac{|B|^2}{\mu_0} \right) \tag{8}$$

where brackets | | are employed to indicate the magnitude of the fields.
The correspondent Hamiltonian, representing the energy of the system, is:

$$H = \int_{Vol} U dV = \frac{1}{2} \int_{Vol} \left( \varepsilon_0 |E|^2 + \frac{|B|^2}{\mu_0} \right) dV \tag{9}$$

Substituting Equations (5) and (6) in Eqn. (9) and integrating, the Hamiltonian can be re-conduced to the form:

$$H = \frac{1}{2}\left( \omega^2 q^2 + p^2 \right) \tag{10}$$

which is identical to the one associated to a harmonic oscillator having unit mass *m*, with position *x* and momentum *p*:

$$H = \frac{1}{2}\left(kx^2 + \frac{p^2}{m}\right) \quad (11)$$

In the analogy between an harmonic oscillator and an electromagnetic field, *p* and *q* of Eqn. (10) are defined in terms of photons annihilation and creation operators [3], [7].

For sake of brevity, here we do not enter in details since an exhaustive discussion on the topic can be found in several books and papers dealing with quantum optics [3], [7], [8], [9].

5.2 NANOPARTICLE SIZING: MATHEMATICAL MODEL

An electromagnetic cavity can thus be considered as a one-dimensional harmonic oscillator.

As yet explained in Chapter 3, if light propagating within the cavity is perturbed along its path due to the presence of a disturb element, i.e. a nanoparticle or a structural imperfection such as surface or side-wall roughness, light is deviated from its natural path and a scattering phenomenon occurs with intensity proportional to the perturbation magnitude. The travelling wave modes of the resonator, i.e. the degenerate clockwise (CW) and counterclockwise (CCW) modes that are orthogonal to each other, will not exchange energy unless there this perturbation will be so significant to couple them each other.

With these assumptions, nanoparticle-microresonator interaction can be modeled as the interaction between a system S constituted by the one-dimensional oscillator, and a system R (called reservoir), constituted by all those matter oscillators (i.e. phonons, other photon modes, etc.) confined in the cavity and interested by the scattering and the damping of cavity eigenmodes occurring when light is perturbed along its path [3].

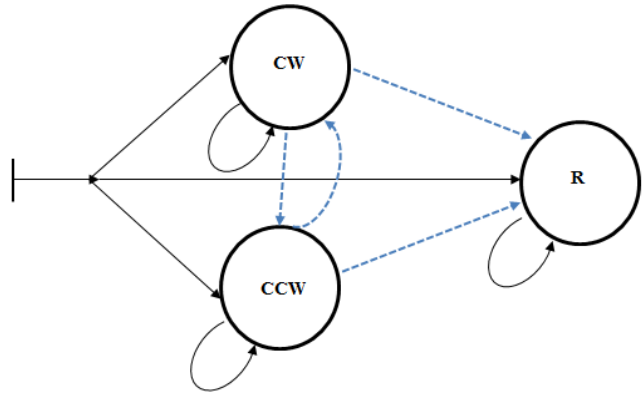

**Figure 1.** Graph representation of the system to be modeled. Dotted arrows refer to the coupled system.

Thus, the system under investigation is well approximated by a *coupled harmonic oscillator system* composed by two WGMs (CW and CCW) with degenerate resonant frequency and by n reservoir modes.

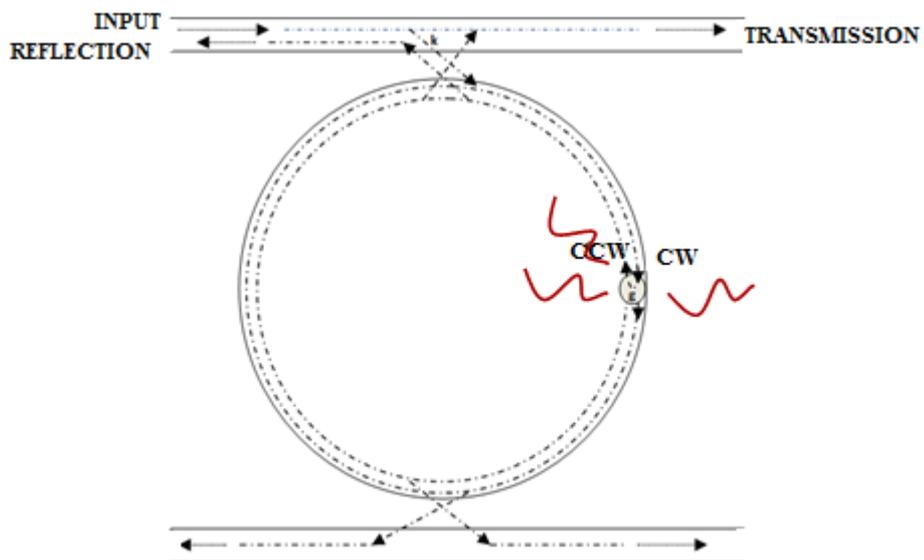

**Figure 2.** Schematic of the system to be investigated.

Under these conditions, the energy of the resonator-reservoir system is described by the total quantized Hamiltonian *H*:

$$H = H_0 + H_1 + H_2 \tag{12}$$

where $H_0$ is the free-Hamiltonian of the unperturbed system accounting for the energy of the free (i.e. uncoupled) field and the reservoir modes, while $H_1$ and $H_2$ are the coupling-Hamiltonians (i.e. expressing the interaction energy) regarding auto- and mutual-coupling of WGMs (CW and CCW) and the mutual-coupling between reservoir modes and WGMs, respectively.

Since the energy of a generic electromagnetic mode $k^{th}$ of frequency $\omega$ with creation and annihilation operators $a_k, a_k^\dagger$ is $\hbar\omega N$, where $N = (a_k^\dagger a_k)$ expresses the number of photons in mode k, the components of the total Hamiltonian are:

$$H_0 = \hbar\omega_0 \left( a_{CW}^\dagger a_{CW} + a_{CCW}^\dagger a_{CCW} \right) + \sum_j^n \hbar\omega_j b_j^\dagger b_j \tag{13}$$

$$H_1 = \hbar g \left( a_{CW}^\dagger a_{CW} + a_{CW}^\dagger a_{CCW} + a_{CCW}^\dagger a_{CW} + a_{CCW}^\dagger a_{CCW} \right) \tag{14}$$

$$H_2 = \sum_j^n \hbar g' \left( a_{CW}^\dagger b_j + a_{CW} b_j^\dagger + a_{CCW}^\dagger b_j + a_{CCW} b_j^\dagger \right) \tag{15}$$

where: $\hbar = h/2\pi$, h= Planck's constant; $a_{CW}^\dagger, a_{CCW}^\dagger$ are the annihilation operators of WGMs CW and CCW, respectively, $a_{CW}, a_{CCW}$ are the creation operators of WGMs CW and CCW, respectively; $b_j^\dagger, b_j$ are the annihilation and creation operators of the $j^{th}$ reservoir mode, propagating at frequency $\omega_j$; $g$ and $g'$ are the coupling coefficients between WGMs and WGMs and reservoir modes, respectively.

The Heisenberg equations of motion of the *isolated system* can be expressed as:

$$\dot{a}_{CW} = \frac{1}{i\hbar}[a_{CW}, H] \tag{16}$$

$$\dot{a}_{CCW} = \frac{1}{i\hbar}[a_{CCW}, H] \tag{17}$$

$$\dot{b}_j = \frac{1}{i\hbar}\left[b_j, H\right] \tag{18}$$

where *i* is the imaginary unit. Since the number of photons in a generic mode *m* can be any non-negative whole number, photons are bosons and they obey Bose-Einstein statistics [3].

That means that:

1) the annihilation and creation operators of photons of the same mode satisfy the bosonic commutation law, i.e. *formalism of second quantization* (the state of a set of identical bosons remains unchanged under permutation of two particles) and
2) they are linear operators.

By substituting the H expression into the Equations, (16)-(18), it comes:

$$\dot{a}_{CW} = -i\left[\omega_0 a_{CW} + g(a_{CW} + a_{CCW}) + \sum_j g' b_j\right] \tag{19}$$

$$\dot{a}_{CCW} = -i\left[\omega_0 a_{CCW} + g(a_{CW} + a_{CCW}) + \sum_j g' b_j\right] \tag{20}$$

$$\dot{b}_j = -i\left[\omega_j b_j + g'(a_{CW} + a_{CCW})\right] \tag{21}$$

Since the system under consideration is coupled to an external waveguide, a coupling and a loss term, $k_{ext}$ and $k_0$ need to be introduced. Properly $k_0$ ($=2/\tau_0$) is the damping rate due to material and radiation losses, $k_{ext}$ ($=2/\tau_{ext}$) is the waveguide-resonator coupling rate. Assuming the resonator to be coupled to two distinct straight waveguides and assuming the $k_{ext}$ term to be equal for both the couplers, the mode decay rate inside the cavity is:

$$t = \frac{1}{\tau_0} + \frac{2}{\tau_{ext}} = \frac{k_0 + 2k_{ext}}{2} \tag{22}$$

according to the definition of Q-factor ($1/Q = 2/\omega\tau$).

Under the additional assumption that the waveguide-resonator coupling is ruled by the law $\kappa v$, where $|k|^2 = k_{ext}$ (as yet reported in Chapter 3) and $v$ is the external source, the Heisenberg equations of motion of the *coupled system* become:

$$\dot{a}_{CW} = -i\left[\omega_0 a_{CW} + g(a_{CW} + a_{CCW}) + \sum_j g'b_j\right] - ta_{CW} - \sqrt{k_{ext}}\,u \qquad (23)$$

$$\dot{a}_{CCW} = -i\left[\omega_0 a_{CCW} + g(a_{CW} + a_{CCW}) + \sum_j g'b_j\right] - ta_{CCW} \qquad (24)$$

$$\dot{b}_j = -i\left[\omega_j b_j + g'(a_{CW} + a_{CCW})\right] \qquad (25)$$

Equation 14, for the reservoir operator $b_j$, can be formally integrated to yield:

$$b_j(t) = e^{-j\omega_j t}b_j(0) - ig'\int_0^t e^{-j\omega_j(t-\tau)}[a_{CW}(\tau) + a_{CCW}(\tau)]d\tau \qquad (26)$$

The first and the second terms of the right side of Eqn. 26 are the free and the forced evolution of the $j^{th}$ reservoir mode, respectively.

Properly, the free evolution of reservoir modes gives rise to a noise operator (obtained by substituting Eqn. 26. in Eqn. 23):

$$f_{a_{WGM}}(t) = -ig'\sum_j e^{-j\omega_j t}b_j(0) \qquad (27)$$

that varies rapidly due to the presence of all reservoir frequencies [3].

Equations (23) and (24) become:

$$\dot{a}_{CW} = -i[\omega_0 a_{CW} + g(a_{CW} + a_{CCW})] - ta_{CW} - \sqrt{k_{ext}}\,u + f_{a_{WGM}} - \Sigma \qquad (28)$$

$$\dot{a}_{CCW} = -i[\omega_0 a_{CCW} + g(a_{CW} + a_{CCW})] - ta_{CCW} + f_{a_{WGM}} - \Sigma \qquad (29)$$

$$\Sigma = \sum_{j} g_{j}^{2} \int_{0}^{t} e^{-j\omega_{j}(t-\tau)} [a_{CW}(\tau) + a_{CCW}(\tau)] d\tau \tag{30}$$

Through a transformation of variables, useful to remove the fast frequency dependence of the harmonic oscillator operators $a_{CW}$ and $a_{CCW}$ (i.e. $a_{CW} = A_{CW} \cdot exp(-j\omega_0 t)$ and $a_{CCW} = A_{CCW} \cdot exp(-j\omega_0 t)$ ), and by adopting the Weisskopf-Wigner approximation, the term $\Sigma$ can be expressed as $\frac{1}{2}\Gamma(A_{CW} + A_{CCW})$ where $\Gamma$ is the cavity damping rate due to Rayleigh scattering occurring in presence of a nanoparticle.

By neglecting the noise operator $f_{aWGM}$, we get:

$$\dot{A}_{CW} = \left[-i\Delta\omega + ig - t - \frac{1}{2}\Gamma\right]A_{CW} + \left[ig - \frac{1}{2}\Gamma\right]A_{CCW} + \sqrt{k_{ext}}u \tag{31}$$

$$\dot{A}_{CCW} = \left[ig - \frac{1}{2}\Gamma\right]A_{CW} + \left[-i\Delta\omega + ig - t - \frac{1}{2}\Gamma\right]A_{CCW} \tag{32}$$

where $\Delta\omega = \omega - \omega_0$ is the laser detuning frequency.

The previous set of equations describes the evolution of a coupled harmonic oscillator system in the case of a slowly varying field amplitude.

A more compact view of this state equations is:

$$\begin{aligned} \dot{A} &= FA + GU \\ \begin{bmatrix} \dot{A}_{CW} \\ \dot{A}_{CCW} \end{bmatrix} &= \begin{bmatrix} f_{11} & f_{12} \\ f_{21} & f_{22} \end{bmatrix} \begin{bmatrix} A_{CW} \\ A_{CCW} \end{bmatrix} + \begin{bmatrix} \sqrt{k_{ext}} \\ 0 \end{bmatrix} u \\ f_{11} &= f_{22} = -i\Delta\omega + ig - t - \frac{1}{2}\Gamma \\ f_{12} &= f_{21} = ig - \frac{1}{2}\Gamma \end{aligned} \tag{33}$$

Following the formalism of coupled mode theory investigated in Chapter 3, through a linear combination of the state equation, in steady-state regime ($\dot{A} = 0$), two uncoupled equations are derived by introducing new variables:

$$A_+ = \frac{\sqrt{k_{ext}} u}{t + \Gamma + i(\Delta\omega - 2g)} \tag{34}$$

$$A_- = \frac{\sqrt{k_{ext}} u}{t + i\Delta\omega} \tag{35}$$

Equations (34) and (35) describe the amplitude of the symmetric and the asymmetric standing wave modes SWMs, respectively, generated by the interaction of the native travelling wave modes (CW and CCW).

If we compare Equations (34) and (35), the symmetric standing mode is affected by a broadening $\Gamma$ of the resonance line width and a detuning $2g$ from the degenerate WGM.

After solving the coupled system in steady-state regime in order to derive the analytical formula of $a_{CW}$, the transmission coefficient $Tt$ as defined in [10] is calculated:

$$Tt = u - \sqrt{k_{ext}} A_{CW} \tag{36}$$

and the system transfer function T, i.e. its transmission spectrum analytical formula, is:

$$T = \left|\frac{Tt}{u}\right|^2 = \left|\left(1 - \frac{k^2 [i(g - \Delta\omega) - t - 0.5\Gamma]}{[i(g - \Delta\omega) - t - 0.5\Gamma]^2 - (ig - 0.5\Gamma)^2}\right)\right|^2 = \\ = \left|1 - \frac{k^2}{2}\left(\frac{1}{i(2g - \Delta\omega) - t - \Gamma} + \frac{1}{-i\Delta\omega - t}\right)\right|^2 \tag{37}$$

having eigenfrequencies:

$$\begin{aligned} \omega_1 &= \omega_0 \\ \omega_2 &= \omega_0 + 2g \end{aligned} \tag{38}$$

That means that a doublet will appear in the system transmission spectrum T or, equivalently, the resonant frequency of the specific examined mode will experience a modal splitting leading to the degeneracy of the two counter-propagating modes CW and CCW.

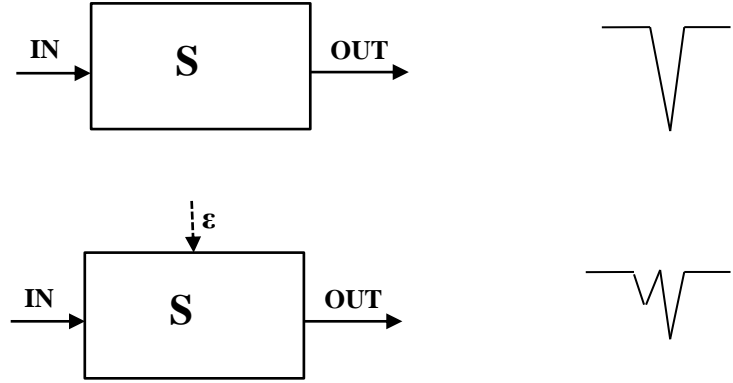

**Figure 3**. Block diagram of the unperturbed and perturbed system S, assumed as the isolated microresonator and the microresonator interacting with a localized perturbation.

As from Equations (37) and (38), the value of the coefficient $g$ (WGMs mutual coupling coefficient) determines the distance between the two dips of the splitting doublet, while the $\Gamma$ value rules the deep and the band broadening of resonances. Both g and $\Gamma$ expressions are related to the cavity modal volume $V_c$ which describes how efficiently the cavity concentrates the electromagnetic field in a restricted space, or, equivalently, it quantifies the spatial localization of energy within the resonator:

$$V_c = \frac{\int_{Vol} \varepsilon(r)|E(r)|^2 d^3 r}{\max\left\{\varepsilon(r)|E(r)|^2\right\}} \quad (39)$$

Numerator of Eqn.(39) represents the energy W of the electric field within the resonator according to Eqn. (9). For a fixed value of this volumetric integral, a smaller mode volume is associated to a larger peak of the electric field. A smaller modal volume can also be obtained by increasing the contrast between resonator and background refractive index.

Properly, the $\Gamma$ and g expression is:

$$g = \frac{-\alpha f(r)^2 \omega_0}{2V_c} \quad (40)$$

$$\Gamma = \frac{-g\alpha^2 f(r)^2 \omega_0^4}{3\pi c_b^3 V_c} \quad (41)$$

where $c_b$ is the speed of light in the surrounding medium. The term $f(r)$ is instead the field distribution function that reaches its maximum at the resonator-background medium interface and its square accounts for the cavity mode functions of both CW and CCW modes.

By assuming [11] $V_c = k*(\lambda/n)^3$ with $k$ = constant, Eqn. (37) has been plotted in Fig. 4 for different $k$ values.

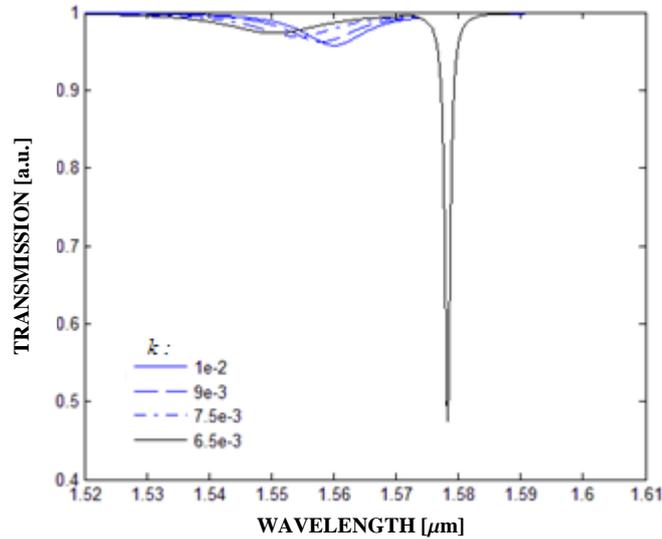

**Figure 4.** Overlapping of resonant cavity transmission spectra obtained by changing the $k$ coefficient in $V_c$.

As from Fig.4, the symmetric mode dip experiences a significant line broadening and shift when the cavity modal volume value decreases. Thus, smaller is the modal volume, wider is the relative shift between splitting dips.

The main hypothesis adopted so far is that a generic perturbation affects light propagation within the cavity.

If this perturbation is due to the presence of a *scatterer* like a nanoparticle NP and if its size is smaller than a certain fraction of the wavelength of the incident light (i.e. $\lambda/8$), the scattering process to be identified is a Rayleigh scattering in the regime of electrostatic approximation. That means that the electrical field E interacting with NP induces a dipole regime in it, thus conferring it an electrostatic polarizability $\alpha$ [12].

If the NP has a spherical shape, the polarizability expression is:

$$\alpha(\lambda) = 4\pi r^3 \left( \frac{n_p^2(\lambda) - n_{bckgr}^2(\lambda)}{n_p^2(\lambda) + 2n_{bckgr}^2(\lambda)} \right) \tag{42}$$

where r is NP radius, $n_p$ is NP refractive index and $n_{bckgr}$ is the background refractive index. The polarizability is the same along the three axes identifying the sphere.

In the following, $\alpha$, $n_p$ and $n_{bckgr}$ are assumed to be constant, i.e. not wavelength dependent.

Since the ratio $\Gamma/g$ is dependent on NP polarizability $\alpha$, according to the formula [5]:

$$\frac{\Gamma}{g} = \frac{\alpha \omega_0^3}{3\pi c_b^3} = \frac{8\pi^2 n_{bckgr}^3 \alpha}{3\lambda^3} \tag{43}$$

the analytical expression of the radius of a spherical NP can be derived by employing Equations (42) and (43):

$$r = \frac{\lambda}{2\pi n_{bckgr}} \sqrt[3]{\frac{3}{4}\left(\frac{n_p^2 + 2n_{bckgr}^2}{n_p^2 - n_{bckgr}^2}\right)\frac{\Gamma}{g}} \tag{44}$$

Equation (44) is a mathematical relation useful to estimate the radius of a spherical NP only by analyzing resonator transmission spectrum.

An additional benefit coming from the employment of this size estimating method is that it is a *self-reference and self-consistent method.* External sources of noise (such as the thermal ones), indeed, will affect at the same time and at the same manner both the splitting dips leaving their relative shift unaltered.

Anyway, an *a priori* knowledge of NP refractive index is necessary to apply this method that can be mainly suited for analytical purposes, i.e. to extract information from resonator transmission spectrum. In the following, anyway, the method is also investigated in order to extract some guidelines for the design of a cavity to be employed for NPs detection.

Due to the analytical nature of the method, high accuracy in parameters extrapolation from resonator transmission curve is necessary and it can be assured by approximating Eqn. (37) with a double Lorentzian curve through a fitting function of the form:

$$L_{fit}(\lambda) = 1 - \frac{A_1 w_1}{(\lambda - \lambda_1)^2 + w_1} - \frac{A_2 w_2}{(\lambda - \lambda_2)^2 + w_2} \qquad (45)$$

In Fig. 5 the overlapping of the theoretical transmission spectrum of a cavity interacting separately with a NP having radius 60nm and 80nm, respectively, is plotted. A Matlab code has been written to implement Eqn. (37), assuming a NP refractive index of 1.59 RIU, similar in value to the one of a polystyrene (PS) NP.

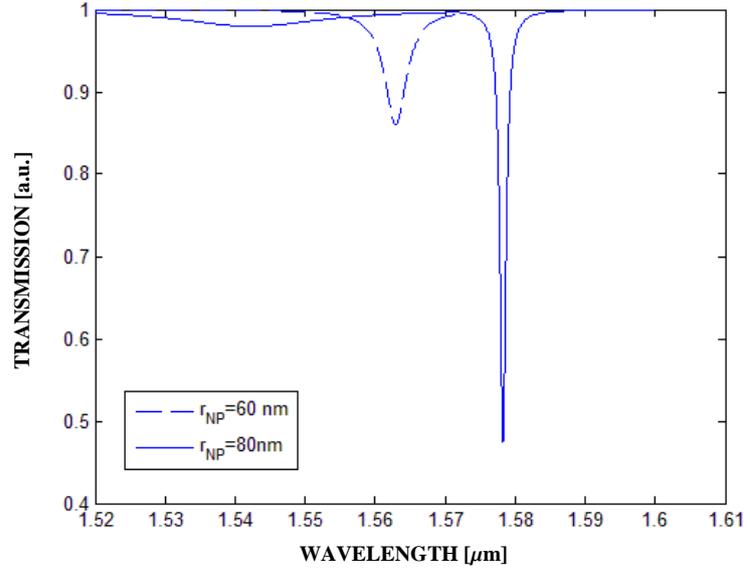

**Figure 5.** Modal splitting occurring due to the interaction of cavity mode and a PS NP with a 60 nm and 80 nm radius.

5.2.1 DETECTION LIMIT ENHANCEMENT

In order to evaluate the resonator detection limit in terms of minimal detectable NP radius, a graphical analysis of resonator transmission spectrum is proposed.

The condition to be satisfied to make a NP detectable is that the splitting is distinguishable. As previously indicated, a condition required to solve the modal splitting is to have a small cavity modal volume. As from Fig. 4, a modal volume reduction lifts to the symmetric mode dip line enlargement and shit.

From a geometrical point of view, the splitting is distinguishable when the doublet dips distance $2g$ is bigger than the sum of the line widths $FWHM_1 + FWHM_2$ of the splitting dips:

$$|2g| > \frac{\omega}{Q} + \Gamma = FWHM_1 + FWHM_2 \tag{46}$$

In Fig.6 a symmetric splitting (i.e. in absence of line broadening) is depicted as predicted with the CMT.

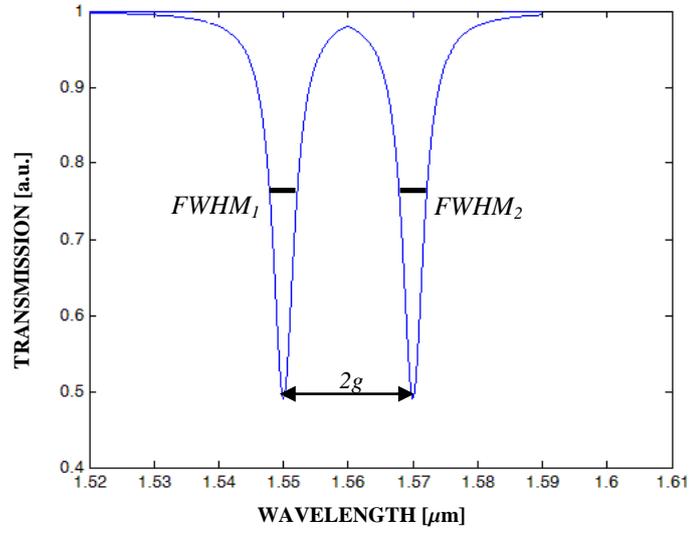

**Figure 6**. Illustration of symmetric mode splitting.

Since it is valid the relation:

$$|g| = \frac{\alpha \omega f(r)^2}{2V_c} \qquad (47)$$

And the line broadening $\Gamma$ can be neglected because its value is much more smaller than $\omega/Q$ (i.e. $\Gamma << \omega/Q$), Equations (46) and (47) become :

$$\frac{\alpha f(r)^2}{V_c} > \frac{1}{Q} \qquad (48)$$

By substituting Eqn. (42) into (48), the lower value of radius for a NP to be detectable through the modal splitting method is:

$$r_{NP} > \sqrt[3]{\frac{1}{4\pi f(r)^2} \frac{V_c}{Q} \left( \frac{n_p^2 + 2n_{bckgr}^2}{n_p^2 - n_{bckgr}^2} \right)} \qquad (49)$$

Eqn. (49) can be suited to find the minimum requirements an optical cavity should satisfy in order to detect a specific nanoparticle having refractive index $n_p$ in a known environment of index $n_{bckgr}$.

The term $V_c/Q$ in Eqn. (49) relates the detection limit of the method to the Purcell factor $F_p$:

$$F_p = \frac{3}{4\pi^2}\left(\frac{\lambda_0}{n_{bckgr}}\right)\frac{Q}{V_c} \tag{50}$$

Which is a figure of merit expressing the radiative rate of an emitter within a cavity with respect to its value in free space and it is known as "spontaneous emission enhancement factor".

It has indeed been demonstrated in the mid 1940's by Purcell that, if an ideal emitter is spatially and spectrally aligned with a mode in an electromagnetic cavity, its spontaneous emission rate is enhanced by the quantity indicates as $F_p$ [13]. This enhancement is favorite by the reduction of cavity modal volume or, equivalently, the increasing of the ratio $Q/V_c$.

Since the $V_c$ value is a direct function of the cavity size, a reduction of cavity dimensions will lift to a modal volume reduction and to an enlargement of the enhancement factor.

Thus it's possible to affirm that the spatial confinement of light in a cavity is typically quantified by the mode volume ($V_c$), whereas its temporal confinement is described by the quality factor ($Q$). These parameters are, indeed, respectively related to the electric field enhancement and the photon lifetime in the resonator.

At the same manner, smaller is the ratio $V_c/Q$ which is inversely proportional to the Purcell factor, smaller is the theoretical minimal radius detectable by employing the splitting method.

In Fig. 7, the theoretical cavity detection limit versus the Purcell factor value is plotted by assuming that the spherical bead to be detected is surrounded by air or it is immersed in a bulk solution (water).

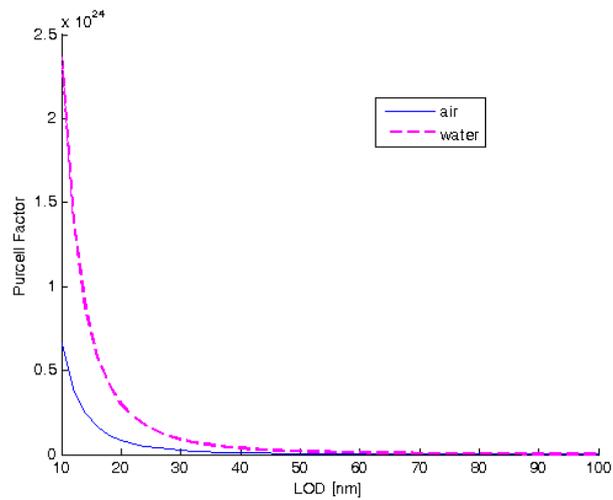

**Figure 7**. Theoretical cavity detection limit as function of the Purcell Factor for two different scenarios.

In next figure, the requirements that a cavity should satisfy in order to be employed for nanoparticle sizing purpose are represented in terms of flat surface. Properly, the one depicted is a color map concerning the modal volume value that the field inside the cavity should have to size a NP, for a certain Q-factor value. The operative environment here considered is *water*.

The investigated LOD is expressed in terms of minimal NP radius to be detected and its value has been fixed in the range [10:100] nm.

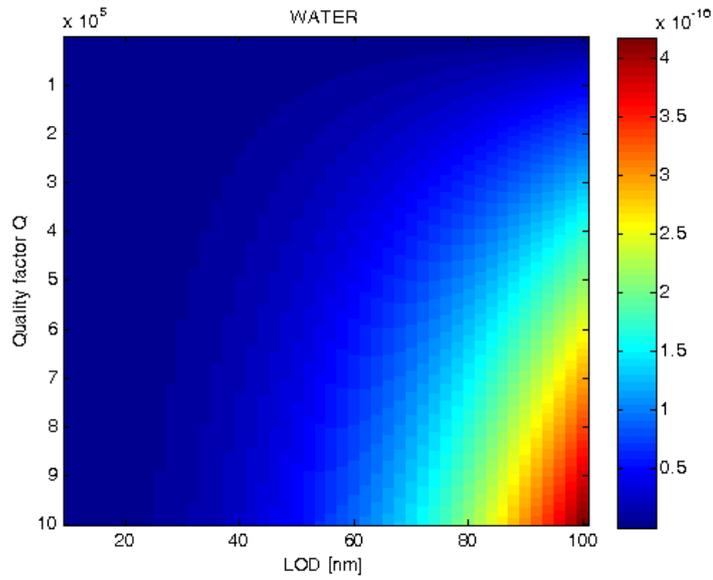

**Figure 8**. Color map expressing modal volume value $V_c$ as function of LOD and Q when the NP is immersed in water. $Vc$ is expressed in m$^3$.

In Fig. 9 (A) and (B), the cavity Q-factor value is assumed in the range $10^3$ - $10^5$ which corresponds to the range of Q-factor values obtained experimentally with planar ring resonators [14]. This class of WGM based optical cavities offers the advantage to be a good compromise between a device having both an high Q-factor and a small modal volume and will be investigated in the following.

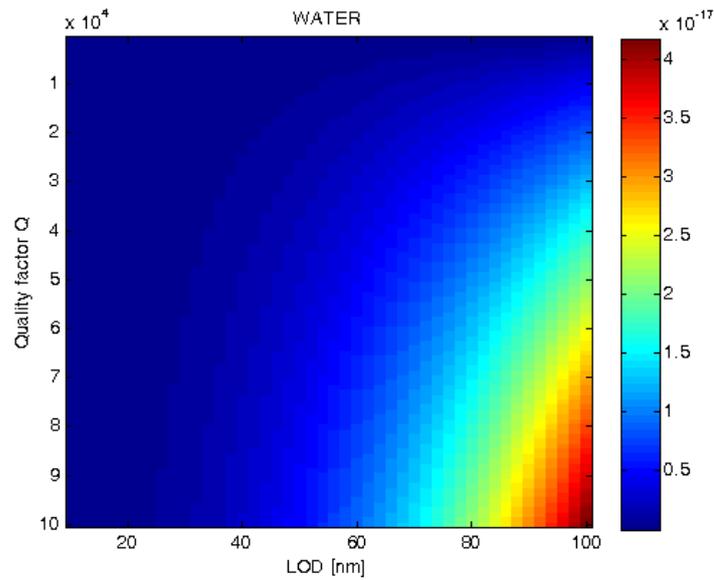

(A)

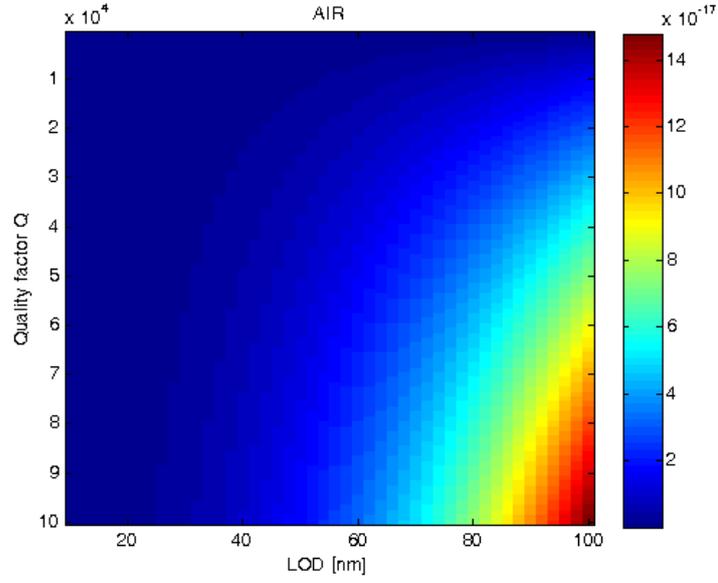

(B)

**Figure 9.** Color map expressing modal volume value $V_c$ as function of LOD and Q when the NP is immersed in (A) water and (B) air for a cavity Q-factor up to $10^5$. $V_c$ is expressed in m$^3$.

As from Fig. 9, the theoretical limit value of modal volume $V_c$ increases slightly when the NP to be detected is immersed in air. This is due to the increment of NP-background index contrast and the subsequent increment of Clausius-Mossotti coefficient $K$ which describes the NP residual polarizability, according to Eqn. (42) which can be re-written as:

$$\alpha = NP_{vol} \cdot K = NP_{vol} \cdot \left( \frac{n_p^2 - n_{bckgr}^2}{n_p^2 + 2n_{bckgr}^2} \right) \quad (51)$$

Where $NP_{vol}$ is the spherical bead volume in the dipole approximation.

The $K$ function curve is plotted in Fig. 10 for a fixed NP refractive index, and a background refractive index ranging from 1 RIU (air) to NP refractive index in order to consider only positive values of $K$.

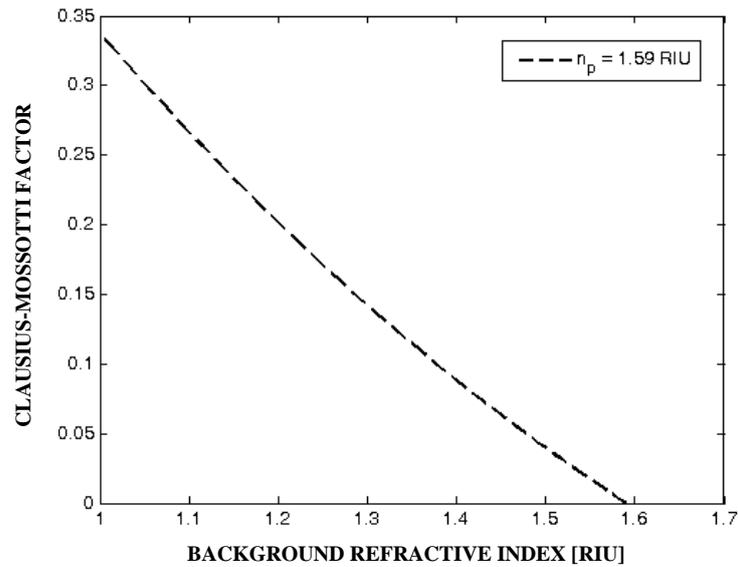

**Figure 10**. Clausius-Mossotti factor (or function) trend for a fixed NP refractive index and a background index ranging from the one of air to the one of the NP.

If the background is water (n= 1.3101 at $\lambda$= 1550 nm), both K and Q decrease (due to dispersion) giving rise to a degradation of LOD. Thus, a cavity having an higher Q factor value should be employed in order to compensate the (1/K) increment and get the same LOD when the operative environment is water. This is also what we expect from theory due to the phenomenon of light attenuation in water.

The theoretical LOD of a cavity in terms of minimum detectable NP radius can be derived from the color-maps by matching the Q-factor and the modal volume value, for a specific scenario.

Anyway, due to the not univocal relationship existing between NP radius analytical expression, NP refractive index and $V_c/Q$ ratio, color maps should only be considered as a guideline for the choice of the most appropriate resonator configuration. The selectivity of the method, i.e. the identification of a specific NP, will be dependent on the surface chemistry employed to activate the sensing layer, as defined in Chapter 2.

## 5.3 CAVITY DESIGN: HYBRID RING RESONATOR

The resonant structure configuration here assumed to validate the NP sizing method based on modal splitting has been chosen according to the design criteria previously indicated, concerning the cavity Q-factor and modal volume $V_c$.

The main characteristics of the cavity will be briefly in the following.

According to subsection 2.6.1, Silicon on insulator (SOI) technology has been chosen for cavity realization. We remember that in SOI structures, indeed, a strong light confinement can be realized and a single mode operation, i.e. only one guided mode for each polarization is supported by the guide, can be reached by employing a waveguide cross-section dimension in the nanometer range. Waveguide dimensions (height of the Si rib and its width) have been chosen according to the properties descripted in Chapter 3.

Thus, waveguide width has been chosen to be 500 nm, while its height to be 220 nm, according also to the low propagation loss of 0.92dB/cm ± 0.12 demonstrated in [15].

Figure 11 depicts the field distribution of the TE-like and TM-like fundamental mode supported by the employed 500x220 SOI waveguide.

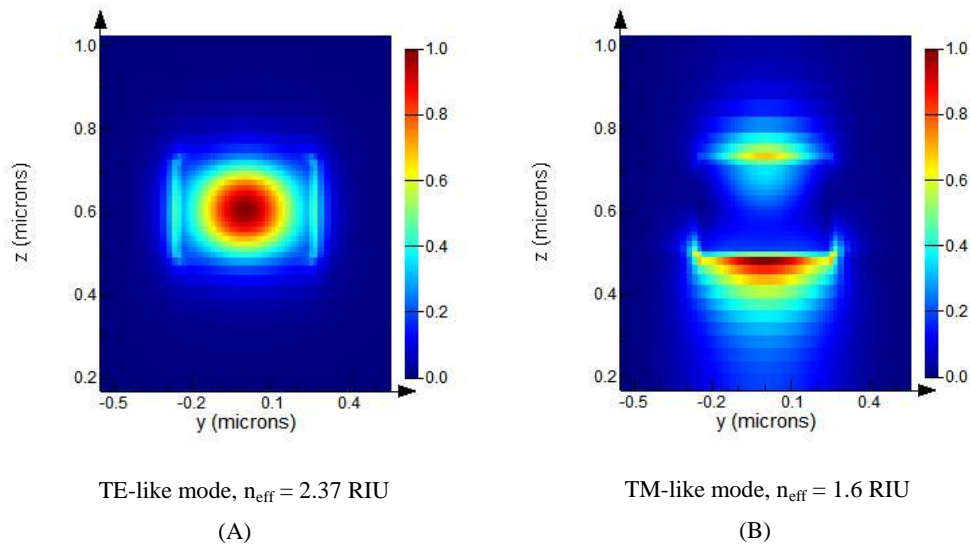

TE-like mode, $n_{eff}$ = 2.37 RIU    TM-like mode, $n_{eff}$ = 1.6 RIU

(A)    (B)

**Figure 11**.(A) TE-like mode field distribution. (B) TM-like mode field distribution.

As from Fig. 11, a stronger confinement in the waveguide core is obtained for the TE-like mode.

Cavity design guidelines are driven by the individuation of a compact, small footprint optical resonator having a low modal volume *Vc* and a Q factor value such that the ratio $V_c/Q$ allows to detect NP having a radius in the interest range of 30 – 100 nm. This range corresponds to the one associated to the size of most intruders such as viruses and toxic NPs [16].

For 500x220 SOI waveguide the critical radius value has been evaluated to be 3 *μ*m by employing a commercial software for electromagnetic field propagation. Critical radius is indeed that value of radius below which bending loss are no more linear with bending radius and they have a dominant effect on resonator loss.

A study concerning the coupling gap has also been effectuated to identify the value of the cross-coupling coefficient $k_{opt}$ in the coupler section for different gaps, according to the coupled mode theory CMT in space domain described in Chapter 3.

Coupling efficiency trend is presented in Fig.12 for a gap ranging from 150 to 400 nm.

Data have been obtained by employing a FEM commercial software. A quadratic fitting algorithm has been implemented in Matlab environment to interpolate data.

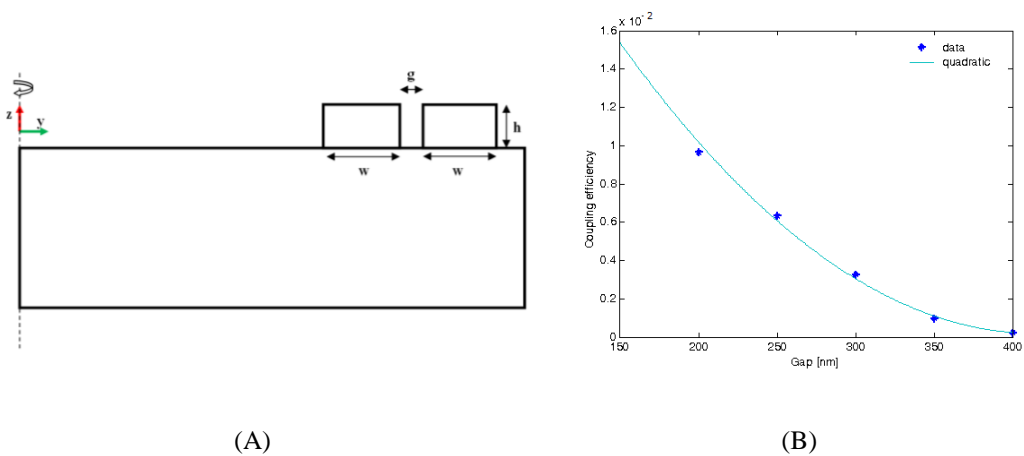

(A)            (B)

**Figure 12.** (A) Cavity cross-section at the coupler area. (B) Coupling efficiency evaluated for a gap ranging from 150 up to 400 nm.

According to what we expect by increasing resonator-waveguide distance, i.e. the gap g, coupling efficiency reduces when the gap increases since the coupling is ruled by the penetration depth of the evanescent tail of the propagating field, which exponentially decays besides the guiding structure.

A reduction of coupling efficiency gives rise to an increase of Q factor value which is one of the requisites to be satisfied in order to solve the doublet splitting. The other requisite is related to the modal volume $V_c$ whose value can be reduced by increasing the index contrast between the resonator and the background (thus employing SOI technology as here assumed) and by reducing resonator sizes, as stated previously.

Since are valid the relations [17], [18]:

$$FWHM = \frac{K\lambda^2}{\pi L n_{eff}} \qquad (52)$$

$$Q = \frac{\lambda}{FWHM} \qquad (53)$$

the theoretical Q-factor value for a TE and TM mode has been calculated and its trend is plotted in Fig. 13 for a cavity having a radius as small as 5 μm. The exact theoretical Q-factor value is indeed function of cavity length L and the operative wavelength λ (here assumed to be 1550 nm). The *K* term in Eqn. (39) is the coupling efficiency, not to be confused with the Clausius-Mossotti function mentioned previously.

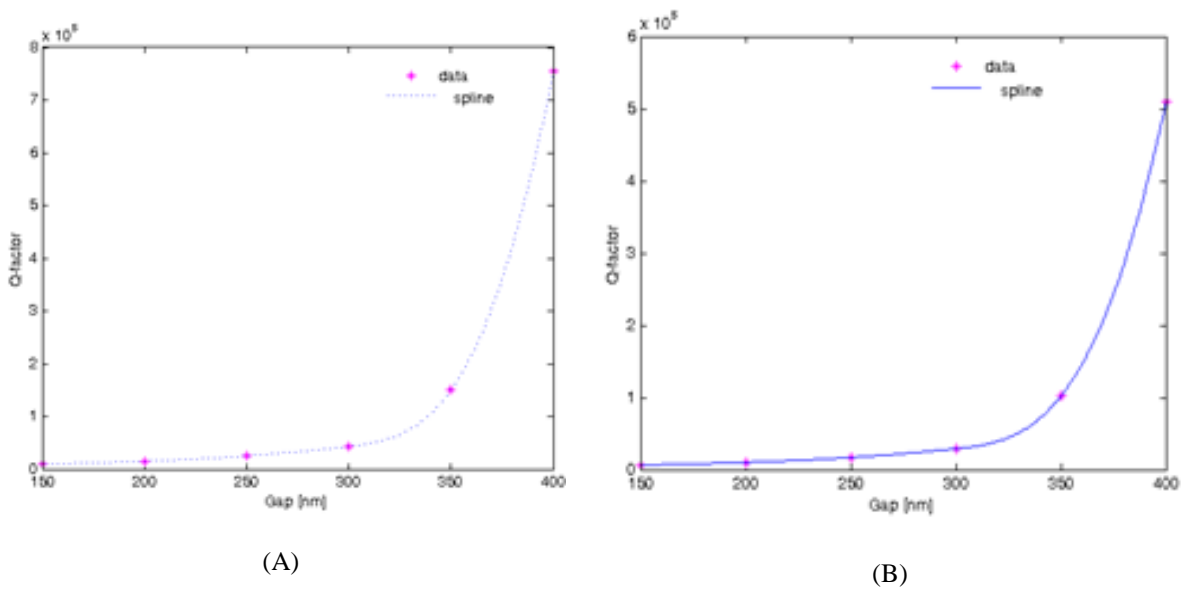

**Figure 13**. Q-factor value vs resonator-waveguide gap for a (A) TE and (B) TM polarized mode.

As from Fig. 13, a slightly lower Q-factor value is reached if light propagating within the cavity is TM polarized rather than TE polarized due to the effective index value characterizing the different polarizations. Theoretically, a Q-factor of the order of $10^5$ can be anyway reached for both polarization conditions when the coupling is really weak. This condition corresponds also to a significant reduction of the resonator extinction ratio ER which is undesirable for sensing applications. To work in proximity of critical coupling condition is indeed one of requisite to be satisfied when a resonant cavity is employed as biosensor. Thus, the coupling gap to be explored has been chosen in order to be far from a really weak coupling condition where back-reflection effects are not negligible giving rise to intermodal coupling [19], [20].

The investigated values of gap are 300 nm and 350 nm. Two classical planar configurations, i.e. a ring and a disk resonator, having a radius of 5 μm have been investigated by moving the coupling gap. Data are reported in Table 1.

**Table 1.**

|  | Gap = 300 nm | | Gap = 350 nm | |
|---|---|---|---|---|
| **RING** | **Q** | **ER [dB]** | **Q** | **ER [dB]** |
|  | $3.67 \cdot 10^4$ | 15.58 | $1.06 \cdot 10^5$ | 13.04 |
| **DISK** | **Q** | **ER [dB]** | **Q** | **ER [dB]** |
|  | $7.74 \cdot 10^4$ | 28.4 | $1.73 \cdot 10^5$ | 23.4 |

Reported data are related to electromagnetic simulations of the cavities excited by a TE polarized beam. A comparison with the Q-factor values of Fig. 13 (A) denotes a good trend matching between simulated data and the predicted ones.

In Fig. 14 the overlapping of the transmission spectrum of the analyzed cavities is depicted. The maximum mode radial number supported by the disk is equal to 3, for both the coupling gaps.

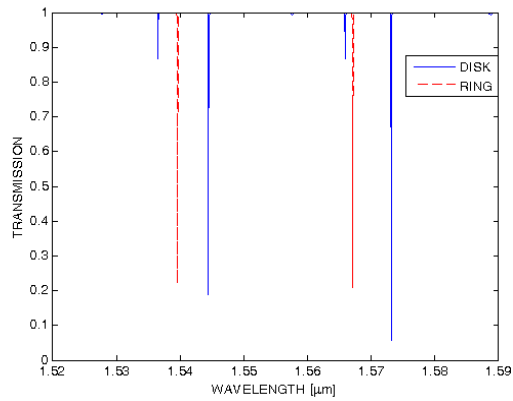

GAP 300 nm

(A)

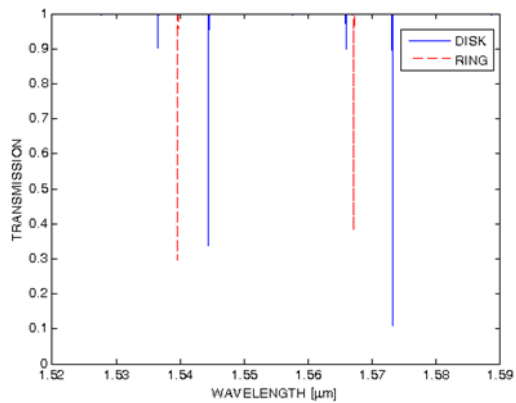

GAP 350 nm

(B)

**Figure 14.** Ring and Disk resonator transmission spectrum for a resonator waveguide gap of (A) 300 nm and (B) 350 nm.

The same FDTD commercial software employed for the electromagnetic simulation of the cavities has been suited to calculate their electric energy density distribution in 3D simulations for all the supported modes (see Fig. 15).

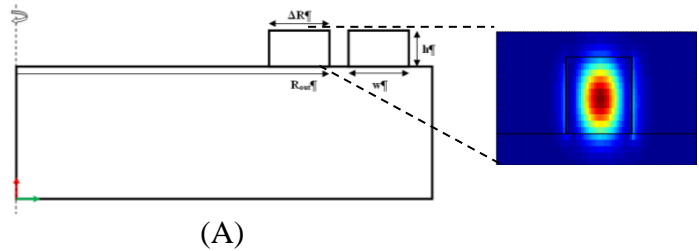

**Figure 15**. *Left*. (A) Ring and (B) disk resonator cross section in the coupling region. *Right*. Electric energy density for all the modes propagating within every cavity.

Really small $V_c$ values (of the order of $10^{-19}$ m$^3$) have been obtained by employing classical shape cavities having small sizes, as depicted in Fig. 16.

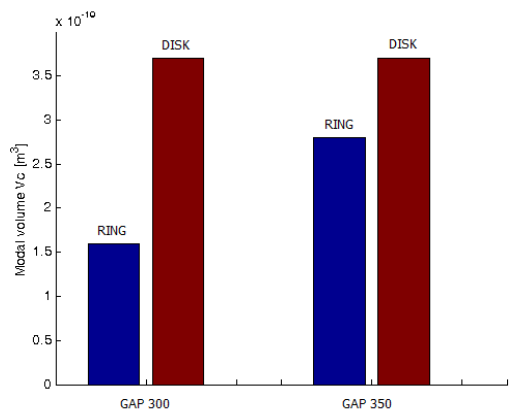

**Figure 16**. Modal volume of ring and disk fundamental mode as function of the coupling gap [nm].

Despite a Q-factor of the order of $10^5$ reached in both configurations when a coupling gap of 350 nm is employed, the ring resonator show an ER value close to 15 dB which is considered the threshold condition for sensing applications [20]. This could thus affect the detection performances of the device. At the same time, the presence of higher order modes (up to the third order) characterizing the disk could lead to intermodal coupling when a scattering element interacts with the cavity. Only a single mode is instead supported by the ring resonator.

A brief resume of analyzed configurations is reported in Table 2.

**Table 2**

| RING | DISK |
|---|---|
| *Small $V_c$* | *Small $V_c$* |
| *High Q* | *High Q* |
| *Low ER* | *High ER* |
| *Single mode* | *Multi-mode* |

Thus, we notice that as the gap increases, coupling decreases and this causes the ER value to decrease and the Q-factor value to increase.

Theoretically, the requirements to solve the splitting doublet are associated only to the Q and $V_c$ value, but also the ER value need to be taken in consideration for a realistic design of a cavity to be employed for sensing applications.

A possible way to keep at the same time the $V_c$ value low and the Q-factor value high is engineering a classical planar ring resonator configuration by moving the microring inner radius $R_{in}$ [21] by limiting the number of higher order modes supported by the cavity.

This engineered resonator is here termed "hybrid resonators" since its engineerization can lead to a microring or microdisk optical behavior in terms of number of guided modes and energy storage (i.e. Q-factor) within the resonator. For sake of brevity, we remand the reader to subsection 5.4.1 for a wide discussion concerning the proposed configuration.

Fixing the resonator outer radius to the value of 5 $\mu$m, the inner radius value has been moved in order to switch from a classical ring configuration to a disk resonator configuration. We remember that classical ring means that the width of ring annulus is equal to the width of waveguide cross-section, i.e. w = 500 nm. A disk resonator is instead obtained by setting $R_{in}$ to zero.

We found that the hybrid structure acts as a disk, i.e. it supports a maximum mode radial order number equal to 3, in the hypothesis of using a resonator having an annulus ΔR which is at least the double of the input waveguide (500x220 nm) in terms of width. That means for ΔR > 1 $\mu$m.

In Fig. 17, the modal field distribution of a cavity having an annulus width of ΔR = 1 $\mu$m (limit value for the propagation of an higher order mode of the second order) is plotted.

**HYBRID**

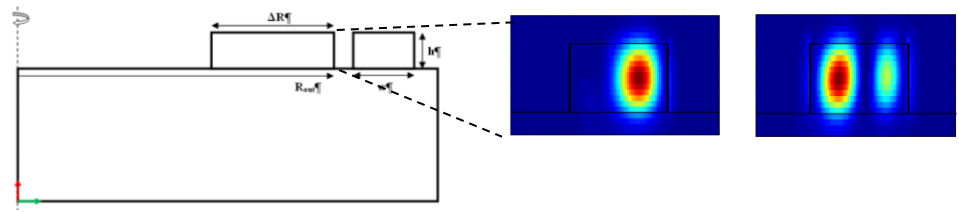

**Figure 17**. Modal field distribution of Hybrid configuration with annulus width of 1 $\mu$m.

Despite this limit configuration, other annulus widths have been investigated. Data reported in Table 3 have been extrapolated from electromagnetic simulations where the geometry of all the resonant cavities is characterized by a waveguide-resonator gap fixed to 300 nm.

**Table 3.**

| Annulus width [nm] | $R_{in}$ [nm] | Q factor | ER [dB] | Finesse |
|---|---|---|---|---|
| 500 | 4500 | $3.67 \cdot 10^4$ | 15.58 | $0.65 \cdot 10^3$ |
| 750 | 4250 | $7.04 \cdot 10^4$ | 27.09 | $1.25 \cdot 10^3$ |
| 1000 | 4000 | $7.54 \cdot 10^4$ | 27.14 | $1.34 \cdot 10^3$ |
| 1500 | 3500 | $7.74 \cdot 10^4$ | 28.2 | $1.37 \cdot 10^3$ |
| 5000 | 0 | $7.74 \cdot 10^4$ | 28.4 | $1.37 \cdot 10^3$ |

Increasing the annulus width beyond the one considered as the limit width (ΔR = 1000 nm) leads the hybrid resonator to act as a disk from an electromagnetic point of view.

When the coupling gap is increased, instead, this condition is reached exactly at the limit annulus width of 1 μm, as from Table 4 where data concerning a 350 nm gap are reported.

**Table 4.**

| Annulus width [nm] | $R_{in}$ [nm] | Q factor | ER [dB] | Finesse |
|---|---|---|---|---|
| 500 | 4500 | $1.06 \cdot 10^5$ | 13.04 | $1.9 \cdot 10^3$ |
| 750 | 4250 | $1.53 \cdot 10^5$ | 23 | $2.34 \cdot 10^3$ |
| 1000 | 4000 | $1.73 \cdot 10^5$ | 23.37 | $3.08 \cdot 10^3$ |
| 5000 | 0 | $1.73 \cdot 10^5$ | 23.4 | $3.08 \cdot 10^3$ |

Transmission curves for a 300 nm gap are depicted in Fig. 18.

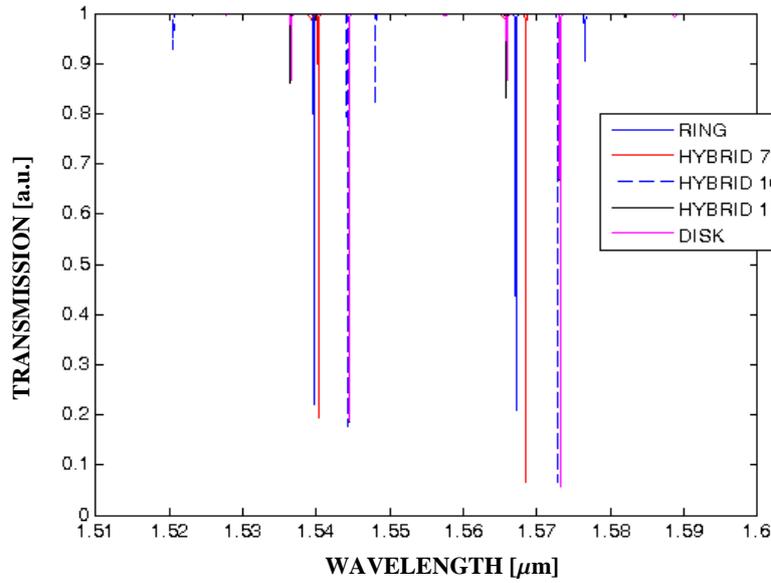

**Figure 18**. Overlapping of transmission spectrum of cavities having a different annulus, but equal outer radius $R_{out}$ fixed at 5 μm.

From theory, FSR value is function of radial distance of fundamental mode from the symmetry axis of the structure (i.e. z-axis) which is usually assumed to be equal to resonator outer radius $R_{out}$, but it is not exactly so. For this reason, FSR value is roughly equal to 28 nm for all the configurations and its exact value depends on the spatial confinement of the propagation path travelled by the fundamental mode, which is slightly different for the analyzed configurations.

Despite the FSR, the quality factor value changes significantly by modifying resonator annulus in terms of inner radius. For a fixed $R_{out}$, a reduction of $R_{in}$ gives rise to a Q-factor increasing since the first order radial mode (i.e. the fundamental mode) is strongly confined only along one sidewall and losses are reduced as it will be discussed in section 5.4.1.

The modal volume value of the fundamental mode for every analyzed configuration is roughly equal to $3 \cdot 10^{-19}$ m$^3$. Its value slightly increases when moving from a ring to a disk configuration, but preserves the same order of magnitude.

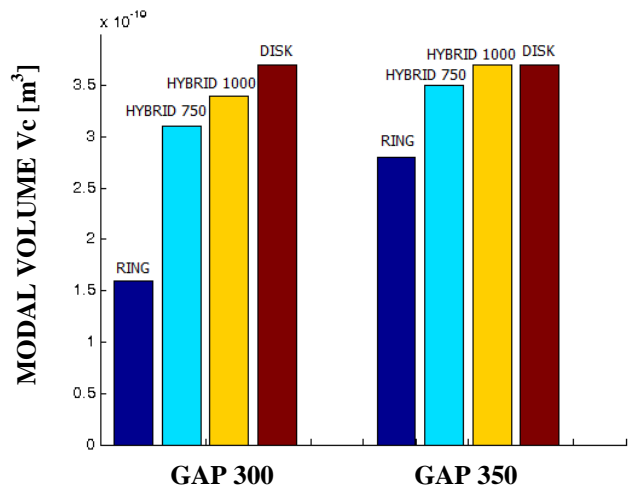

**Figure 19**. Resonators modal volume for two different gap values.

Fixing the annulus width to a specific value, and moving the gap, an increasing of Q-factor is observed from electromagnetic simulations as predicted from theory. In Fig. 20, the overlapping of the transmission spectrum curves of a hybrid resonator having annulus width of 1 μm with a resonator-waveguide gap ranging from 200 to 350 nm is depicted.

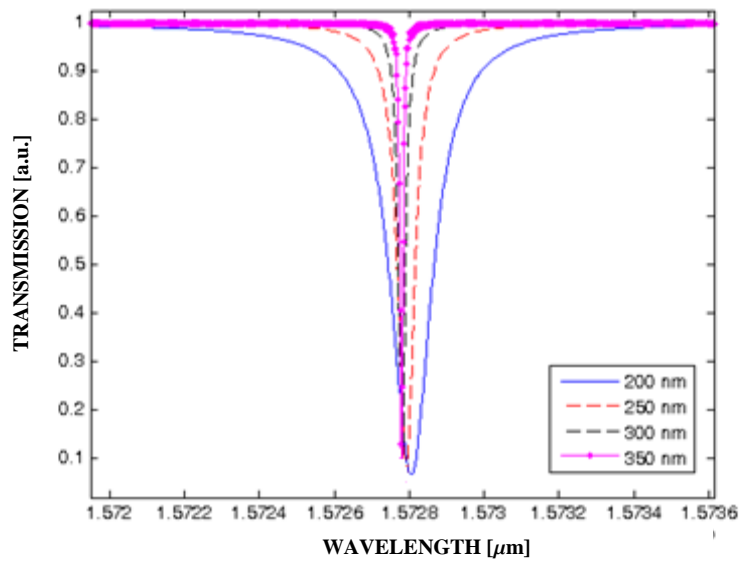

**Figure 20**. Enlargement of transmission spectrum overlapping in proximity of fundamental mode resonance condition.

The correspondent Q-factor values are instead plotted in Fig. 21. Data here reported are consistent with the ones of Fig. 13 (A) calculated from theory.

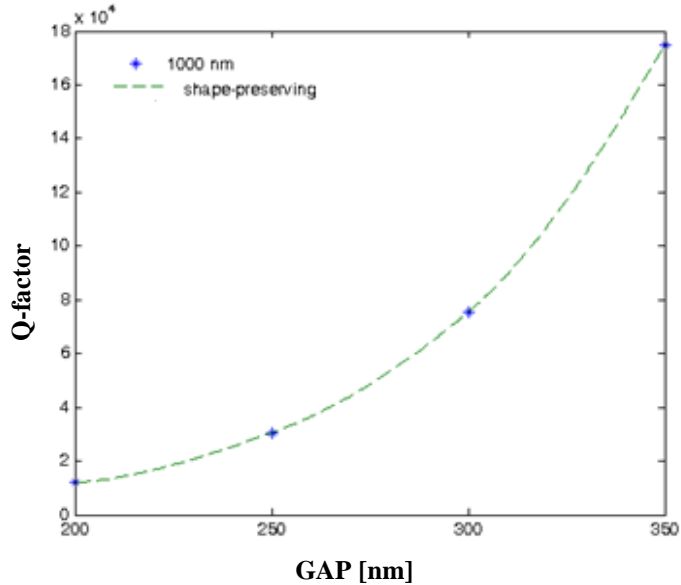

**Figure 21.** Hybrid resonator Q-factor value as function of coupling gap.

Both gap values of 300nm and 350nm allow to reach an high Q-factor keeping the $V_c$ value low due to the reduced size of the resonator. Since this is a rule of thumb for all the different examined configurations, we will investigate different cavities response to the presence of a spherical NP moving the coupling gap in the range 300-350 nm, for a fixed outer radius.

It is indeed possible to identify only some pre-requisites that the cavity should satisfy for NPs detection. Selectivity of the detection method and specificity of the molecular interactions for NPs binding, detection and sizing are dependent on the surface chemistry employed, as reported in Chapter 2. Usually, anyway, electrostatic interactions are suited to immobilize NPs on the sensor surface giving rise to a lack of selectivity.

5.3.1 SINGLE NANOPARTICLE DETECTION AND SIZING

Due to the characteristics indicated in the previous section, we start the analysis by employing a hybrid resonator based microcavity having an annulus of 750 nm and a coupling gap of 300nm. For this configuration, a Q-factor value of roughly $7 \cdot 10^4$ and a FSR of 28 nm have been demonstrated by employing a commercial software implementing the FDTD algorithm. Also, the cavity supports higher order modes up to the second radial order.

The same FDTD based software has been used as tool for the evaluation of the resonator modal volume $V_c$, whose value is equal to $3 \cdot 10^{-19}$ as reported in the bar plot of Fig. 19. A refractive index monitor and a field profile monitor have been employed to collect data concerning material permittivity and electric field in order to solve Eqn. (28) in a 3D environment.

Thus a $Q/V_c$ ratio roughly of $10^{23}$ [m$^{-3}$] has been obtained and the theoretical resonator LOD (Eqn. (36)) has been calculated to be roughly equal to 24.9 nm for a field distribution at the interface close to unity, a background refractive index equal to unity (air) and an NP refractive index equal to 1.59 RIU.

The latter value is the one associated with polystyrene NP, widely employed to emulate the presence of virus or proteins to be detected, as previously indicated.

Different NPs (in terms of size and nature) have been placed within the resonator outer periphery, i.e. where the optical field propagating as whispering gallery mode is more confined and the field distribution function achieves its maximum value.

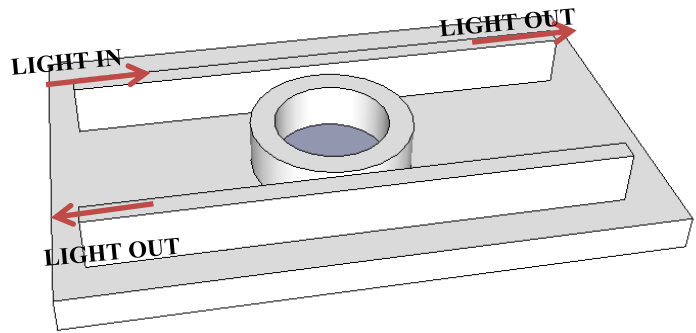

(A)

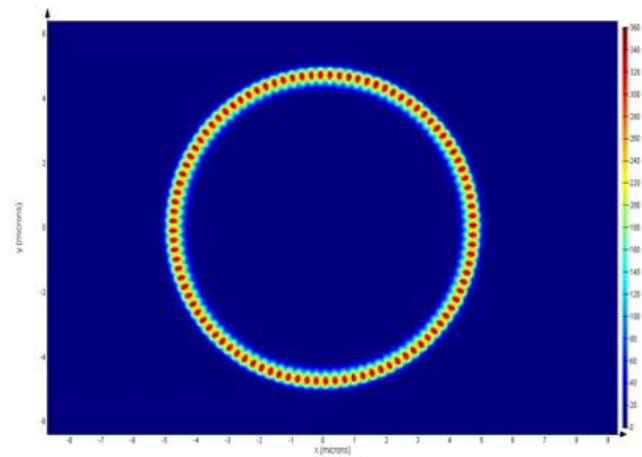

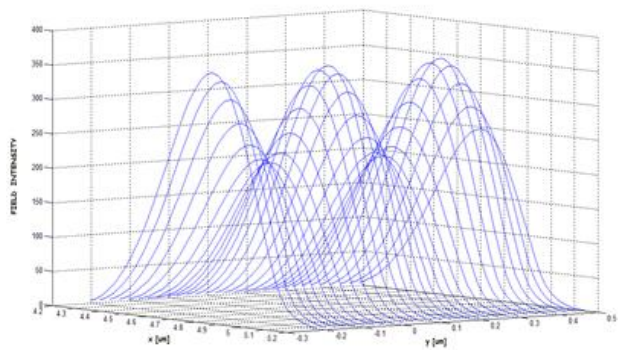

**Figure 22.** (A) Hybrid resonator geometry. (B) Electric field intensity at resonance and (C) first order radial mode field intensity at the resonator outer interface in an unperturbed condition.

For a fixed NP refractive index (n=1.59 RIU), NPs having a radius in the interest range [30:100] nm have been assumed as testing elements.

The cavity electromagnetic response in presence of the scattering NP has been then simulated to validate the theoretical investigation.

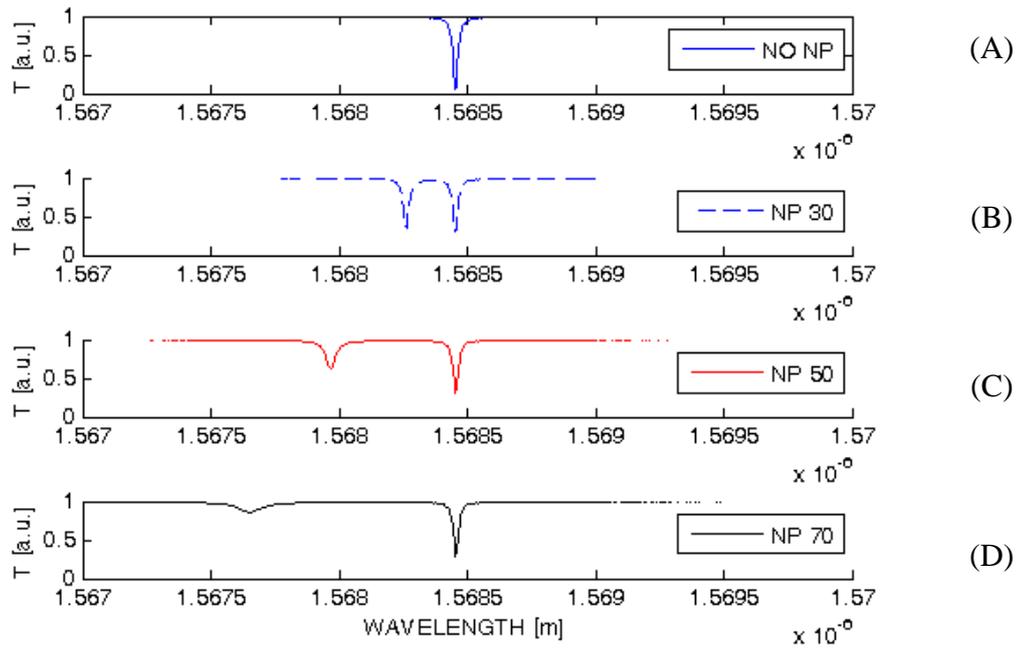

**Figure 23.** Resonator transmission spectrum in an unperturbed condition (A) and in presence of a spherical scatterer (B-D). All the curves are referred to the unperturbed condition.

A Matlab predictive code based on Eqn. (26) has also been implemented to evaluate the response of the cavity to the presence of a single NP. The theoretical transmission spectrum has been calculated by moving both NPs refractive index and nominal radius value and a Monte Carlo simulation has been performed by adding a source of noise to the estimated NP radius in order to emulate a more realistic scenario where unmodeled dynamics exist. This source of noise has been assumed to have a Gaussian distribution, with mean value equal to NP nominal radius and variance $\sigma^2 = 0.1$.

A comparison of attained results in terms of mean value and standard deviation of predicted and simulated cavity response is presented in Fig. 23. Nanoparticle refractive index has been moved in the range 1.5 : 1.59 RIU, while the radius in the range 30 : 100 nm as previously indicated.

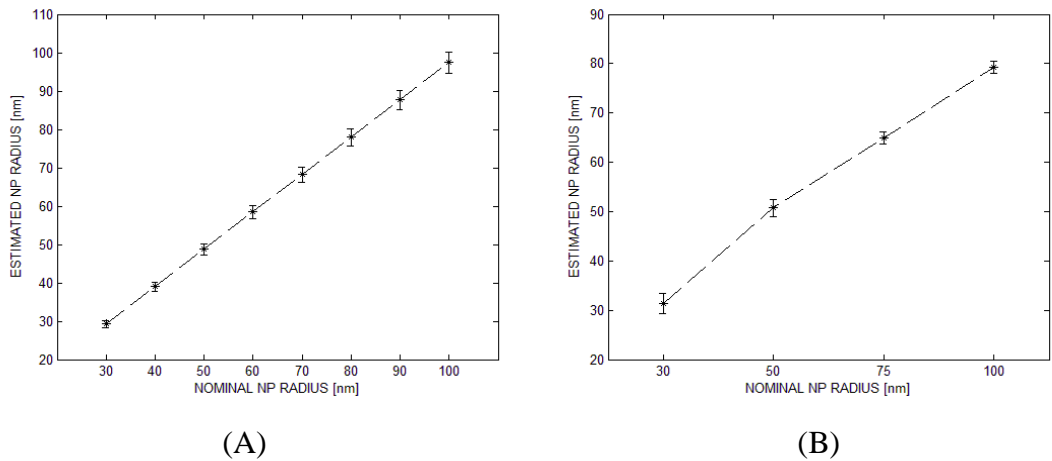

**Figure 24.** (A) Nanoparticle radius estimation calculated theoretically calculated in a MATLAB environment, (B) FDTD results.

As from Fig. 24, the designed cavity response to the presence of a single NP fits the predicted data when NP radius is small. An estimation error increasing with NP radius is visible that reaches the 20% value for a 100 nm nominal radius NP. A plausible explanation of this effect is associated to the interaction between the second order radial mode propagating within the cavity and the splitting doublet. More the NP radius increases, more the splitting doublet is broadened due to the relation between NP polarizability and splitting doublet spacing, i.e. $g \propto \alpha$, up to coupling to the higher order mode and exchanging energy with it. Also the higher order mode, indeed, suffers from the splitting phenomenon as depicted in Fig. 25 where the transmission spectrum of the hybrid resonator interacting with a 100 nm radius PS NP is plotted.

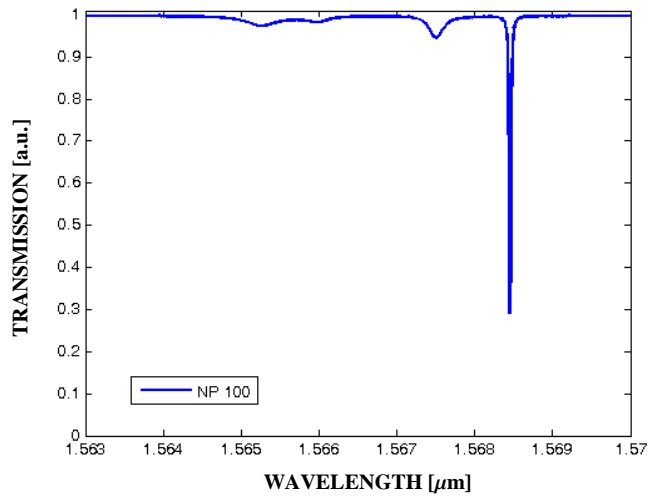

**Figure 25.** Transmission spectrum of the resonator interacting with a 100 nm radius polystyrene NP.

In absence of higher order radial mode, i.e. by employing a ring resonator based cavity, better results in terms of estimation error are obtained when NP nominal size is increased. A comparison between the hybrid and the ring resonator response is presented in Fig. 26.

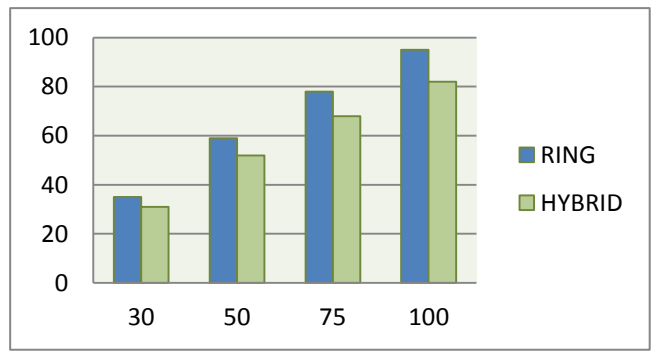

**Figure 26.** Comparison of hybrid and ring resonator in NP sizing operation.

The ring resonator improves the detection of NPs having a big size in terms of reduction of the error associated to the size estimation, but lacks of accuracy in sizing of small NPs.

A comparison between the two hybrid configurations supporting only the second order radial mode shows that the mode spacing characterizing the hybrid resonator with 1 μm wide annulus (see Fig. 27) is higher than the 750 nm one.

This peculiarity is suited to verify if an improvement in big size NPs detection could be reached by employing a different hybrid cavity.

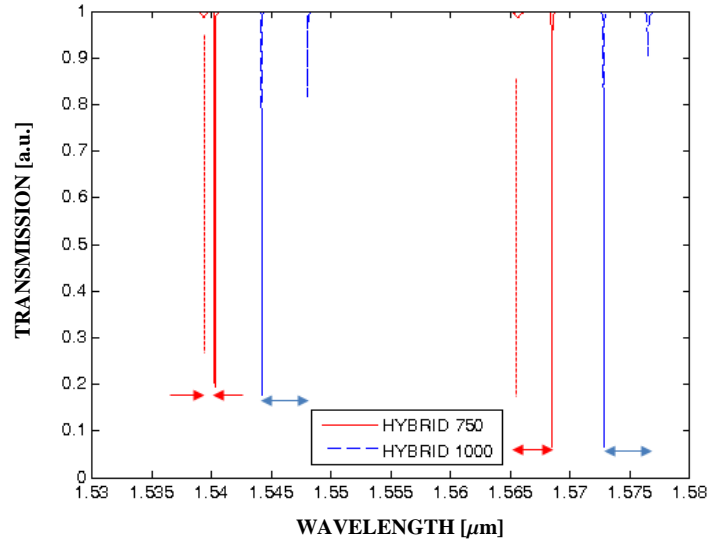

**Figure 27**. Overlapping of transmission spectrum curve of two hybrid resonators having a different annulus width.

Thus electromagnetic field simulations have been run to evaluate the cavity response in presence of a PS NP. Data are showed in Fig. 28 where the errorbar is indicating the estimating error.

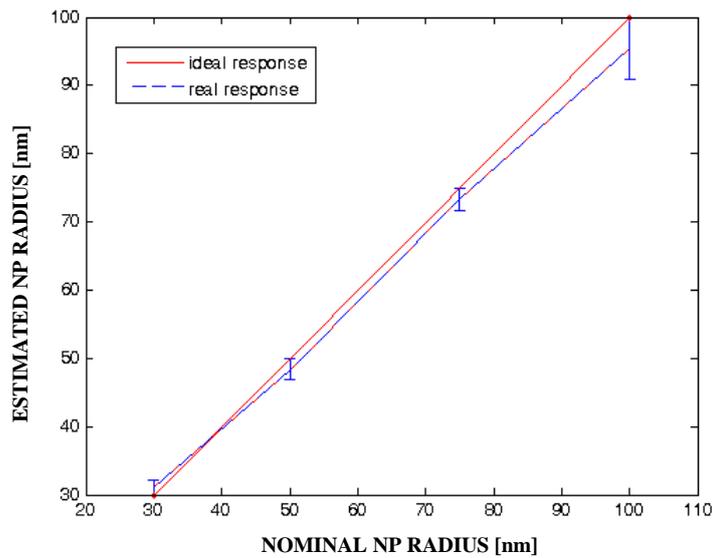

**Figure 28**. PS nanoparticle sizing by employing a hybrid cavity having a 1μm wide annulus.

An improvement in the estimation of NPs radius has been thus achieved with a hybrid resonator having an annulus width of 1 μm, coupled in a lateral 4 port configuration to two distinct waveguides.

Despite so far investigated cavities having spherical or toroidal shape and having an outer radius value about ten times larger than the one here chosen and a Q-factor one or two orders of magnitude higher than the one of the hybrid cavity, the designed resonator shows a good sensitivity to NPs detection and a LOD of 30 nm in terms of minimum detectable radius as theoretically predicted.

5.4 DISCUSSION

As reported in Chapter 4, splitting property of WGM in presence of a *structural defect* within the cavity could also be employed for NPs investigation.

The introduction of a notch on the outer periphery of a microring resonator in four ports configuration has, indeed, also been suited as possible investigative method for NPs detection (see Fig. 29 (A)) according to [27]. As theoretically predictable, the presence of a notch induces a modal splitting due to the local perturbation experienced by light propagating according to WGM physics. If a NP is placed inside the notch, the splitting doublet yet existing will be subjected to a red shift in presence of a dielectric NP, or it will shift toward the blue region of the electromagnetic spectrum if NP constituent material is a metal. This is due to the value assumed by the real part of the refractive index of dielectrics and metals which is lower than unity for these latters.

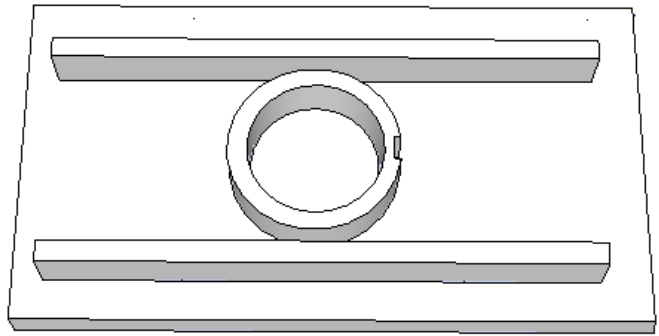

(A)

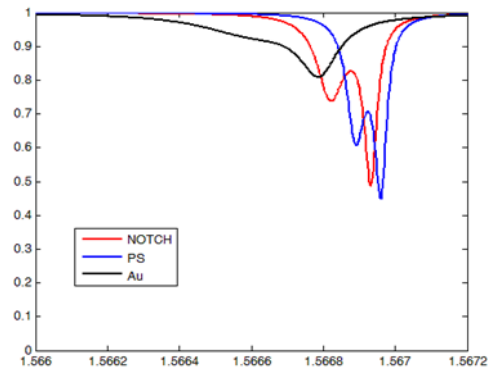

(B)

**Figure 29**. (A) Schematic of a notch ring resonator. (B) Resonator spectral response in presence of a PS and Au NP.

A similar study has been performed by employing the hybrid configuration (ΔR = 750 nm). In both the configurations, notch size has been chosen in order to weakly perturb mode propagation (a depth of 100 nm has been assumed).

As previously demonstrated, the Q-factor and the extinction ratio ER of the hybrid configuration are higher than the ones of ring configuration for an unperturbed cavity (i.e. in absence of defects). Thus, the splitting phenomenon is enhanced as visible by comparing Fig. 29 (B) with Fig. 30 (B).

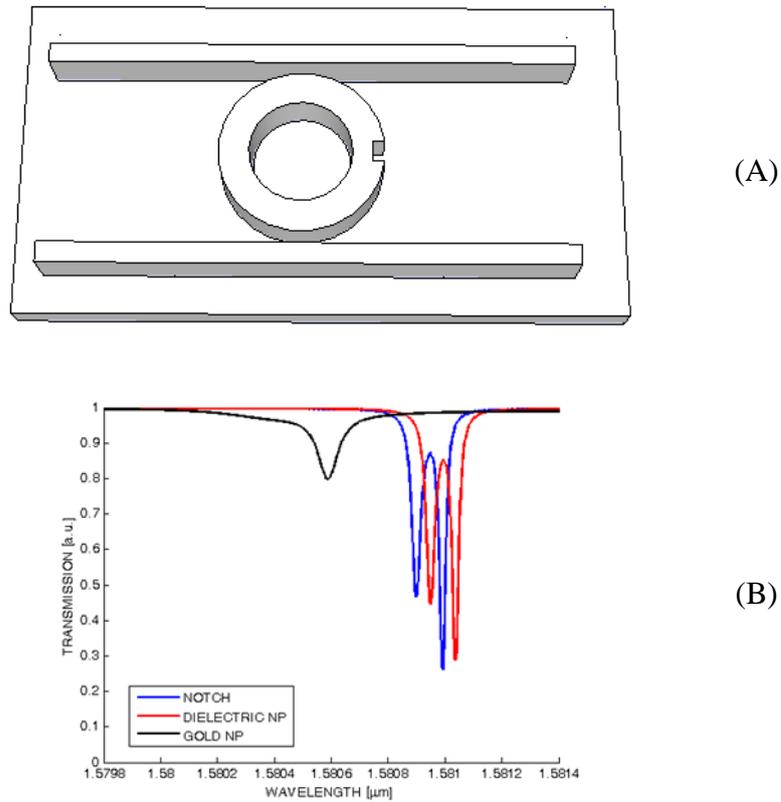

**Figure 30**. (A) Schematic of a notch hybrid ring resonator. (B) Resonator spectral response in presence of a PS and Au NP.

The introduction of a structural defect within the cavity and the insertion of the NP within the defect only leads to the identification of NP nature in terms of matter properties, i.e. dielectric or metal. This identification relies on the evaluation of the shift direction experienced by the splitting doublet yet generated by the presence of the structural imperfection. As from Figures 29 and 30, discriminating the nature of the NP does not require the employment of a resonator having stringent conditions on the Q-factor and $V_c$ value.

### 5.4.1 HYBRID RING RESONATORS ADVANTAGES

The choice of employing an hybrid configuration as sensitive element for NPs detection relies on different potential advantages it could lead to. Not only a very small modal volume is achievable (as in the microring or microdisk configuration), but also a high mode confinement along the outer cavity interface due also to the choice of the technology.

Another benefit with respect to the classical ring configuration relies on the reduction of sidewall roughness loss associated with the interaction between the inner sidewall and the mode energy; this way, only the outer sidewall loss will affect light propagation.

If, indeed, the ring is no more a circularly closed bus waveguide, but its width is bigger than the one of the coupling waveguide, the mode propagating within the resonator will be confined only along a sidewall, experiencing a loss reduction. This way, a higher tolerance during the etching phase of fabrication process is allowed.

As yet reported in Chapter 3, sources of loss affecting resonator Q-factor and extinction ratio ER are different, including substrate leakage, bending loss in the waveguide, coupling loss between waveguides and ring resonator and radiation loss at etched waveguide sidewalls. Although a strong confinement of the mode in a Si wire can be reached, the exponential tail of the mode decaying into the substrate causes substrate leakage. This loss can be reduced by increasing mode confinement, i.e. increasing waveguide section in the limit of single mode operation that means to employ a wire having a width less than 600 nm in SOI technology [28].

*Bending loss* can be optimized by optimizing the bending radius, i.e. by detecting the radius critical value below which loss have no more a linear behavior.

*Coupling loss* and coupling efficiency, as stated before, are related to the gap existing between the straight and the circular waveguide and can be well managed by employing an add-drop or a racetrack configuration.

*Scattering or radiation loss* due to the etching process at resonator sidewalls, instead, can be reduced by employing etchless silicon photonic waveguides [29] which are characterized by ultra-smooth sidewalls induced by selective oxidation. Alternatively, an opportune design of the device can lead to a decrease of the interaction between the inner sidewall and the mode energy. This way, only the outer sidewall loss will affect light propagation. The latter is one of the ideas ruling the choice of the hybrid configuration as resonant cavity biochemical sensor.

In addition, the inner part of the hybrid resonator can be filled with the analyta to be investigated and a good sensitivity in terms of resonance wavelength shift can be achieved, as described in the followings.

5.4.2 BIOLOGICAL SENSING: GLUCOSE CONCENTRATION

The designed cavity has also been tested in terms of sensitivity and detection limit for glucose concentration detection purposes. A liquid solution containing glucose disperse in a certain concentration ranging from 0 to 180 g/l has been assumed as testing sample due to the availability of experimental data concerning the optical properties of a glucose solution.

The following dependence between the average refractive index of the solution $n_{g/l}$ and the glucose concentration C has been assumed:

$$n_{g/l} = a(\lambda)C + b(\lambda) \qquad (54)$$

where a and b are two parameters related to the change of refractive index as function of glucose concentration ($a(\lambda) = dn/dC$) and the refractive index of the buffer solution (i.e. water), respectively. They are both function of the incident wavelength.

Water refractive index at $\lambda = 1550$ nm, which is the wavelength of interest, is equal to 1.3101 RIU and coincides with the $b(\lambda)$ value, while the $a(\lambda)$ value is assumed to be $1.189 \cdot 10^{-4}$ according to [30] and [31].

The resulting refractive index of samples containing a glucose concentration ranging from 0 to 180 g/l is reported in Table 5.

**Table 5.**

| C [g/l] | n | C [g/l] | n |
|---|---|---|---|
| 0 | 1.3101 | 100 | 1.32199 |
| 20 | 1.312478 | 120 | 1.324368 |
| 40 | 1.314856 | 140 | 1.326746 |
| 60 | 1.317234 | 160 | 1.329124 |
| 80 | 1.319612 | 180 | 1.331502 |

Two different detection schemes have been employed. First the presence of a fluid filling the cavity inner part has been simulated in order to test the detection ability of the device to act contemporary as sample container and sensor.

The employment of a 1000 nm wide annulus cavity filled with water shows that the fundamental mode, whose evanescent tail does not directly interact with water, experiences a shift of 20.47 pm, showing a sensitivity of 670 pm/RIU. The higher order mode, instead, is wavelength shifted of 703 pm when the medium filling the cavity is water rather than air. Resonator transmission curve is reported as an overlapping curve in Fig. 31 (B).

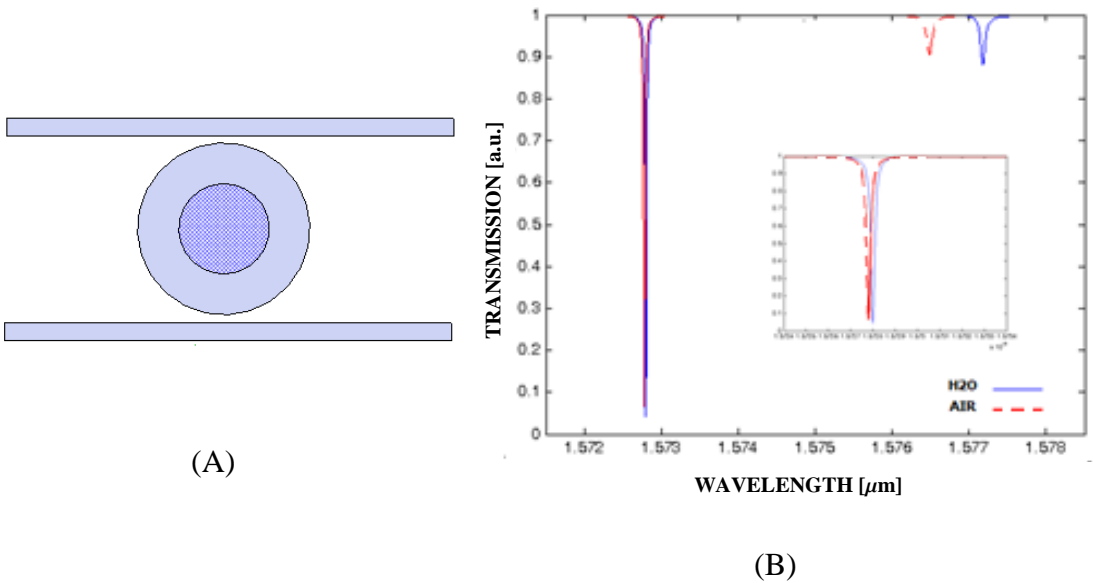

(A)

(B)

**Figure 31**. (A) Schematic of the inner part of cavity filled with water. (B) Transmission curve

Then surface sensing has been employed as sensing scheme. It has indeed been assumed the whole cavity surface to be activated and functionalized via the GOD enzymes in order to potentially detect only glucose (β–D-glucose) as analyte. GOD immobilization on commercial Si wafers has been demonstrated by employing conventional methods [32]. GOD refractive index has been assumed to be equal to 1.45 RIU which is the average RI value of proteins at λ = 1550 nm [33]. Effective index change as function of the immobilized layer thickness has been investigated, as reported in Fig. 32.

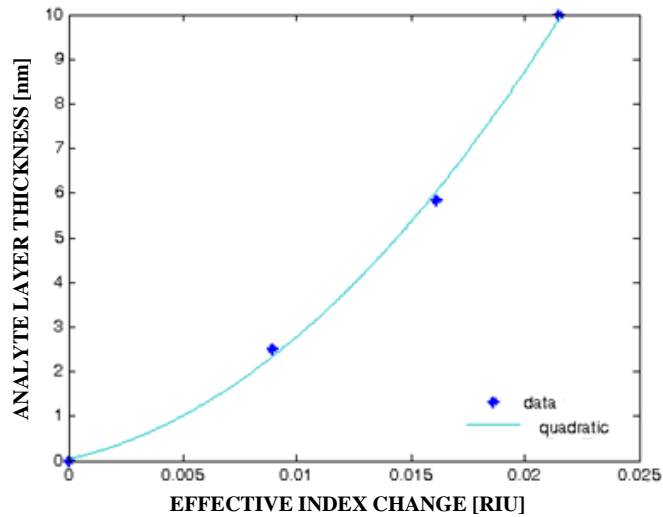

**Figure 32.** Cavity effective index change as function of thickness of immobilized receptors layer.

Following anyway theoretical and experimental investigations, a receptors layer thickness of 10 nm has been assumed. With these hypothesis, FDTD simulations have been run and a linear redshift of resonant wavelength has been observed as depicted in Fig. 33. Also cavity Q-factor and ER value degrades from $1.7 \cdot 10^5$ to $1.08 \cdot 10^5$ and from 23.37 dB to 21.97 dB, respectively.

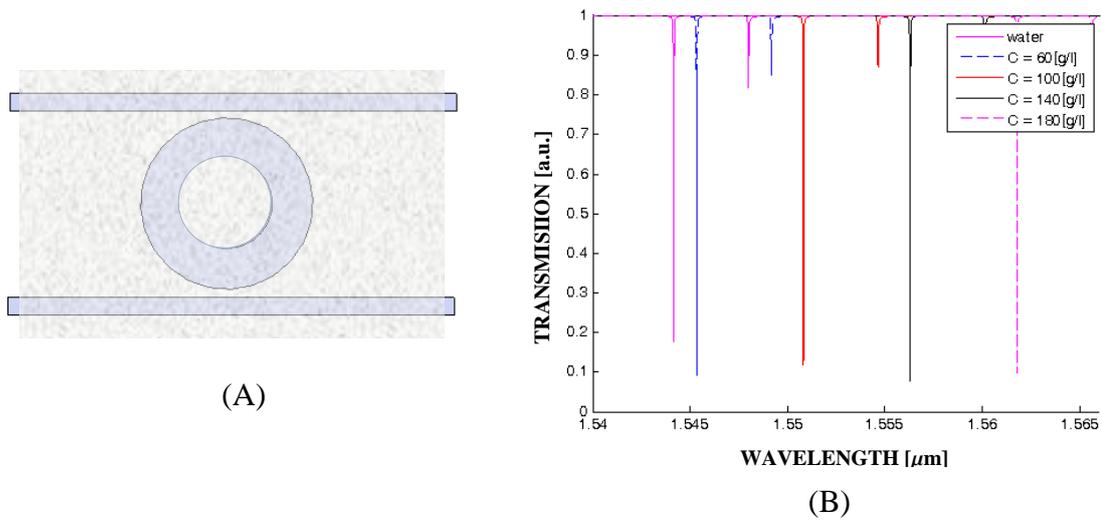

(A)

(B)

**Figure 33**. (A) Schematic of the cavity. (B) Overlapping of resonator transmission curve obtained changing glucose concentration in solution.

The device analytical sensitivity, as defined in Chapter 2, has been calculated to be 553 nm/RIU by fitting data with a linear algorithm. This way also the sensor calibration curve (see Fig. 33) has been obtained. The minimum detectable effective index change has been estimated to be $10^{-6}$ RIU with a wavelength shift of 2 pm. This value of $10^{-6}$ RIU defines the sensor LOD.

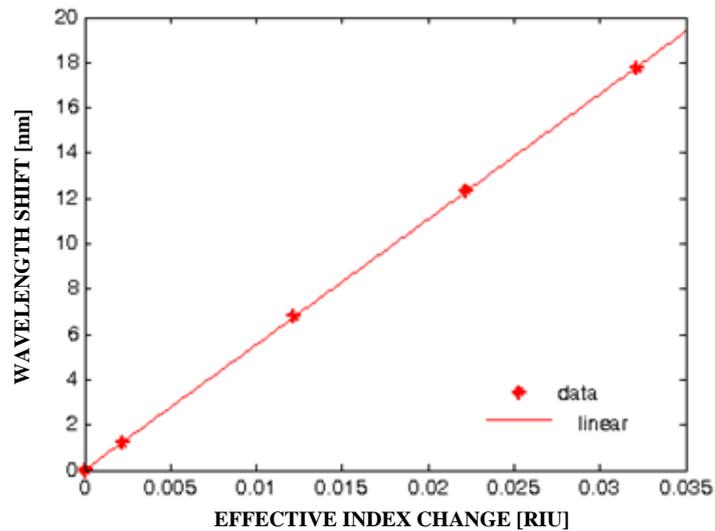

**Figure 34.** Glucose sensor calibration curve. Data have been fitted with a linear function y = $p_1$x + $p_2$, where $p_1$ =553.4 and $p_2$ = 0.014.

The designed hybrid resonator offers thus the advantages of high Q-factor and low modal volume *Vc* that have been suited for biochemical sensing. The versatility of the proposed cavity, indeed, allows to detect a single NP with a size down to 30 nm and to investigate glucose concentration in solution, showing a sensitivity of more than 500 nm/RIU, a LOD of $10^{-6}$ RIU and a potential selectivity induced by the molecular immobilization technique.

In the examined scenario, GOD immobilization on the sensor surface is the critical step for optimizing device performances. If the analyta to be detected is indeed contained in a complex solution such as blood, a big relevance is assumed by the employment of the opportune molecular receptors.

### 5.4.3 CHALLENGES: ON CHIP PLATFORM

Due to the versatility of the proposed hybrid cavity, it could be employed as sensitive element in a more complex platform. Its sensitivity make it a good candidate for biosensing applications devoted to the refractometric detection of the concentration of analytes in a complex solution. A possible on chip configuration for a high throughput and multianalyte detection relies on the employment of an array of resonant cavities interacting with a single nano-fluidic channel, as reported in Fig. 35. The surface of every resonator should be chemically modified with the receptors that specifically bind to the interest analyta in the complex solution.

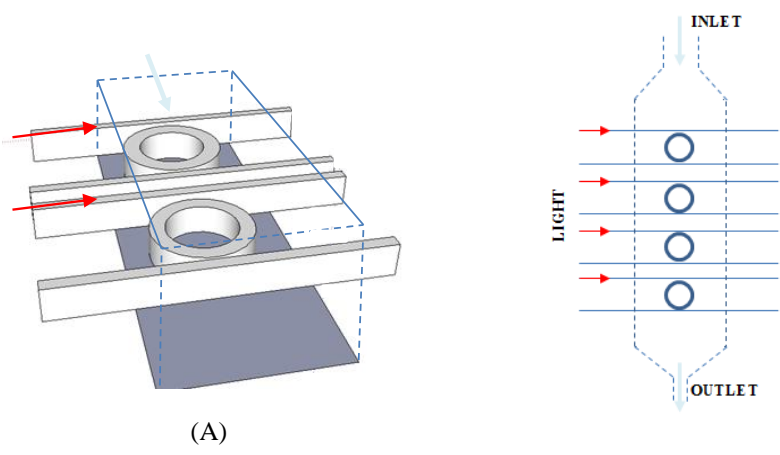

(A)                                    (B)

**Figure 35.** Schematic of the proposed configuration in (A) prospective and (B) top view.

As evident from Fig. 35, a pressure driven flow based pumping technique has been hypothesized. It relies on creating a pressure gradient by modifying the channel section according to Bernoulli's principle. The outlet channel section should also be chosen in order to slow the solution flow and so to increase the time the analyta interacts with the functionalized resonator surface. A real time analysis of molecular interactions indeed takes several seconds as reported in different experimental works [34].

At the same time, the proposed cavity could be suited to detect the presence of NPs by addressing the fluid to be investigated toward an opportune sensing area on the resonator surface. Properly, a microfluidic apparatus aimed to drive nanoparticles toward the sensing area of the microresonator should be carefully designed. The pumping mechanism capable to flow the fluid sample containing beads into the cavity could rely on the employment of a vertical optofluidic structure composed by two planes interacting via an auxiliary channel. The latter should indeed flow the sample from the primary channel (situated on the top plane) into the sensing area (in the lower plane). Since particle size plays an important role in the determination of their toxicity, a flat sheet membrane could be employed in order to filter beads (nanoparticles or viruses) flowing from the primary microfluidic channel to the resonator. This will allow only particles having a size smaller than the one of membrane pores to flow through the membrane and to reach the sensing area. Also, channel dimensions (as stated by fluid dynamics laws) should be set up in order to get an opportune sample flow rate inside the microfluidic channels and regulate the interaction time between nanoparticles dissolved in the sample and the microresonator.

The size of the vertical channel should also be designed in order not to perturb the microresonant cavity optical properties, i.e. its effective index and modal distribution.

A low sample volume consumption is achievable due to the reduced cavity size.

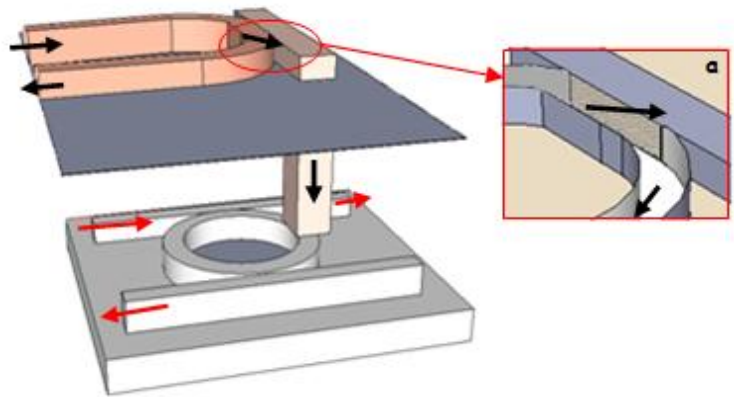

**Figure 36.** Schematic of the optofluidic proposed cavity. In the inset *a* the permeation of the fluid through the membrane is depicted.

5.5 CONCLUSIONS

A method for NPs detection and sizing has been analyzed. It relies on the coupling of CW and CCW travelling wave modes propagating within a WGM based cavity occurring when light is perturbed along its path by the presence of a scattering element, i.e. a NP or surface roughness. On the basis of the analytical formulae modeling the resonator-NP interaction, some criteria for designing a planar cavity for single NP detection have been deduced and a cavity with a high Q-factor, a low modal volume $V_c$ and a high extinction ratio ER value has been designed and tested for NPs detection and sizing by employing an FDTD method based commercial software. NPs having a radius in the range 30 : 100 nm have been employed as testing elements and the estimation error in NP sizing is in the order of 2%. The sensitivity of the cavity has also been investigated in terms of glucose concentration detection in solution. The refractometric detection scheme has been employed and an analytical sensitivity of about 550 nm/RIU and a LOD of $10^{-6}$ RIU have been measured. Thus the designed cavity is a versatile device for biosensing applications, i.e. for NP detection and sizing and analyta concentration detection, and could be employed in a biosensing platform on chip

## 5.6 REFERENCES

1. F. Vollmer, S. Arnold, D. Keng, "Single virus detection from the reactive shift of a whispering-gallery mode," P. Natl. Acad. Sci. USA 105(52) pp. 20701-20704, 2008.
2. C. Ciminelli, C.M. Campanella and M.N. Armenise, "Hybrid optical resonator for nanostructured virus detection and sizing," Medical Measurements and Applications Proceedings (MeMeA), 2011 IEEE.
3. L. Mandel and E. Wolf, "Optical coherence and quantum optics," Cambridge University Press, 1995.
4. A. Mazzei, S. Götzinger, L. de S. Menedes, G. Zumofen, O. Benson, V. Sandoghdar, "Controlled coupling of counter-propagating whispering-gallery-modes by Rayleigh scatterer", Phys. Rev. Lett., Vol. 99, Issue 17, 2007.
5. J. Zhu, S.K. Ozdemir, Y.F. Xiao, L. Li, L. He, Da-Ren Chen and L. Yang, "On-chip single nanoparticle detection and sizing by mode splitting in an ultrahigh-Q microresonator", Nature Photonics, Vol.4, 2010.
6. X. Yi, Y-F. Xiao; Y-C. Liu; Bei-Bei Li, You-Ling Chen; Y. Li and Q. Gong, "Multiple-Rayleigh-scatterer-induced mode splitting in a high-Q whispering-gallery-mode microresonator", Physical Review A, vol. 83, Issue 2, 2011.
7. P. Meystre and M. Sargent III, "Elements of quantum Optics," 4th Edition, Springer 2007.
8. S. Haroche and J.-M. Raimond, "Exploring the Quantum: Athoms, Cavities and Photons," Oxford University Press, 2006.

9. W.M. Itano, C. Monroe, D.M. Meekhof, D. Leibfried, B.E. King and D.J. Wineland, "Quantum harmonic oscillator: state, synthesis and analysis,"

10. T. J. Kippenberg, S. M. Spillane, K. J. Vahala, "Modal coupling in travelling-wave resonators", Optics Letters, Vol.27, No.19, 2002

11. K. Srinivasan, M.Borselli, O. Painter, A. Stintz and S. Krishna, "Cavity Q, mode volume, and lasing threshold in small diameter AlGaAs microdisks with embedded quantum dots," Optics Express, Vol. 14, No. 3, 2006.

12. L. Tsang, J.A. Kong and K.H. Ding, "Scattering of electromagnetic Waves: Theories and Applications, " John Wiley & Sons, 2000.

13. G.S. Solomon, Z. Xie, W. Fang, J.Y. Xu, A. Yamilov, H. Cao, Y. Ma and S.T. Ho, "Large spontaneous emission enhancement in InAs quantum dots coupled to microdisk whispering gallery modes," Phys. Stat. Sol. (b), Vol. 238, No. 2, pp. 309–312, 2003.

14. H.K. Hunt and A.M. Armani, "Label-free biological and chemical sensors", Nanoscale, 2(9), pp. 1544–1559, 2010.

15. A. Samarelli, M. Gnan, R.M. De La Rue and M. Sorel, "Low propagation loss photonic wire and ring resonator devices in silicon-on insulator using hydrogen silsesquioxane electron-beam resist," Proceedings of the European Conference on Integrated Optics (ECIO), Eindhoven, The Netherlands, 11-13 June 2008.

16. C. Buzea, I.I. Pacheco Blandino and K. Robbie, "Nanomaterials and nanoparticles: Sources and toxicity," Biointerphases, Vol. 2, Issue 4, pp. MR17 - MR172, 2007.

17. D.G. Rabus, "Integrated Ring Resonators. The Compendium," Springer, 2007.

18. A. Yariv, "Critical Coupling and Its Control in Optical Waveguide-Ring Resonator Systems," IEEE Photonics Technology Letters, Vol. 14, No. 4, 2002.

19. J.H. Chow, M.A. Taylor, T.T-Y. Lam, J. Knittel, J.D. Sawtell-Rickson, D.A. Shaddock, M.B. Gray, D.E. McClelland and W.P. Bowen, "Critical coupling control of a microresonator by laser amplitude modulation," Optics Express, Vol. 20, Issue 11, pp. 12622-12630, 2012.

20. W. Bogaerts, P. De Heyn, T. Van Vaerenbergh, K. DeVos, S.K. Selvaraja, Tom Claes, P. Dumon, P. Bienstman, D. Van Thourhout, and R. Baets, "Silicon micoring resonators," Laser and Photonics Review, Vol. 6, No. 1, pp. 47–73, 2012.
21. C.M. Campanella, "Project and fabrication of an optical cavity resonator based biosensor," Tesi di Laurea Specialistica in Ingegneria dell'Automazione, Relatrice: Prof. C.Ciminelli (Politecnico di Bari); Correlatori: Prof. R.M. De La Rue, M. Sorel (The Glasgow University, UK).
22. I.H. El-Sayed, X. Huang, M.A. El-Sayed "Surface Plasmon Resonance Scattering and Absorption of anti-EGFR Antibody Conjugated Gold Nanoparticles in Cancer Diagnostics: Applications in Oral Cancer" Nano Letters, Vol. 5, No. 11, 2005.
23. R.K. Visaria, R.J. Griffin, B.W. Williams, E.S. Ebbini, G.F. Paciotti, C.W. Song, J.C. Bischof, "Enhancement of tumor thermal therapy using gold nanoparticle–assisted tumor necrosis factor-α delivery", Molecular Cancer Therapeutics, Vol. 5, 2006.
24. Kubo, A. Diaz, Y. Tang, T.S. Mayer, I. Choon Khoo, T.E. Mallouk, "Tunability of the refractive index of gold nanoparticle dispersions," Nano Letters, Vol. 7, No. 11, 2007.
25. C. Ciminelli, C.M. Campanella, R. Pilolli, N. Cioffi and M.N. Armenise, "Optical sensor for nanoparticles, "Optical sensor for nanoparticles," 13th International Conference on Transparent Optical Networks (ICTON), 26-30 June 2011, Stockholm (Sweden).
26. E.D. Palik, "Handbook of optical constants of solids," Elsevier Science , USA, 1991.
27. S. Wang, K. Broderick, H. Smith, and Y. Yi, "Strong coupling between on chip notched ring resonator and nanoparticle," Appl. Phys. Lett. 97, 051102, 2010.
28. W. Bogaerts, V. Wiaux, J. Wouters, S. Beckx, J. Van Campenhout, D. Taillaert, B. Luyssaert, P. Bienstman, D. Van Thourhout, and R. Baets, "Low-Loss SOI Photonic Wires and Ring Resonators Fabricated With Deep UV Lithography," IEEE Photonics Technology Letters, Vol. 16, No. 5, pp. 1328-1330, 2004.

29. J. Cardenas, C.B. Poitras, J.T. Robinson, K. Preston, L. Chen, and M. Lipson "Low loss etchless silicon photonic waveguides," Optics Express, Vol. 17, Issue 6, pp. 4752-4757, 2009.

30. Yen-Liang Yeh, "Real-time measurement of glucose concentration and average refractive index using a laser interferometer," Optics and Lasers in Engineering, Vol. 46, pp. 666–670, 2008

31. D. Daly and G.C. Lein, "Optical measurement of glucose content of the aqueous humor," Applied Diagnostics Report, http://www.maths-in-industry.org/miis/31/1/Glucose.pdf

32. A. Curulli, A. Cusmà, S. Kaciulis, G. Padeletti, L. Pandolfi, F. Valentini and M. Viticoli, "Immobilization of GOD and HRP enzymes on nanostructured substrates," Surface and Interface Analysis, Vol. 38, Issue 4, pp. 478–481, 2006.

33. J. Vörös, "The Density and Refractive Index of Adsorbing Protein Layers," Biophysical Journal, Vol. 87(1), pp. 553–561, 2004.

34. Y. Sasuga, T. Tani, M. Hayashi, H. Yamakawa, O. Ohara, and Y. Harada, "Development of a microscopic platform for real-time monitoring of biomolecular interactions," Genome Research, Vol. 16(1), pp. 132–139, 2006.

# Chapter 6.
# INTERFEROMETRIC REFLECTANCE IMAGING SENSOR BASED NANOPARTICLE DETECTION

Microscopes are usually used to image micron sized objects as their name suggests. In recent years, several microscopy techniques have been developed to characterize nano-scale particles [1-5]. In particular, the Interferometric Reflectance Imaging Sensor (IRIS) has been successfully demonstrated for material specific detection, sizing and shape determination of single NPs [6-10]. However, this technique has been limited to end-point measurement due to the sample preparation process and its measurement procedure. In this chapter we demonstrate that this technique can also be implemented in real-time enabling sensitive measurements to be performed in order to understand important dynamic behavior of biological processes.

This chapter is organized as follows: first the basic principle of the IRIS is discussed. Then a way to improve the detection capability of the technique is proposed. It relies on the employment of solid immersion lens (SIL) technology and on the confocal detection. After developing the theoretical foundations of the proposed method, the optimized operation conditions of the sensor are identified in terms of detection limit.

6.1 BASICS OF INTERFEROMETRIC MICROSCOPY FOR NANOPARTICLE DETECTION

The main idea forming the basis of IRIS for NP detection is based on coherent mixing of the weak scattered light from NPs with a strong reference light through using an optimized layered dielectric substrate on which the nanoparticles are either immobilized specifically or adsorbed on the surface due to electrostatic interaction. During the measurement this substrate is illuminated with a coherent or incoherent light source (i.e. a laser or an LED, respectively ); then the scattered light from NPs and reference light originated from the substrate are detected by a single element or an array based photoreceiver (i.e. a photodiode or CCD, respectively).

A schematic of a setup based on incoherent illumination and projection of the image into an array detector is depicted in Fig.1.

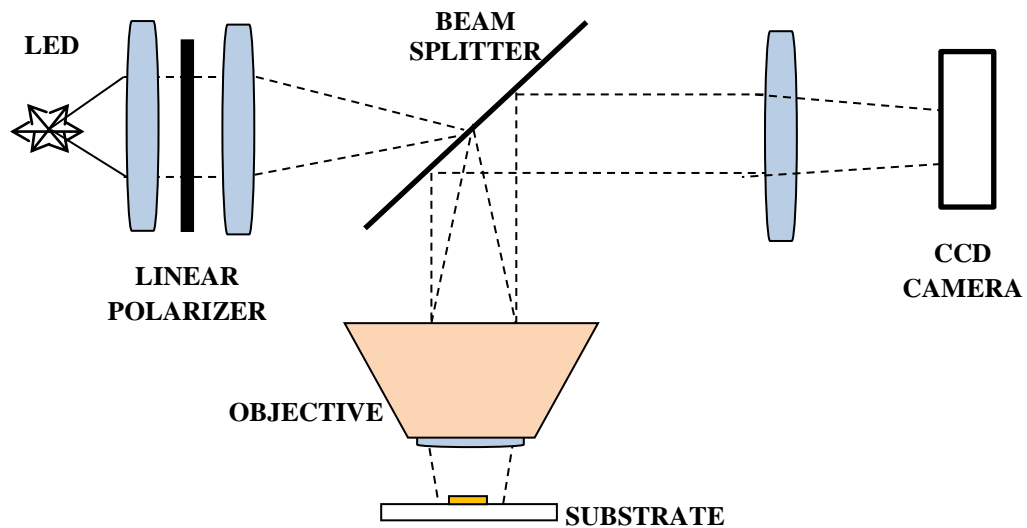

**Figure 1.** Schematic of setup based on incoherent illumination.

As previously indicated (see Chapter 5) the intensity of light scattered by a small spherical NP in the electrostatic approximation is proportional to the 6$^{th}$ power of particle radius. Thus, intensity of scattered light at the detector $I_{det}$ is directly dependent on particle size, due to relations:

$$I_{det} \propto |E_{sca}|^2 \propto \alpha^2 \propto K^2 r^6 \qquad (1)$$

In Eqn. (1), $\alpha$ is NP polarizability, K is the Clausius-Mossotti coefficient and r is NP radius, as defined in the previous Chapter.

Due to the strong size dependence ($r^6$), the scattered light intensity quickly vanishes for small NPs below the noise floor of the camera or detector.

In the IRIS, this difficulty is overcome by mixing the weak field scattered by NP with a stronger reference field reflected at the reference interface (for example, the substrate over which the NP is located).

In Fig. 2, the basic principle of interferometric detection is shown.

Since the total field can be expressed as the sum of scattered $E_{sca}$ and reference field $E_{ref}$, the intensity at the detector can be found as follows:

$$I_{det} \propto |E_{ref} + E_{sca}|^2 \propto |E_{ref}|^2 + |E_{sca}|^2 + 2|E_{ref}||E_{sca}|\cos\theta \qquad (2)$$

Where $\theta$ is the phase difference between the reference and the scattered field.

The third term represents the interferometric mixing term which is proportional to $r^3$. Thus, the strong dependence of the scattered field on the size of the particle is reduced and the weak scattered field amplitude ($|E_{sca}|$) is also amplified by the strong reference field amplitude ($|E_{ref}|$).

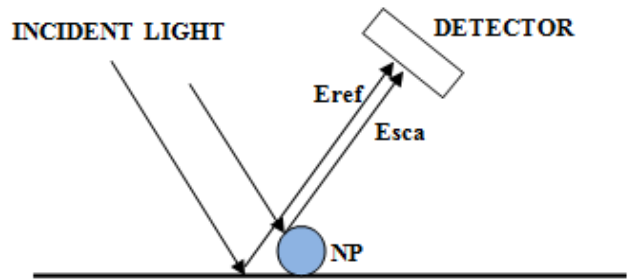

**Figure 2.** Interferometric detection.

In this study, we devise several methods to improve the current state-of-the-art interferometric microscopy for NP detection. Studies conducted so far on this interferometric microscope are based on *widefield illumination and detection* (Fig. 1), i.e. the sample is illuminated by an extended uniform light beam and the corresponding signal is detected from an extended area of the sample [6-10].

We first demonstrate that an improvement of the set-up can be realized by adopting a different microscope scheme based on the confocal detection of the interferometric term shown in Eqn. (2). In a confocal IRIS a laser beam is focused on a NP and resulting scattered and reference light is collected and focused on a point photodetector as shown in Fig 3.

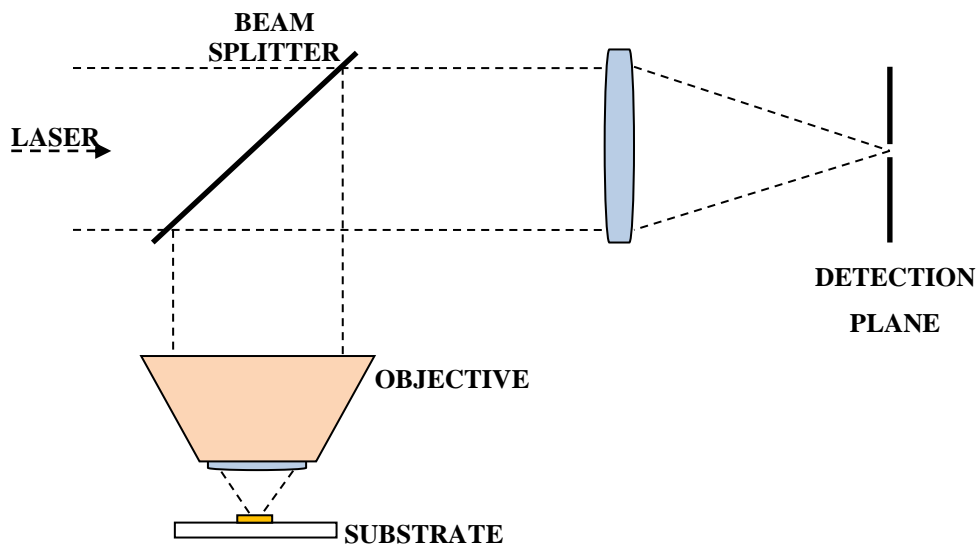

**Figure 3.** Confocal microscopy scheme.

The advantage of using confocal scheme is that the spatial resolution and detection bandwidth for per particle can be improved significantly in principle in order to investigate fast bio-processes occurring at the single nanoparticle limit. Secondly, we study the details of how a SIL can improve the detection limit and sensitivity of this NP detection scheme through increasing the numerical aperture (NA) of the microscope. The SIL attached to the backside of the chip thus also allows performing dynamic detection in a real-time as the optical aberrations are eliminated between the objective and its focal plane where NPs are captured.

We here perform a rigorous theoretical analysis of such configuration of the IRIS in order to optimize the operating conditions and understanding its limits.

The configuration employed to model the system is a 4f configuration, i.e. two lens are interposed between the object plane and the image plane as depicted in Fig. 4 [11].

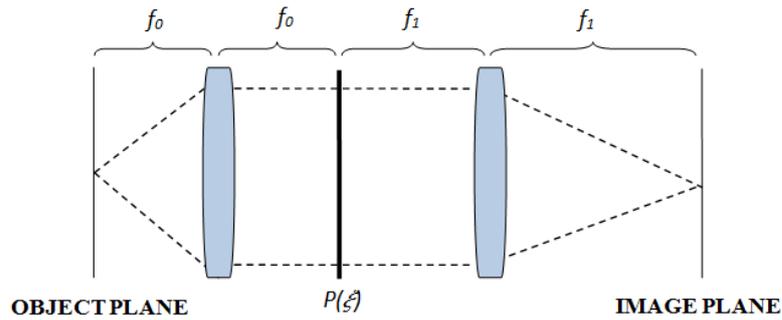

**Figure 4**. Four *f* configuration.

The analytical approach followed is at first the evaluation of the focused incident field; then the interaction between the incident light and the single NP in the electrostatic approximation (see Chapter 5), i.e. by assuming the NPs acting as dipole scatterers, has been calculated.

After this primary study, the Angular Spectrum Representation (ASR) is employed as the mathematical instrument useful to model the reference and the scattered field in the object space (the one containing the NPs and the substrate) and then to propagate them into the image space (the one containing the detector). This last step completes the mathematical model of this imaging system.

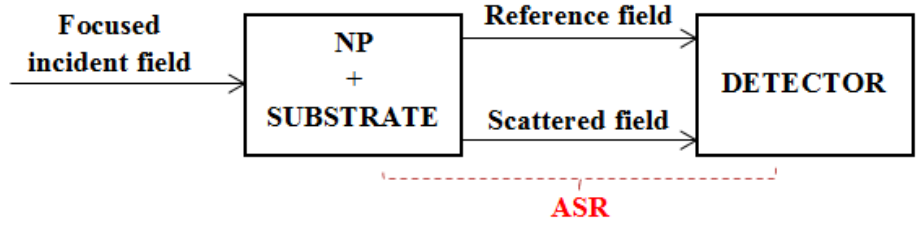

**Figure 5.** Schematic of the analytical approach followed during this study.

Here a brief description of ASR is given to clarify the analytical method that lays in the center of this investigation. A more detailed explanation can be found in [11-14].

6.2 ANGULAR SPECTRUM REPRESENTATION

The ASR is an analytical technique useful to describe electromagnetic fields in homogeneous media as superposition of plane and evanescent waves.
This method is widely employed to model the propagation of a beam through a lens or a planar or spherical interface and it is a fundamental instrument in this study.
Assuming the electric field at a point r=(x,y,z) in the object space, i.e. E(r) = E(x,y,z), the ASR enables to calculate the field in a plane z= constant (usually in the plane (x,y, z=0)) perpendicular to the propagation axis.
The procedure employed to derive the ASR of the E-field consists, first, in defining the Fourier representation of E(r):

$$\hat{E}(k_x,k_y;z) = \frac{1}{4\pi^2} \int\int_{-\infty}^{\infty} E(x,y,z)e^{-i(k_x x + k_y y)} dxdy \qquad (3)$$

Where (x,y) are the Cartesian coordinates in the transverse plane, while ($k_x$, $k_y$) are the corresponding spatial frequencies.

In the hypothesis that the medium in the transverse plane is linear, isotropic, homogeneous and source-free, the time-harmonic optical field has to satisfy the vector Helmholtz equation:

$$\left(\nabla^2 + k^2\right)E(r) = 0 \tag{4}$$

Here, $k$ is the wavenumber, i.e. $k=(2\pi/\lambda)*n)$, and $n$ is the index of refraction of the medium, i.e. $n=(\mu\epsilon)^{1/2}$.

Since the expression of the spatial frequency along z-axis is:

$$k_z = \sqrt{(k^2 - k_x^2 - k_y^2)} \tag{5}$$

With Im$\{k_z\} \geq 0$, after inserting Eqn. (3) in (4) and by employing (5), the following relation between Fourier spectrum of E-field in two different planes is derived:

$$\hat{E}(k_x, k_y; z) = \hat{E}(k_x, k_y; 0)e^{\pm ik_z z} \tag{6}$$

The exponential term in Eqn. (6) is called *propagator in reciprocal space* since it allows to calculate the Fourier spectrum of the E-field in a z-constant plane (that we assume as *image plane*) by knowing the F-spectrum in another plane at z=0 (*object plane*).

The sign ± is associated to the propagation direction: if the wave is propagating in the half space with z>0, the + sign has to be considered; while the − sign denotes a wave propagating into the half-space z<0.

Defining the inverse Fourier transform of the E-field as:

$$E(x, y, z) = \int\int_{-\infty}^{\infty} \hat{E}(k_x, k_y; z)e^{i(k_x x + k_y y)} dk_x dk_y \tag{7}$$

And inserting Eqn.(6) in (7), the ASR of the E-field is found:

$$E(x,y,z) = \int\int_{-\infty}^{\infty} \hat{E}(k_x,k_y;0) e^{i(k_x x + k_y y \pm k_z z)} dk_x dk_y \qquad (8)$$

The term Ê is called *angular spectrum* since it describes the amplitude and the propagation directions of the different waves (plane and evanescent) composing the field.

The wavenumber $k_z$ can indeed be real or imaginary, turning the propagator in reciprocal space into an oscillatory or exponentially decaying function defining either a plane or an evanescent wave, respectively, as resumed in Fig. 6.

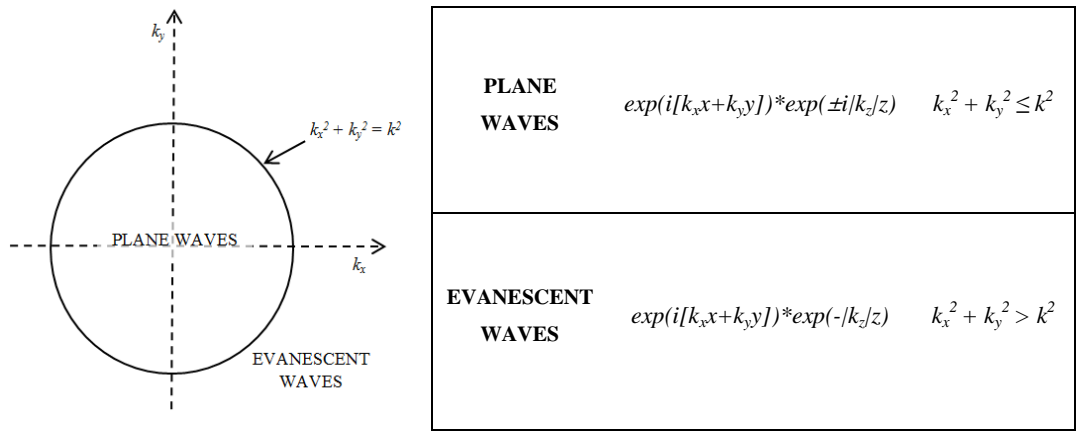

**Figure 6.** Graphical representation of the nature of a wave according to the condition satisfied by the spatial frequencies.

## 6.3 ANGULAR SPECTRUM REPRESENTATION OF A FAR FIELD

Mapping a field, whose distribution is known in a plane z =0 (object plane), into another plane z = $z_0$ (image plane)(where $z_0 \gg 0$), requires to evaluate the field in a point $r_\infty$ far from the object plane.

The field to be calculated is named far-field and will be indicated in the followings as $E_\infty$.

By defining r as the distance of the observation point $r_\infty$ from the origin, i.e. $r^2 = x^2 + y^2 + z^2$, a dimensionless unit vector $s = (s_x, s_y, s_z)$ can be employed in the direction of $r_\infty$ in order to calculate the double integral of Eqn. (6), i.e. $s = (x/r, y/r, z/r)$:

$$E_\infty(s_x, s_y, s_z) = \lim_{kr \to \infty} \iint\limits_{(k_x^2 + k_x^2) \leq k^2} \hat{E}(k_x, k_y; 0) e^{ikr(\frac{k_x}{k} s_x + \frac{k_y}{k} s_y \pm \frac{k_z}{k} s_z)} dk_x dk_y \qquad (9)$$

The integration range is reduced to plane waves since evanescent waves do not contribute to the fields at infinity due to their exponential decaying behavior.
By employing the method of stationary phase and the rules of geometrical optics [12], [15], Eqn. (9) can be reduced to:

$$E_\infty(k_x, k_y) = -\frac{i 2\pi k_z e^{ikr}}{r} \hat{E}(k_x, k_y; 0) \qquad (10)$$

By expressing the Fourier spectrum Ê in terms of far-field, Eqn. (8) becomes:

$$E(x, y, z) = \frac{ire^{-ikr}}{2\pi} \iint\limits_{(k_x^2 + k_x^2) \leq k^2} \frac{1}{k_z} E_\infty(k_x, k_y) e^{i(k_x x + k_y y \pm k_z z)} dk_x dk_y \qquad (11)$$

Equation (11) is the angular spectrum representation of the far-field in the image plane.
This description has a purely theoretical nature and it is not related to a realistic optical problem, i.e. neither the way the fields propagate from the object to the image space, neither their nature have been defined so far.
Thus, a brief description of the propagation of a paraxial optical field through a focusing optical element, i.e. a lens, will be given.

## 6.4 FIELD FOCUSED BY AN APLANATIC LENS

An aplanatic lens (see Appendix 2) will be assumed as focusing element, i.e. a lens that corrects on-axis wave-front aberrations in an imaging system. Some Geometrical Optics (GO) rules are followed to describe an aplanatic lens, known as *sine condition* and *intensity law rules*. The main assumption of GO is that energy is transported along light rays.

The sine condition rule states that every optical ray emerging from (or converging to) the point of focus F of an aplanatic lens intersect its conjugate ray* on a sphere, called *Gaussian reference sphere*, having radius *f* equal to the focal length of the lens.

The intensity law is instead related to the principle of conservation of energy. It states that the energy incident on an aplanatic lens equals the energy that leaves the lens.

According to these rules, whose more exhaustive explanation can be found in [12],:

- An aplanatic lens can be modeled as a spherical surface having radius *f*;
- Incident rays are refracted by the reference sphere and it results more convenient to express their components in spherical coordinate system $(\theta, \phi)$;
- The lens modifies the incident field into the refracted field according to the intensity law:

$$|E_r| = |E_{inc}| \sqrt{\frac{n_{inc}}{n_r}} \sqrt{\cos(\theta)} \qquad (12)$$

In Eqn. (12) magnetic permeability has been neglected since its value is unitary at optical frequency in all media. $\theta$ is the angle between the refracted ray and the optical axis, $n_{inc}$ is the refractive index of the medium in which the incident field is propagating, while $n_r$ is the index of the medium after the lens.

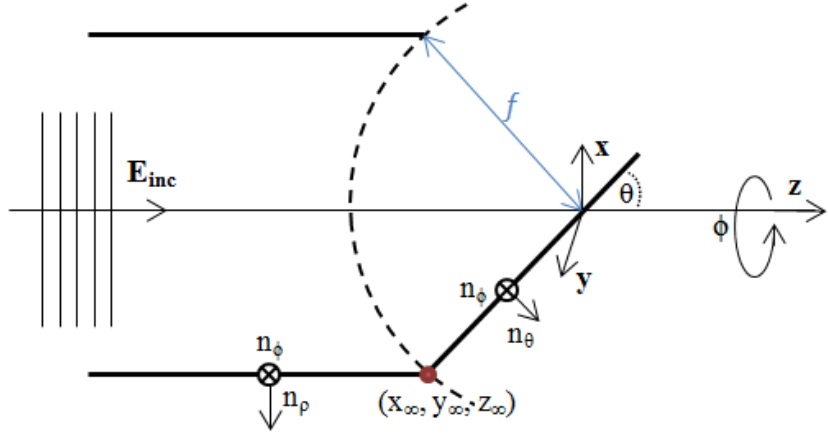

**Figure 7.** Aplanatic system described through geometrical optics.

Since it results more convenient to model refraction at reference sphere by splitting the incident field $E_{inc}$ into two components, denoted as $E_{inc}{}^s$ and $E_{inc}{}^p$, unit vectors of a cylindrical coordinate system $n_\phi$ and $n_\rho$ are used to characterize the field components:

$$E_{inc} = E_{inc}\bar{n}_\phi + E_{inc}\bar{n}_\rho = E_{inc}{}^s + E_{inc}{}^p \qquad (13)$$

Properly, s stands for s-polarization (the E-field component is parallel to the interface) and p for p-polarization (the E-field component is perpendicular to the k-vector and to the interface).

By employing the following matrix transformation to express unit vectors of cylindrical and spherical coordinates in Cartesian coordinates:

$$\begin{bmatrix} \bar{n}_\rho \\ \bar{n}_\phi \\ \bar{n}_\theta \end{bmatrix} = \begin{bmatrix} \cos\phi & \sin\phi & 0 \\ -\sin\phi & \cos\phi & 0 \\ \cos\theta\cos\phi & \cos\theta\sin\phi & -\sin\theta \end{bmatrix} \begin{bmatrix} \bar{n}_x \\ \bar{n}_y \\ \bar{n}_z \end{bmatrix} \qquad (14)$$

And the following relations ($\rho$ and $\varphi$ are the coordinates at the focal plane):

$$\begin{bmatrix} k_x \\ k_y \\ k_z \end{bmatrix} = k \begin{bmatrix} \sin\theta\cos\phi \\ \sin\theta\sin\phi \\ \cos\theta \end{bmatrix} \quad (15)$$

$$\frac{1}{k_z} dk_x dk_y = k\sin\theta d\theta d\phi \quad (16)$$

$$\begin{bmatrix} x \\ y \end{bmatrix} = \rho \begin{bmatrix} \cos\varphi \\ \sin\varphi \end{bmatrix} \quad (17)$$

the ASR of the focal field, i.e. the field in proximity of the focus F, can be expressed as

$$E(\rho,\varphi,z) = \frac{ikfe^{-ikf}}{2\pi} \int_0^{\theta_{max}} \int_0^{2\pi} E_\infty(\theta,\phi) e^{ikz\cos\theta} e^{ik\rho\sin\theta\cos(\phi-\varphi)} \sin\theta d\phi d\theta \quad (18)$$

Where $E_\infty$ is the field on the reference sphere of the focusing lens. The integration over $\theta$ is limited to the semi-angular aperture $\theta_{max}$ of the lens, which defines objective numerical aperture NA* according to the relation:

$$NA = n\sin\theta_{max} \quad (19)$$

Where n is the refractive index of the medium interposed between the focusing lens and its focus, as depicted in Fig. 8.

---

*Numerical Aperture NA. It is a parameter used in optics to define the acceptance cone of a microscope, i.e. its ability to gather light. Its value influences also the resolution of the microscopy system.

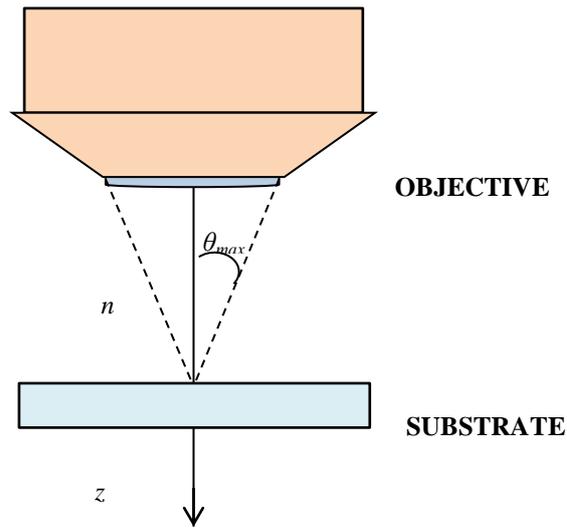

**Figure 8**. Objective numerical aperture.

## 6.5 FIELD FOCUSED NEAR A PLANAR INTERFACE

Many applications in optics, including confocal microscopy, are based on the employment of a laser beam strongly focused on a planar interface.

When a field impinges on an interface, according to the incidence angle and the nature of the media of incidence and transmittance, a portion of the field is transmitted, while another portion is reflected.

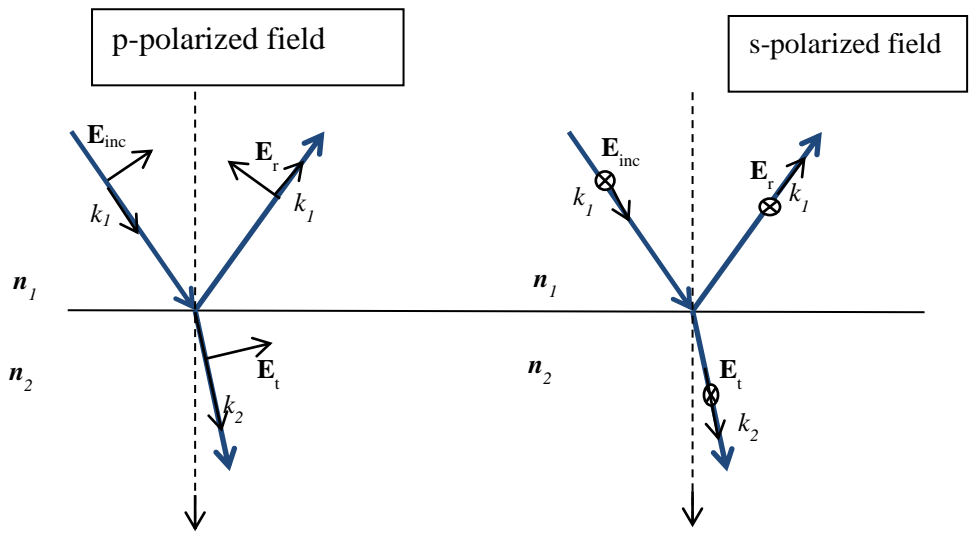

**Figure 9**. Reflected and transmitted component of the E-field, for both polarizations.

Reflection and transmission at the interface here have be modeled through Fresnel coefficients [16], defined for every polarization as:

$$r^s = \frac{k_{z_1} - k_{z_2}}{k_{z_1} + k_{z_2}}$$

$$r^p = \frac{n_2^2 k_{z_1} - n_1^2 k_{z_2}}{n_2^2 k_{z_1} + n_1^2 k_{z_2}}$$

$$t^s = \frac{2 k_{z_1}}{k_{z_1} + k_{z_2}}$$

$$t^p = \frac{2 n_2^2 k_{z_1}}{n_2^2 k_{z_1} + n_1^2 k_{z_2}} \frac{n_1}{n_2}$$

(20)

In Eqn. (20), $r^s$ and $r^p$ are the reflection coefficients for the s- and p-polarized component of the incident field, respectively; while $t^s$ and $t^p$ are the equivalent transmission Fresnel coefficients. The terms $k_{z1}$ and $n_1$ are, respectively, the z-component of the k-vector and the refractive index of the incidence medium, while $k_{z2}$ and $n_2$ are associated to the transmittance medium, as depicted in Fig. 9. When more than one interface is present, i.e. the field is propagating though different material layers, generalized Fresnel coefficients are employed. Their analytical expression and their derivation can be found in [16].

Here we report reflection and transmission coefficients of a single layer of thickness *d* since these formula will be of major interest to the NP detection problem later:

$$r^{(p,s)} = \frac{r_{1,2}^{(p,s)} + r_{2,3}^{(p,s)} e^{i 2 k_{z_2} d}}{1 + r_{1,2}^{(p,s)} r_{2,3}^{(p,s)} e^{i 2 k_{z_2} d}}$$

$$t^{(p,s)} = \frac{t_{1,2}^{(p,s)} t_{2,3}^{(p,s)} e^{i 2 k_{z_2} d}}{1 + r_{1,2}^{(p,s)} r_{2,3}^{(p,s)} e^{i 2 k_{z_2} d}}$$

(21)

In the followings it is assumed that the incident filed $E_{inc}$ is polarized along the x-axis, the propagation direction is z and the interface is located at $z = z_0$ as depicted in Fig.10.

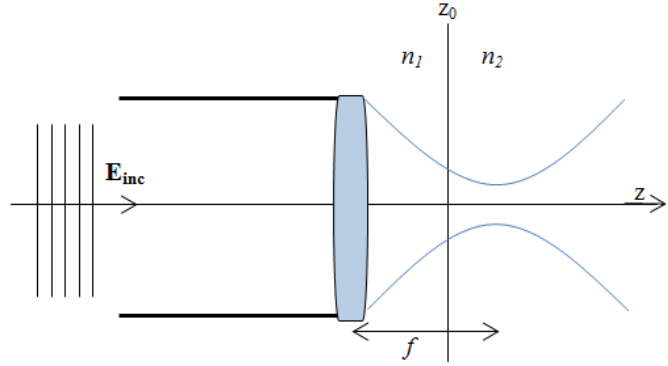

**Figure 10**. Incident field focused near an interface between two media ($n_1$ and $n_2$).

Thus:

$$E_{inc}(x,y,z) = \begin{bmatrix} E_0 \bar{n}_x \\ 0 \\ 0 \end{bmatrix} \qquad (22)$$

By expressing the E-field in cylindrical coordinates, according to Eqn. (14) [where the inverse matrix needs to be calculated], we get:

$$E_{inc}(\rho,\phi) = E_0 \left( \cos\phi \bar{n}_\rho - \sin\phi \bar{n}_\phi \right) \qquad (23)$$

Which represents the field impinging on the lens. The intensity law and an additive coordinates transformation (cylindrical->spherical) are employed to calculate the field refracted by the lens:

$$E_{refr}(\theta,\phi) = E_0 \left( \cos\phi \bar{n}_\theta - \sin\phi \bar{n}_\phi \right) \sqrt{\frac{n_0}{n_1}} \sqrt{\cos\theta} \qquad (24)$$

Refracted field can be expresses in Cartesian vector components in terms of spatial frequencies:

$$E_{refr}(k_x, k_y) = E_0 \begin{bmatrix} k_y^2 & +\dfrac{k_x^2 k_{z_1}}{k_1} \\ -k_x k_y & +\dfrac{k_x k_y k_{z_1}}{k_1} \\ 0 & -\dfrac{k_x}{k_1}(k_x^2 + k_y^2) \end{bmatrix} \sqrt{\dfrac{n_0}{n_1}} \dfrac{\sqrt{\dfrac{k_{z_1}}{k_1}}}{k_x^2 + k_y^2} \qquad (25)$$

In Eqn. (25), the first terms in the square brackets specify the s-polarized field, the second terms the p-polarized one. $n_0$ and $n_1$ are the refractive indexes of the medium before and after the lens, respectively.

Reflected $E_r$ and transmitted $E_t$ field are calculated by employing the boundary condition $z = z_0$ and the Fresnel reflection and transmission coefficients for a single interface (Eqn. (20).

$$E_r(k_x, k_y) = -E_0 e^{i 2 k_{z_1} z_0} \begin{bmatrix} -r^s k_y^2 & +r^p \dfrac{k_x^2 k_{z_1}}{k_1} \\ r^s k_x k_y & +r^p \dfrac{k_x k_y k_{z_1}}{k_1} \\ 0 & +r^p \dfrac{k_x}{k_1}(k_x^2 + k_y^2) \end{bmatrix} \sqrt{\dfrac{n_0}{n_1}} \dfrac{\sqrt{\dfrac{k_{z_1}}{k_1}}}{k_x^2 + k_y^2} \qquad (26)$$

$$E_t(k_x, k_y) = E_0 e^{i(k_{z_1} - k_{z_2}) z_0} \begin{bmatrix} t^s k_y^2 & +t^p \dfrac{k_x^2 k_{z_2}}{k_2} \\ -t^s k_x k_y & +t^p \dfrac{k_x k_y k_{z_2}}{k_2} \\ 0 & +t^p \dfrac{k_x}{k_2}(k_x^2 + k_y^2) \end{bmatrix} \sqrt{\dfrac{n_0}{n_1}} \dfrac{k_{z_2}}{k_{z_1}} \dfrac{\sqrt{\dfrac{k_{z_1}}{k_1}}}{k_x^2 + k_y^2} \qquad (27)$$

In Equations (26) and (27), the field is dependent on the amplitude $E_0$ of the incident field (paraxial by hypothesis) and on the defocus $z_0$, i.e. the distance between the interface and the focus point F.

The exponential term in Eqn. (27) is known as *spherical aberration function*. Spherical aberration is indeed introduced in the imaging system when an interface between two materials having mismatching indices is present, as it is well described in [17] and defocus is not null.

The ASR of Equations (25)-(27) allows evaluating the field distribution near a plane interface illuminated by a strongly focused laser beam.

Results found so far are related to the field in the object space, i.e. the field that in the investigated system S has to be propagated through a set of lens toward the image space to be visualized on a CCD camera.

In Fig. 11 (A) a sketch of the set-up to be investigated is reported. The geometrical equivalent structure is depicted in Fig. 11 (B).

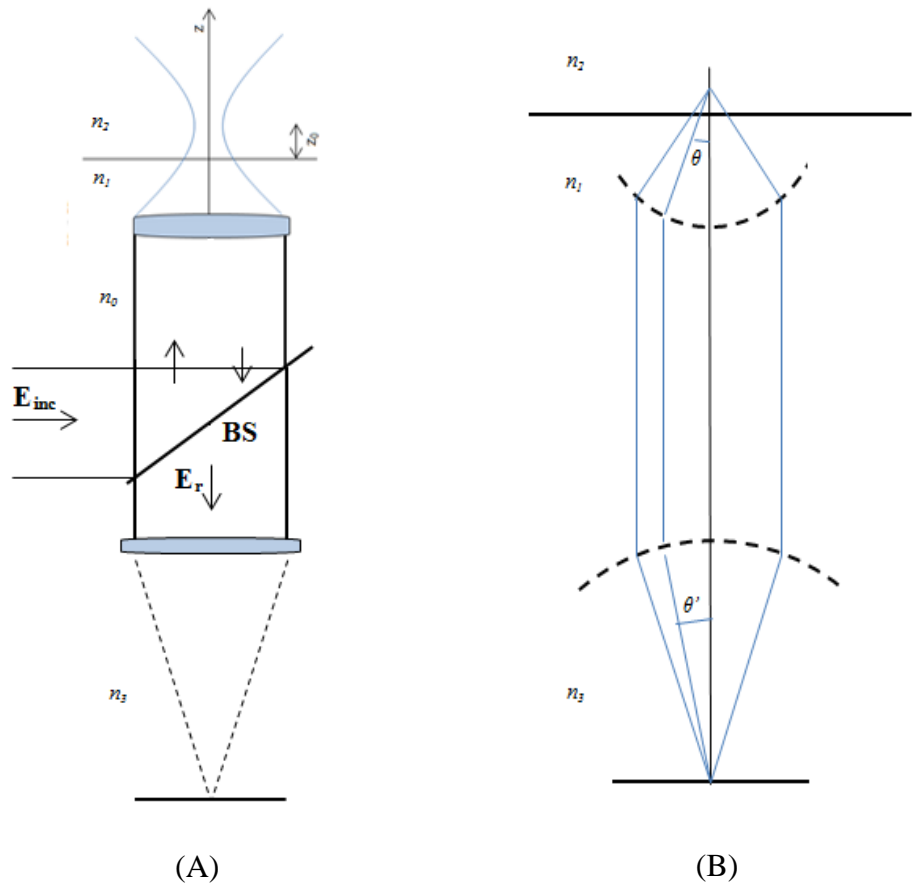

(A)                          (B)

**Figure 11.** (A) Schematic of the investigated set-up. (B) Equivalent geometry.

As illustrated, a linearly polarized field impinges on a beam-splitter where it is reflected toward a high NA lens objective and focused on an interface between two dielectric media. The field reflected by the interface is then collected by the same lens and propagates toward the beam-splitter which transmits it to a second lens, having focal length f', useful to focus the field on the photodetector.
In next paragraph, the analytical model of the propagating field is presented according to [12].

6.6 FIELD IN THE IMAGE SPACE

As indicated previously, Eqn.(26) describes a field that, after passing a focusing lens having focal radius f, is reflected on an interface between two media having mismatching indices.
Starting from the representation of the same field in spherical coordinates, and considering its refraction at the same lens f, the following expression is obtained :

$$E(\rho,\phi) = E_0 e^{i2k_{z_1}z_0} \left( -r^p \cos\phi \, \overline{n}_\rho - r^s \sin\phi \, \overline{n}_\phi \right) \tag{28}$$

Which represents the field propagating as a collimated beam in the negative z-direction.
The field is then refracted at the second lens (*tube lens*) f' and with a new azimuth angle $\theta$' and focused on the detector.
A transformation of coordinates (cylindrical->spherical) and the intensity law are applied to the field of Eqn.(24).
The resultant field, expressed in Cartesian field components, is

$$E_r(\theta',\phi) = -E_0 e^{i2k_{z_1}z_0} \begin{bmatrix} -r^s \sin^2\phi & +r^p \cos\theta'\cos^2\phi \\ r^s \sin\phi\cos\phi & +r^p \cos\theta'\cos\phi\sin\phi \\ 0 & -r^p \sin\theta'\cos\phi \end{bmatrix} \sqrt{\frac{n_0}{n_3}} \sqrt{\cos\theta'} \tag{29}$$

That constitutes the far-field to be introduced in Eqn. (18) in order to calculate the field on the focal plane (the plane of the detector).

$$E(\rho,\varphi,z) = \frac{ik_3 f' e^{-ik_3 f'}}{2\pi} \int_0^{\theta'_{max}} \int_0^{2\pi} E_r(\theta',\phi) e^{-ik_3 z \cos\theta'} e^{ik_3 \rho \sin\theta' \cos(\phi-\varphi)} \sin\theta' d\phi d\theta' \quad (30)$$

In order to simplify the expression and solve the double integral of Eqn. (30), some useful relations are employed:

$$\frac{\sin\theta}{\sin\theta'} = \frac{f'}{f} \quad (31)$$

$$k_{z_3} = k_3 \cos\theta' = k_3 \sqrt{1 - \left(\frac{f}{f'}\right)^2 \sin^2\theta} \quad (32)$$

$$\sqrt[n]{1 - \left(\frac{f}{f'}\right)^2 \sin^2\theta} \cong 1 - \frac{1}{n}\left(\frac{f}{f'}\right)^2 \sin^2\theta \quad (33)$$

$$\sin^2\phi = \frac{1}{2}(1-\cos 2\phi)$$
$$\cos^2\phi = \frac{1}{2}(1+\cos 2\phi) \quad (34)$$

Equation (26) becomes:

$$E_r(\theta,\phi) = \frac{1}{2} E_0 e^{i2k_1 \cos\theta z_0} \begin{bmatrix} (r^s - r^p) - (r^s + r^p)\cos 2\phi \\ (r^s + r^p)\sin 2\phi \\ 0 \end{bmatrix} \sqrt{\frac{n_0}{n_3}} \quad (35)$$

Due to the relations:

$$\int_0^{2\pi} \cos(n\phi) e^{ix\cos(\phi-\varphi)} d\phi = 2\pi(i)^n J_n(x)\cos(n\varphi)$$
$$\int_0^{2\pi} \sin(n\phi) e^{ix\cos(\phi-\varphi)} d\phi = 2\pi(i)^n J_n(x)\sin(n\varphi) \quad (36)$$

where $J_n(x)$ is the Bessel function of $n^{th}$ order, the insertion of Eqn. (35) into the inner integral of manipulated Eqn. (30) gives:

$$\int_0^{2\pi} E_r(\theta,\phi)e^{ix\cos(\phi-\varphi)}d\phi = E_0 e^{i2k_1\cos\theta z_0}\pi \begin{bmatrix} A_1 \\ A_2 \\ 0 \end{bmatrix}\sqrt{\frac{n_0}{n_3}} \quad (37.a)$$

$$A_1 = J_0(x)(r_s - r_p) + J_2(x)(r_s + r_p)\cos 2\varphi$$
$$A_2 = -J_2(x)(r_s + r_p)\sin 2\varphi \quad (37.b)$$

Here $x = k_3 \rho \sin\theta\, f/f'$. The ratio $f/f'$ can also be expressed as function of the transverse magnification $M = (n/n')(f'/f)$.

The field in the image plane can thus be expressed in the ASR formalism by employing the diffraction integrals $I_{0t}$ and $I_{2t}$ as:

$$E(\rho,\varphi,z) = E_{00}\frac{ik_3 f^2 e^{-ik_3(f'+z)}}{2f'}\sqrt{\frac{n_0}{n_3}}\begin{bmatrix} I_{0t} + \cos 2\varphi I_{2t} \\ -I_{2t}\sin 2\varphi \\ 0 \end{bmatrix} = \begin{bmatrix} E_{x\_ref} \\ E_{y\_ref} \\ 0 \end{bmatrix} \quad (38.a)$$

$$I_{0t} = \int_0^{\theta_{max}} f_\omega(\theta)J_0(\xi)(r_s - r_p)\sin\theta\cos\theta\, e^{i\left(2k_1 z_0 \cos\theta + \frac{1}{2}k_3 z\left(\frac{f}{f'}\right)^2 \sin^2\theta\right)}d\theta$$

$$I_{2t} = \int_0^{\theta_{max}} f_\omega(\theta)J_2(\xi)(r_s + r_p)\sin\theta\cos\theta\, e^{i\left(2k_1 z_0 \cos\theta + \frac{1}{2}k_3 z\left(\frac{f}{f'}\right)^2 \sin^2\theta\right)}d\theta$$

(38.b)

Where the incident field has been assumed to be a fundamental Gaussian beam, defined as $E_0 = E_{00} f_\omega(\theta)$ where $E_{00}$ is the field amplitude and $f_\omega(\theta)$ is the apodization function, defined as:

$$f_\omega(\theta) = e^{-\frac{1}{f_0^2}\frac{\sin^2\theta}{\sin^2\theta_{max}}} \quad (39)$$

The term $f_0$ is known as filling factor and it is equal to the ratio between beam waist radius and the aperture radius of the lens.

In Fig. 12 the field distribution in the image plane of a strongly focused spot has been plotted, i.e. Eqn. (35), by assuming a filling factor $f_0 = 2$, an NA = 1.4 with n = 1.518 for the primary focusing objective lens, a green light source ($\lambda = 525$ nm), and a magnification M = 50. The correspondent $\theta_{max}$ value is 67,26°, that is the critical angle for total internal reflection. The considered situation is the one of glass-air interface, i.e. $n_1 = 1.518$ RIU > $n_2 = 1$.

$z_0 = -\lambda/4$

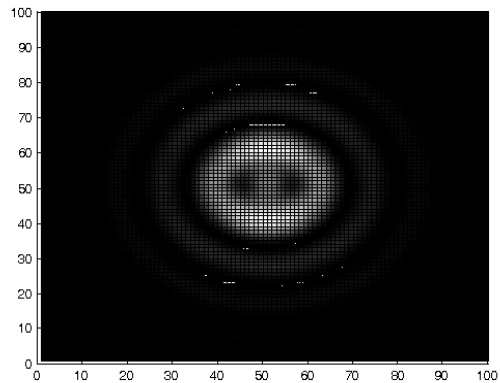

$z_0 = 0$

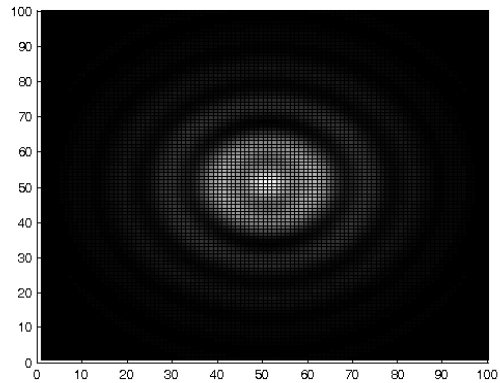

$$z_0 = +\lambda/4$$

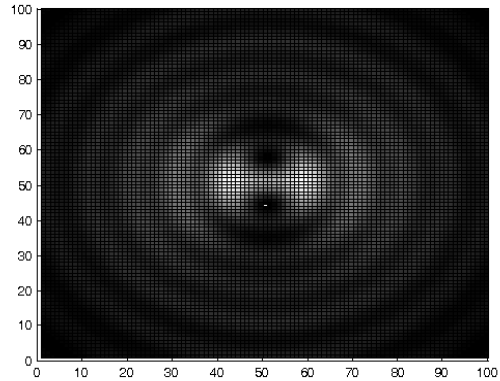

**Figure 12**. Images obtained by moving the defocus $z_0$ in the range [-$\lambda$/4, 0, $\lambda$/4]. Label on both axes is the number of pixels. The field is indeed calculated for every ($\rho,\varphi$) point.

The results obtained so far represent the case where there is no NPs in the focus of the microscope, i.e. the reference field in the absence of particle. Next we will focus on the image if there is a NP in the focus of the microscope.

6.7 BACK-SIDE ILLUMINATION MODELING

Despite standard front side-illumination, back-side allows us to perform real-time measurements since the imaging operation can be performed while the sample is flown in a solution.
An additive benefit associated to the employment of back side illumination comes from the opportunity to use solid immersion lens (SIL) technology to improve the detection limit and sensitivity of the imaging system [11].

Also, in the model a layered substrate rather than a single layer has been assumed in order to be able to optimize the interferometric response of the NP through using a spacer layer to tune the phase between the scattered and reference field [9], [10], i.e. the cosine term in Eqn. (2).

The imaging process to be modeled is also known as subsurface imaging modeling. In our case, it consists of modeling the field scattered by a spherical NP located not in the same space of the illuminating-collecting objective, but separated from this by a layered interface.

The scenario to be modeled, concerning the object space, is the one represented in Fig. 13.

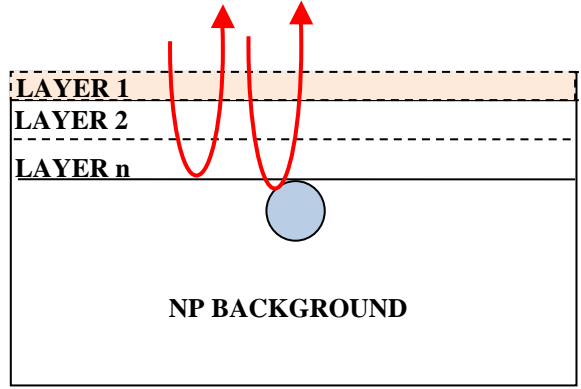

**Figure 13.** Nanoparticle backside illumination schematic.

Since the size of NPs to be detected is much smaller than the wavelength of the incident light, NP can be modeled as a scatterer in the electrostatic approximation. Three main general steps have been followed in the modeling operation, according to [12]:

- Calculation of the excitation field $E_{exc}$, or dipole driving field, in the object space
- Calculation of the interaction between $E_{exc}$ and the dipolar particle, that could be modeled with a linear relationship expressing the dipole moment $\mu_n = (\mu_x, \mu_y, \mu_z)$ as:

$$\mu_n = n_{bck}^{2} \overset{\leftrightarrow}{\alpha} E_{exc}(r_n) \tag{40}$$

Where the double arrow symbol $\overset{\leftrightarrow}{}$ stands for tensor ($\alpha$ is indeed the polarization tensor), $n_{bck}$ is the refractive index of the medium surrounding the dipole, $r_n$ defines the coordinates of the dipolar particle, i.e. $r_n = (x_n, y_n, z_n)$ in the object space.

- Calculation of the response of dipole in image space:

$$E(r) = \frac{k^2}{\varepsilon_0} \ddot{G}_{PSF} \mu_n \qquad (41)$$

Where $k$ is the wave vector and $G_{PSF}$ is the dyadic Green function or detection point spread function (PSF)*.

### 6.7.1 DIPOLE EXCITATION FIELD

In order to calculate dipole excitation field, the ASR of the transmitted component (Eqn. (27)) of the incident field has been calculated at the specific dipole position $r_n = (0, 0, z_n)$:

$$E_t^x(0,0,z_n) = E_{00} \frac{ifk_{bck}\tilde{n}e^{-ik_1 f}}{2} \int_0^{\theta_{max}} f_\omega(\theta) \sin\theta \cos^{\frac{3}{2}}\theta e^{ik_{bck}C_t z_n} \left[T_p C_t + T_s\right] d\theta$$

$$\tilde{n} = \left(\frac{n_1}{n_{bck}}\right)^2 \qquad (42)$$

$$C_t = \sqrt{1 - \tilde{n}\sin^2\theta}$$

The dipole has been assumed to be located at the origin of the object space and shifted in the optical axis by $z_n = d + R$ (equal to the sum of the thickness d of the intermediate layer and dipole radius R).
As from Eqn. (22), the linearly polarized incident field has a component only along the x-axis on the origin where the dipole is located and, consequently, the dipole is aligned along the same axis. Transmitted field components along y and z results, indeed, zero.

---

*Point spread function. It is the diffraction pattern due to a circular aperture, as brought to a focus by a lens. It defines the *resel*, i.e. the resolution element transverse to the optical axis. It is thus a measure of resolution since two self-luminous points viewed by a microscope could appear separate only if they are far enough apart for their PSFs to be distinct. The PSF has the shape of the Airy function for a circular aperture

$T_p$ and $T_s$ are generalized Fresnel transmission coefficients accounting for the presence of a layered substrate. For a single layer of thickness d, their expression has been defined in Eqn. (18).

6.7.2 DIPOLE SCATTERED FIELD

By employing the field of Eqn. (42), dipole moment $\mu_n$ is identified in its only not null component. Polarizability is a scalar quantity depending on the physical properties of the NP and its dielectric environment as mentioned earlier.

$$\mu_x = n_{bck}^2 \alpha E_t^x(0,0,z_n) \tag{43}$$

The E-field expression employed to define the scattered field is the one considering a dipole located near a planar interface, observed in the far-field zone.

$$E_{sca}(\theta,\phi) = \frac{k_{bck}^2}{4\pi\varepsilon_0 n_{bck}^2} \frac{e^{ik_{bck}r}}{r} \mu_x \begin{bmatrix} \cos\phi\cos\theta \Phi_2 \hat{n}_\theta \\ -\sin\phi \Phi_3 \hat{n}_\phi \end{bmatrix} \tag{44}$$

With $\mu_x$ defined in Equations (42-43), $r$ the observation point, $\Phi_2$ and $\Phi_3$ the vector potentials of the dipole [12], defined as:

$$\begin{aligned}\Phi_2 &= -T^{pp} \frac{n_1}{n_{bck}} e^{ik_1[z_n\tilde{s}+h\cos\theta]} \\ \Phi_3 &= T^{ss} \frac{\cos\theta}{\tilde{s}} e^{ik_1[z_n\tilde{s}+h\cos\theta]}\end{aligned} \tag{45}$$

With:

$$\tilde{s} = \sqrt{\left(\frac{n_{bck}}{n_1}\right)^2 - \sin^2\theta} \tag{46}$$

Where $h$ in Eqn. (45) is the total high of the layered substrate; $h = d$ if a single layer of thickness $d$ is present.

The terms $T^{pp}$ and $T^{ss}$ are Fresnel generalized transmission coefficient, for p- and s-polarized field respectively, calculated by assuming as incidence medium the one containing the dipole, not the objective.

The field expressed in Eqn. (44) is the one that is then collected by the primary objective lens $f$, refracted and transmitted to the tube lens $f'$.

The same modeling steps employed to calculate the reference field of Eqn. (38) have been followed for the dipole field and main results are inserted in next subsection.

### 6.7.3 DIPOLE FIELD IN THE IMAGE SPACE

Refraction of the field at the first and second lens, f and f' respectively, imposes two changes of coordinate system and the application of the intensity law.

In order to find the final expression of the dipole field in the image space, the diffraction integrals coming from the mathematical analysis of the problem are here defined:

$$I_{0t} = \int_0^{\theta_{max}} A\sqrt{\cos\theta}\sin\theta(1+\cos\theta)J_0(\xi)e^{-ik_3 z(1-\frac{1}{2}\left(\frac{f}{f'}\right)^2 \sin^2\theta)}d\theta$$

$$I_{2t} = \int_0^{\theta_{max}} B\sqrt{\cos\theta}\sin\theta(1-\cos\theta)J_2(\xi)e^{-ik_3 z(1-\frac{1}{2}\left(\frac{f}{f'}\right)^2 \sin^2\theta)}d\theta \tag{47}$$

Where $\xi$ is the argument of the Bessel functions $J_0$ and $J_2$ and is equivalent to:

$$\xi = k_3 \rho \sin\theta \frac{f}{f'} \tag{48}$$

While the expression of $\rho$ and $\varphi$ comes from Eqn. (17), and the one of A and B is:

$$A = \Phi_2 \cos\theta + \Phi_3$$
$$B = -\Phi_2 \cos\theta + \Phi_3 \tag{49}$$

The field in the image space of a x-oriented dipole is thus equivalent to:

$$E(\rho,\varphi,z) = \frac{n_3 k_0^3 e^{i(k_1 f - k_3 f')}}{8\pi i} \frac{f}{f'} \sqrt{\frac{n_0}{n_3}} \mu_x \begin{bmatrix} I_{0t} + \cos 2\varphi I_{2t} \\ \sin 2\varphi I_{2t} \\ 0 \end{bmatrix} = \begin{bmatrix} E_{xx} \\ E_{yx} \\ 0 \end{bmatrix} \tag{50}$$

### 6.7.4 TOTAL FIELD IN THE IMAGE SPACE

The intensity of field collected by the detector is, for the dipole and reference field (Eqn.s (38) and (50)):

$$I_{dip} = \sum\sum (|E_{x\_dip} + E_{x\_ref}|^2 + |E_{y\_dip} + E_{y\_ref}|^2) \tag{51}$$

$$I_{back} = \sum\sum (|E_{x\_ref}|^2 + |E_{y\_ref}|^2) \tag{52}$$

The normalized intensity, or contrast, results to be:

$$I_{norm} = \frac{I_{dip} - I_{back}}{I_{back}} \tag{53}$$

### 6.8 RESULTS

The presence of different substrates have been simulated in order to increase the contrast obtained when a specific spherical NP, characterized by a refractive index and a radius, has to be detected.

6.8.1 DIELECTRIC NPs

A refractive index of 1.5 RIU has been assumed to characterize the dielectric NP immersed in two potential fluids, water and air.

Objective numerical aperture $NA_{obj}$ has been assumed to be 0.75 which is equivalent to the total NA in absence of a numerical aperture increasing lens (NAIL).

The first investigated configuration is constituted by a substrate of sapphire/silicon oxide having, at the interest wavelength ($\lambda = 525$ nm), a refractive index of 1.77, 1.45 RIU, respectively [18].

Oxide thickness has been moved in the range 25 : 250 nm with a step size of 5 nm.

NP radius has been first moved in the range 15 : 30 nm to identify the detection limit.

An overlapping of resulting contrast curves is plotted in Fig. 14.

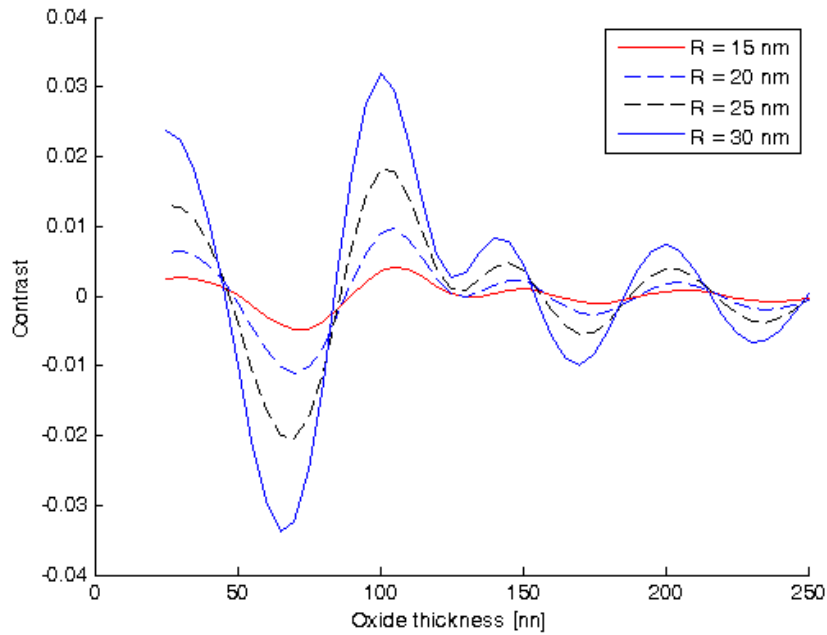

**Figure 14.** Contrast curves VS oxide thickness for different NP radii.

Assuming a noise level of 1%, the detection limit of this configuration, expressed in terms of minimum NP radius detectable, is slightly larger than 20 nm.

In order to evaluate a potential improvement in terms of resolution and detection limit, a similar study has been conducted by employing a SIL. In this scenario, the effective NA ($NA_{eff}$) is equal to $n_{SIL}$* $NA_{obj.}$ (Eqn. 19) assuming the objective space has the same refractive index of the SIL. As the spatial resolution and the collection efficiency of the microscope is proportional to ($1/NA_{eff}$) and $NA_{eff}^2$, it is expected that the performance of the microscope can be improved significantly. The investigated configuration is the one depicted in Fig. 15.

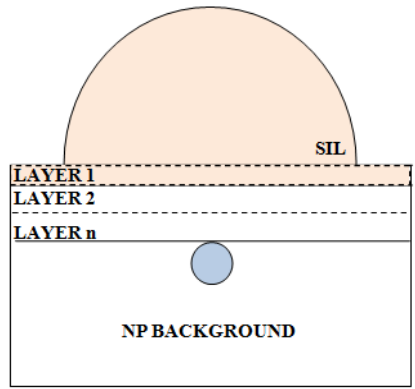

**Figure 15.** Schematic of the back-side NP detection configuration.

In Fig. 16, different contrast curves for different NPs radius have been overlapped always for a sapphire/silicon oxide substrate.

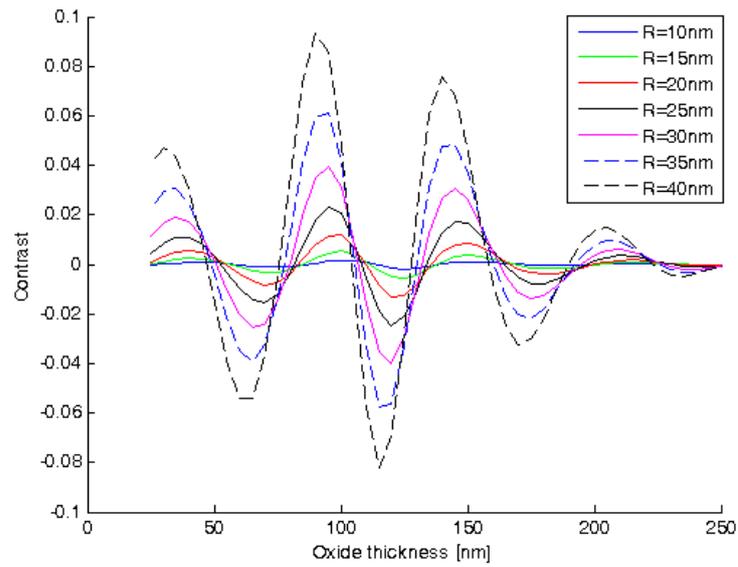

**Figure 16.** Contrast curves calculated for a sapphire/SiO$_2$ substrate

The optimum oxide thickness value, i.e. the one that gives the maximum contrast value, results to be close to 95 nm despite the fact that the exact thickness slightly varies around 95 nm for NPs with different sizes. That means that an optimized thickness could be suited to identify a specific NP size.

At this fixed oxide thickness value, the contrast curve has been calculated as function of NP radius ranging from 20 to 50 nm with a step size of 2 nm as depicted in Fig. 17.

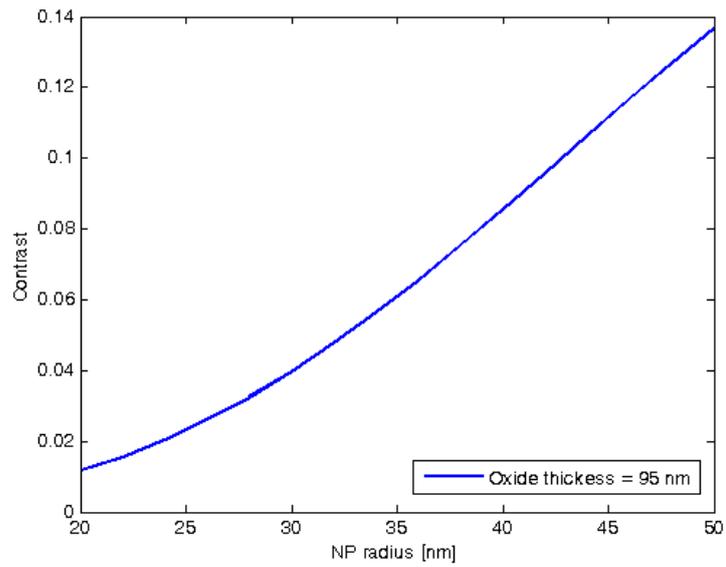

**Figure 17.** For a fixed oxide thickness, the contrast value is plotted versus the radius of the investigated nanoparticle.

With a level noise close to 1%, the detection limit expressed as NP radius is about 15 nm, as from Fig. 16.

When NP to be detected is immersed in air, a slightly different contrast I is obtained (Fig. 18).

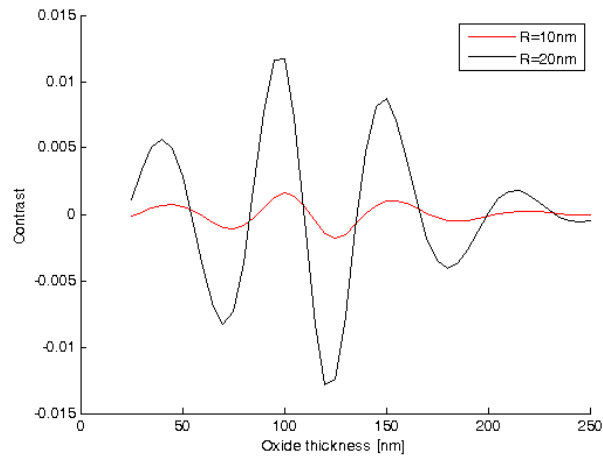

**Figure 18.** Contrast curves calculated for a sapphire/SiO2 substrate for an NP in air.

Considering water as NPs background medium is anyway the realistic scenario to be investigated since it emulates a real-time NPs detection in a flowing solution.

A secondary investigated configuration is the one composed by a quartz/silicon nitrite/silicon oxide substrate. Here, the thickness of both $Si_3N_4$ and $SiO_2$ has been moved to find the optimum point, i.e. the pair of thickness values that maximizes the contrast (see Figures 19 and 20).

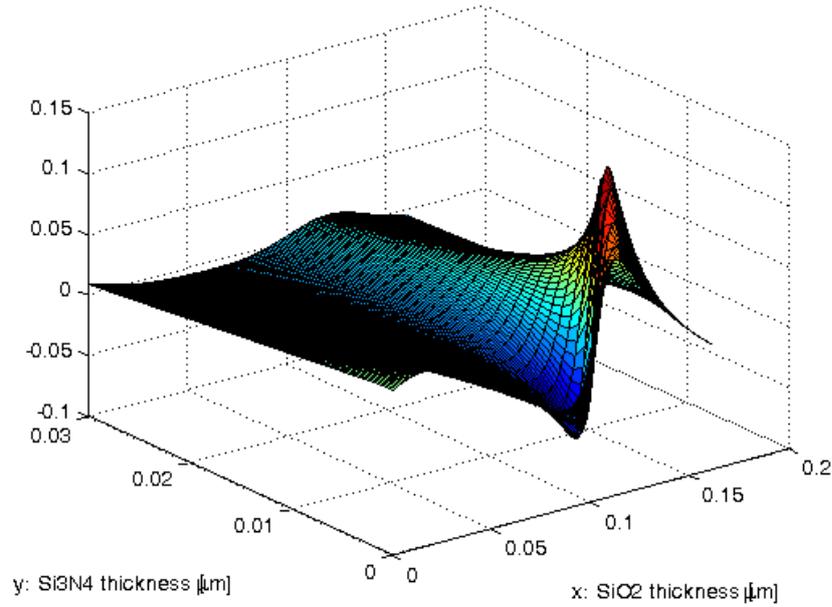

**Figure 19.** Contrast surface for a fixed NP radius (= 25 nm).

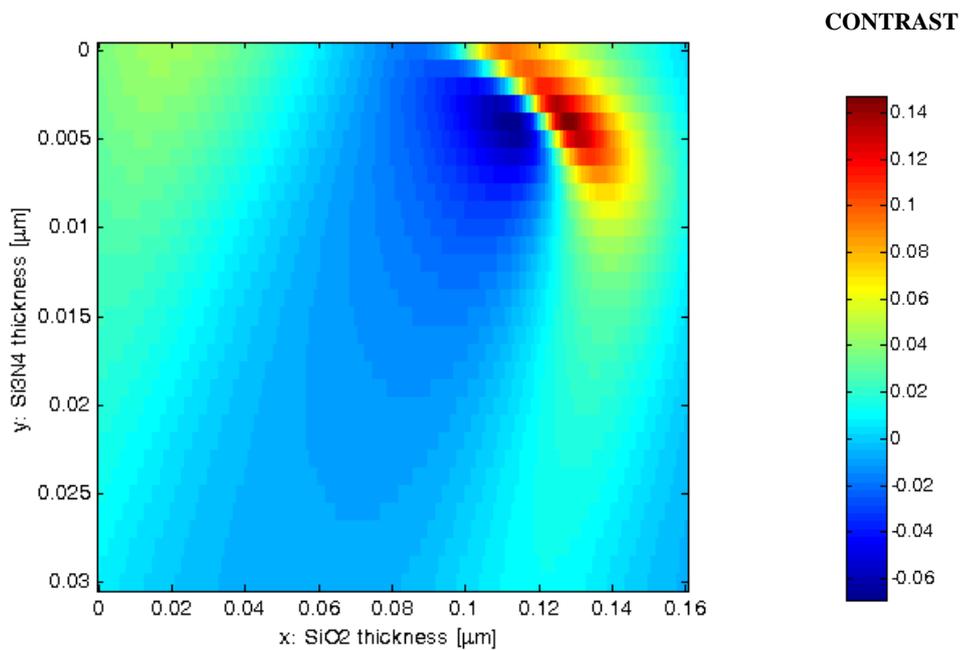

**Figure 20.** Color map expressing contrast as function of $Si_3N_4$ and $SiO_2$ thickness.

A contrast roughly equal to 15% has been obtained for the pair $(h_{Si3N4}, h_{SiO2})$ = (4 nm, 128 nm) for an NP radius of 25 nm. This is much higher than the correspondent contrast obtained with sapphire/silicon oxide simulated substrate (roughly equal to 2%) as discussed earlier.

In terms of practical point of view, the thickness of the silicon nitride causes problems in terms of controllability and repeatability during the fabrication of the chip. Therefore we also investigated a more practical configuration which provides benefit of reduced number of fabrication steps.

Silicon Nitrite and silicon oxide have been substituted by silicon oxi-nitrite SiOxNy. With a Nitrogen percentage ranging from 0 to 80% and an SiOxNy thickness ranging from 50 to 200 nm, different contrast curves have been obtained fixing the NP radius:

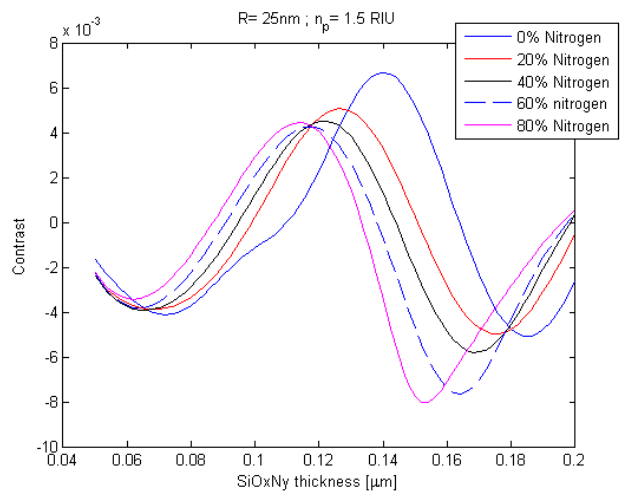

(A)

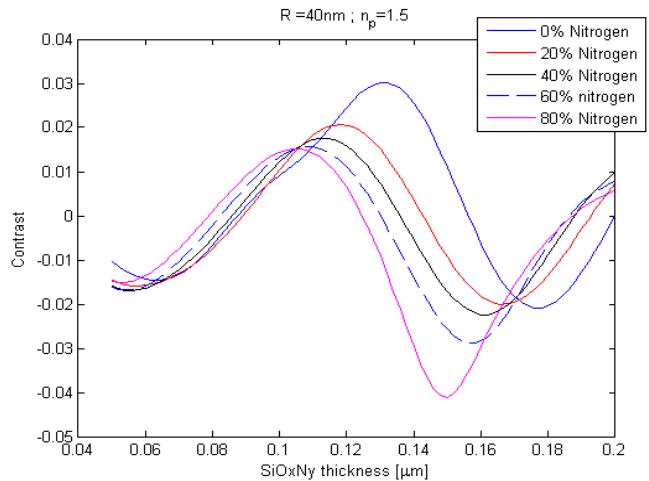

(B)

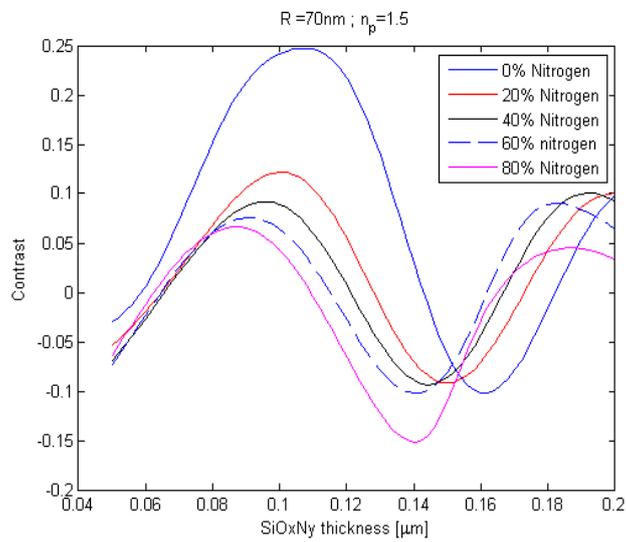

(C)

**Figure 21.** Contrast curve for a spherical dielectric NP having radius (A) 25 nm, (B) 40 nm, (C) 75 nm.

As from fig. 21 (A), LOD now degrades with respect to the previously investigated layered substrates.

Also fused silica/silicon nitride/silicon oxide based layered substrate has been investigated in order to estimate the layers thickness capable to improve the imaging setup. A contrast level approximately of 20% can be achieved for a dielectric 25 nm radius NP as depicted in Fig. 22.

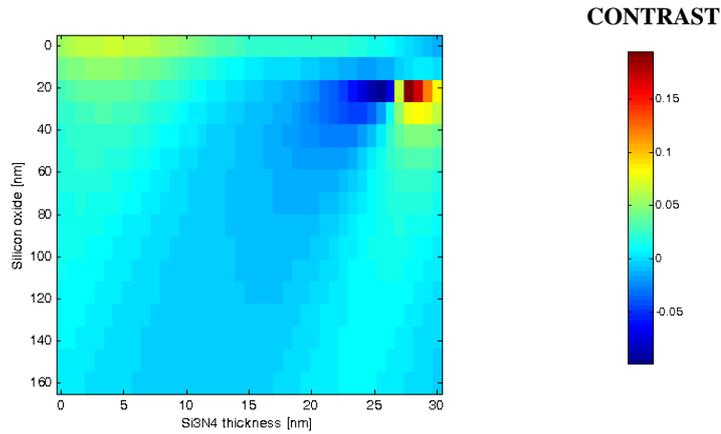

**Figure 22.** Color map expressing contrast as function of $Si_3N_4$ and $SiO_2$ thickness.

A comparison of different substrate configurations investigated is presented in the following table in terms of contrast. Data are associated to a 25 nm radius dielectric NP. A SIL has also been employed, while NP background medium has been considered to be water.

**Table 1.**

| SUBSTRATE | | CONTRAST |
|---|---|---|
| Sapphire/$SiO_2$ | | 2% |
| Quartz/$Si_3N_4$/ $SiO_2$ | | 15% |
| Quartz/$SiO_xN_y$ | N = 0% | 0.7% |
| | N = 20% | 0.5% |
| | N = 40 % | 0.4% |
| | N = 60%, 80% | 0.4% |
| Fused silica/ $Si_3N_4$/ $SiO_2$ | | 19% |

A high contrast value has been obtained for both the layered substrates characterized by the presence of Silicon Nitride. The configuration suiting fused silica as top layer gives not only the highest contrast value, but also this value corresponds to a silicon oxide and silicon nitride thickness more easy to be realized through standard fabrication steps.

A comparison with experimental results should confirm the validity of the model.

## 6.9 CONCLUSIONS

According to the microresonant optical cavities issue of nanoparticles shape discrimination in a complex solution, a sensor based on interferometric reflectance imaging (IRIS) has been here modeled. The main advantage coming from the employment of the interferometric microscopy for nanoparticle detection has been pointed out as an increasing of three orders of magnitude of the light intensity signal read at the photodetector. To maximize the interferometric term collected by the detector, a layered substrate has been assumed.

After describing some mathematical instruments useful to develop the model of the setup to be then built and tested, including the angular spectrum representation of a far field and some geometrical optics rules, the modeling of the IRIS in confocal, sub-surface illumination scheme has been presented. The choice of the illumination configuration relies on the capability of the confocal scheme to improve the setup spatial resolution and the image contrast in order to investigate fast bio-processes occurring at the single nanoparticle limit. By restricting the observed volume (light is focused to a point), the confocal scheme indeed increases the image contrast because light reflected from points adjacent to the studied one does not enter the detector. An additional expedient to increase the spatial resolution, lateral and longitudinal, has been hypothesized in the model, i.e. the employment of a numerical aperture increasing lens (NAIL). A set of layered substrates has been tested in simulation through a Matlab script modeling the confocal-subsurface illumination setup in order to individuate the substrate conferring the highest contrast to the final image in terms of both material and thickness of every single layer.

The model has been developed to be employed in a more complex project leaded by Prof. M. Selim Ünlü and Prof. Bennett Goldberg with the final aim of realizing a microscopy setup useful to collect data coming from the imaging of a vesicular stomatitis virus (VSV) in solution flowing through an opportunely designed microfluidic platform.

Experimental data should be available by May 2013.

## 6.10 REFERENCES

1. A. Gaiduk, M. Yorulmaz, P. V. Ruijgrok, and M. Orrit, "Room-temperature detection of a single molecule's absorption by photothermal contrast," Science pp. 353-356, Vol. 330, 2010

2. L. Kador, T. Latychevskaia, A. Renn and U. P. Wild, "Absorption spectroscopy on single molecules in solids,". J. Chem. Phys. 111, 8755, 1999.

3. I.H. El-Sayed , X. Huang and M.A. El-Sayed, "Surface Plasmon Resonance Scattering and Absorption of anti-EGFR Antibody Conjugated Gold Nanoparticles in Cancer Diagnostics: Applications in Oral Cancer," Nano Letters, pp 829–834, 5 (5), 2005.

4. S. Wang, X. Shan, U. Patel, X. Huang, J. Lu, J. Li and N. Tao, "Label-free imaging, detection, and mass measurement of single viruses by surface plasmon resonance," Proc. Natl. Acad. Sci. U. S. A., pp. 16028–16032, 107, 2010.

5. F.V. Ignatovich and L. Novotny, "Real-time and background-free detection of nanoscale particles," Phys. Rev. Lett. 96, 013901, 2006.

6. M.S Ünlü, "IRIS: Interferometric Reflectance Imaging Sensor - Multiplexed Assays and Single Virus Detection," Laser Science (LS) Conference, (LTh3I), Rochester, NY, October 14, 2012.

7. E. Özkumur, J.W. Needham, D.A. Bergstein, R. Gonzalez, M. Cabodi, J.M. Gershoni, B.B. Goldberg, and M.S Ünlü, "Label-free and dynamic detection of biomolecular interactions for high-throughput microarray applications," PNAS, pp. 7988–7992 , vol. 105, no. 23, 2008.

8. A. Yurt, G.G. Daaboul, J.H. Connor, B.B. Goldberg and M.S Ünlü, "Single nanoparticle detectors for biological applications," Nanoscale, pp-715-726, 2012.

9. G. G. Daaboul, A. Yurt, X. Zhang, G.M. Hwang, B.B. Goldberg, and M.S. Ünlü, "High-Throughput Detection and Sizing of Individual Low-Index Nanoparticles and Viruses for Pathogen Identification," Nano Letters, pp. 4727-473, Vol. 11, 2010

10. G. G. Daaboul, C.A. Lopez, A. Yurt, B.B. Goldberg, G.H. Connor and M.S. Ünlü, "Label-Free Optical Biosensors for Virus Detection and Characterization," IEEE Journal of Selected Topics in Quantum Electronics, pp. 1422-1433, Vol. 18, No. 4, 2012.
11. J.C. Mertz, "Introduction to Optical Microscopy," Roberts and Company Publishers: Greenwood Village, CO, 2009
12. L. Novotny and B. Hecht, "Principles of Nano Optics," Cambridge University Press: Cambridge, 2006.
13. L. Mandel and E. Wolf, "Optical coherence and quantum optics," Cambridge University Press, 1995.
14. A.N. Vamivakas, R.D. Younger, B.B. Goldberg, A.K. Swan, M.S. Ünlü, E.R. Behringer, S.B. Ippolito, "A case study for optics: The solid immersion microscope," Am. J. Phys., 76 (8), 2008.
15. M. Born and E. Wolf, "Principles of Optics. Electromagnetic theory of propagation, interference and diffraction of light", Cambridge University Press, 7$^{th}$ Edition, 1999.
16. W.C. Chew, "Waves and fields in inhomogeneous media," IEEE Press Series on Electromegnetic Waves, 1990
17. P.Torok, P. Varga, Z. Laczik and G.R. Booker, "Electromagnetic diffraction theory of light focused through a planar interface between materials of mismatched refractive indices: an integral representation,"
18. refractiveindex.info

# Chapter 7.
# CONCLUSION

The research activity reported in the present dissertation has been focused on the modeling of two platforms belonging to different branches of optics science to be employed in the sensing field. Properly the investigation field is the one related to the detection of tiny particles, such as nanoparticles composing the particulate matter or virulent agents, with the aim of monitoring human and environment health status.

The approach followed in the modeling operation of both the platforms relied on the study of the interaction of an electromagnetic wave with a small particle in the hypothesis of dealing with a Rayleigh scatterer, as described in the previous Chapters.

The first designed configuration relies on the employment of a high quality factor microresonator for NP detection purposes (see Chapter 5). Nanoparticles as small as 30 nm in radius have been detected in a simulative environment through the employment of an FDTD algorithm based commercial software with an estimation error of the order of 2% with respect to the NP nominal radius. If compared to other cavities proposed in literature [1], [2] which employ whispering gallery mode based resonant structures, the novelty of the designed device resides in its geometry, small footprint and planar configuration that make it a good candidate for the realization of a compact, portable and reliable on-chip platform.

| LOD (NP radius) | CONFIGURATION | REFERENCE |
|---|---|---|
| 140 nm | microsphere | Arnold et al. [1] |
| 12,5 nm | microtoroid | Lu et al. [2] |
| 30 nm | hybrid | Present work |

The geometrical structure of the designed cavity is the one of an annulus coupled to two distinct bus waveguides in a four ports *planar coupling configuration*. The latter is, indeed, the most promising coupling configuration for the practical packaging of systems containing WGM resonators and could lead to the realization of a portable on-chip complex platform such as a micro-Total Analysis System (μTAS) or lab-on-chip.

Also, the choice of a *cavity footprint* on chip as small as *10 μm$^2$* could be an interesting starting point for the development of an implantable or portable bio-assay platform allowing for the refractometric detection of the concentration of the interest analyte in a complex solution.

A possible on chip configuration for a high throughput and multianalyte detection relies on the employment of an array of resonant cavities interacting with a single nano-fluidic channel. The surface of every resonator should be obviously chemically activated and functionalized by depositing a layer of receptors having a high binding affinity with the analyta.

Due to the open issues of discriminating tiny particles shape in a complex solution and getting a high throughput platform, the second investigated configuration for single NP detection relied on the employment of a microscopy setup which is intrinsically capable of overcoming these issues.

The starting point for this study has been the IRIS (Interferometric Reflectance Imaging Sensor) platform developed at the Boston University by the Ünlü group [3].

An improvement of the yet existing configuration has been investigated in the present study in order to realize a real time, sub-surface detection of single molecules and NPs. A setup based on confocal illumination aided by the employment of a numerical aperture increasing lens (NAIL) has been modeled to confer an ultra-high resolution to the microscopy system. In order to increase the contrast, a multilayer substrate has been assumed so that the light intensity collected at the photo-detector is maximized.

A set of layered substrates has been tested in simulation through a Matlab script modeling the confocal-subsurface illumination setup in order to individuate the substrate conferring the highest contrast to the final image in terms of both material and thickness of every single layer.

A contrast level approximately of 20% has been achieved for a dielectric 25 nm radius NP with a fused silica/silicon nitride/silicon oxide substrate in aqueous environment.

A comparison of the investigated detection techniques and platforms is proposed in the following table.

|  | WGM – BASED DETECTION | MICROSCOPY-BASED DETECTION |
|---|---|---|
| LOW LOD | ✓ | ✓ |
| SIZE DISCIMINATION (SORTING) | ✓ | ✓ |
| SHAPE RECOGNITION |  | ✓ |
| SELECTIVITY | ✓ | ✓ |
| HIGH THROUGHPUT | ✓ | ✓ |
| PORTABILITY | ✓ |  |
| IN VIVO DIAGNOSIS |  | ✓ |

They both allows for nanoparticles size discrimination with a LOD down to few tens of nanometers. Their selectivity can be assured with an opportune substrate surface activation and functionalization. The achievement of a high throughput is a property intrinsically showed by microscopy-based detection techniques, while an opportune design of microfluidic aiding channels composing the WGMs based platform need to be planned in order to assure high throughput.

Despite the above mentioned features characterizing both the NPs detection techniques, the microscopy-based technique allows for NPs shape recognition and in vivo diagnosis, while the peculiarity of portability (associated to all miniaturized devices) is the advantage coming from the employment of a WGMs based platform for nanoparticles detection.

## 7.1 REFERENCES

1   S. Arnold, D. Keng, S.I. Shopova, S. Holler, W. Zurawsky, F. Vollmer, "Whispering gallery mode carousel – a photonic mechanism for enhanced nanoparticle detection in biosensing," Optics Express, Vol. 17(8), pp. 6230-6238, 2009.

2   T. Lu, H. Lee, T. Chen, S. Herchak, J.-H. Kim, S.E. Fraser, R.C. Flagan, and K. Vahala, "High sensitivity nanoparticle detection using optical microcavities," P. Natl. Acad. Sci. USA, Vol. 108(15), pp. 5976-5979, 2011.

3   M.S Ünlü, "IRIS: Interferometric Reflectance Imaging Sensor - Multiplexed Assays and Single Virus Detection," Laser Science (LS) Conference, (LTh3I), Rochester, NY, October 14, 2012.

# APPENDIX 1: RING RESONATORS THEORY

To analytically investigate the properties of a ring resonator, the standard configuration of a single mode optical cavity laterally coupled to a single bus waveguide is here considered, as depicted in Fig. 1.

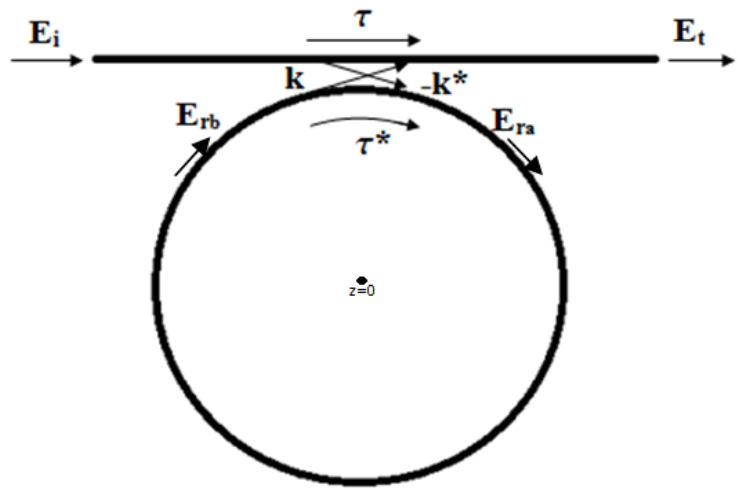

**Figure 1.** Ring resonator coupled to a single bus waveguide.

The terms $E_i$, $E_{rb}$, $E_t$, $E_{ra}$ are the complex amplitudes of the mode propagating within the cavity. They are normalized such that their squared magnitude corresponds to the modal power, i.e. $P_x = |E_x|^2$ with x = generic evaluation point.

The value of the conjugate complex coupling coefficients $\kappa$ and $\kappa^*$ and of the transmission coefficients $\tau$ and $\tau^*$, depends on the nature of the coupling. The terms $\kappa$ and $\tau$ are also named cross-coupling and self-coupling coefficient, respectively.

The spectral response at the through output port will be derived and the most frequently used resonator figures of merit will be defined in the following.

We assume that a single mode having a specific polarization state is supported within the ring cavity and that the back-reflection at the coupling section is negligible. The device is assumed to be symmetric with respect to the plane z = 0. The interaction between the electric fields with amplitude $E_i$, $E_{rb}$, $E_t$, $E_{ra}$ can be described through the following matrix relation:

$$\begin{bmatrix} E_t \\ E_{ra} \end{bmatrix} = \begin{bmatrix} \tau & k \\ -k^* & \tau^* \end{bmatrix} \begin{bmatrix} E_i \\ E_{rb} \end{bmatrix} \quad (1)$$

Assuming a lossless coupling, the relation to be verified is $|\kappa^2| + |\tau^2| = 1$ in order to assure that the determinant of the transfer matrix modelling the coupler is unitary. It means that the coupling coefficients κ and κ* coincide and $|\tau^2| = |\tau^{*2}|$. The single mode supported by the cavity has a complex propagation constant $\gamma = \beta + j\alpha$, where $\beta$ is the phase constant (real) and $\alpha$ is the attenuation coefficient (positive and real) accounting for the total losses within the cavity.

A single round trip along the circumference, $L = 2\pi R$ (where $R$ is the ring radius), leads to

$$E_{rb} = e^{i\gamma L} E_{ra} = a e^{i\theta} E_{ra} \quad (2)$$

where $a = e^{-\alpha L}$ and $e^{j\theta} = e^{j\beta L}$ are the attenuation and phase shift the mode is subjected to, respectively.

By employing Eqn. (1), the following expressions are obtained:

$$E_t = \frac{-a + \tau e^{-i\theta}}{-a\tau^* + e^{-i\theta}} \quad (3)$$

$$E_{rb} = \frac{-ak^*}{-a\tau^* + e^{-i\theta}} \quad (4)$$

From the above equations the transmitted power, also indicated as resonator transmission spectrum $T = |E_t|^2/|E_i|^2$, can be expressed as

$$P_t = T = |E_t|^2 = \frac{a^2 + |\tau|^2 - 2a|\tau|\cos(\theta + \varphi_t)}{1 + a^2|\tau|^2 - 2a|\tau|\cos(\theta + \varphi_t)} \quad (5)$$

where $\tau = |\tau|e^{j\varphi_t}$, with $|\tau|$ representing the amplitude of the complex self-coupling coefficient and $\varphi_t$ is the coupler phase.

The light pulse propagating through the bus waveguide couples to the ring, i.e. resonates, only when the wavelength of the exciting optical signal equals the optical length of the cavity, i.e. when:

$$\lambda = \frac{L n_{eff}}{m} \quad (6)$$

where $m$ is an integer number denoting the mode order, $L$ is the geometrical cavity length and $n_{eff}$ is the mode effective index. The phase term has been here neglected.

Equivalently in vacuum:

$$\beta_m = \frac{2\pi m - \varphi_t}{L} \quad (7)$$

At resonance, i.e. when the cosine term is unitary, the transmitted power reaches its minimum value that is:

$$T_{min} = \frac{(a - |\tau|)^2}{(1 - a|\tau|)^2} \quad (8)$$

Equation (8) goes to zero if the internal loss $a$ equals the outer loss $|\tau|$ (considered as uncoupled portion of light). This condition is known in literature as *critical coupling* and it is visible in the resonator transmission spectrum as a dip to zero.

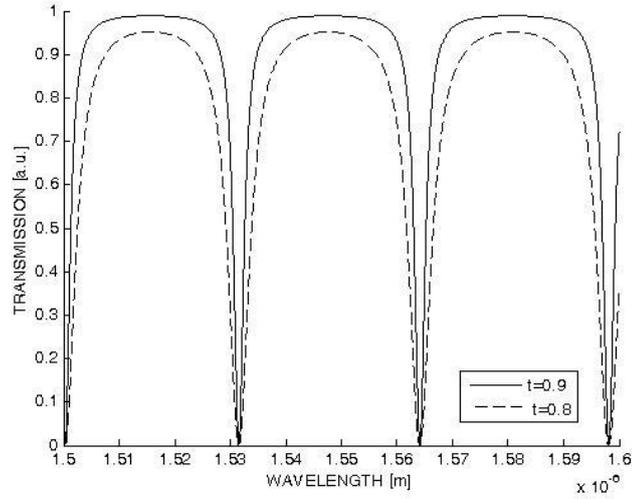

**Figure 2.** Ring resonator transmission spectrum.

Due to the periodic nature of resonances, it is possible to define a figure of merit called Free Spectral Range FSR useful to define the distance between adjacent resonances:

$$\beta_{m+1} = \frac{2\pi(m+1) - \varphi_t}{L} \approx \beta_m + \frac{\partial \beta}{\partial \lambda} \Delta \lambda \qquad (9)$$

where $\Delta \lambda$ is the difference between the vacuum wavelengths corresponding to the two resonant conditions $m$ and $m+1$.

It results that the difference in propagation constant for two adjacent resonant modes is given by:

$$\Delta \beta = \beta_{m+1} - \beta_m = \frac{\partial \beta}{\partial \lambda} \Delta \lambda = \frac{2\pi}{L} \qquad (10)$$

from which

$$\Delta \lambda \cong \frac{2\pi}{L} \frac{\partial \lambda}{\partial \beta} \qquad (11)$$

$\beta$ can also be expressed as $\beta = k_0 n_{eff} = \frac{2\pi n_{eff}}{\lambda}$, where $k_0$ is the vacuum wavenumber, $n_{eff} = \frac{\beta}{k_0}$ ($k_0$) is the waveguide effective index and $\lambda$ the vacuum wavelength. Therefore:

$$\frac{\partial \beta}{\partial \lambda} \cong k \frac{\partial n_{eff}}{\partial \lambda} - \frac{\beta}{\lambda} \qquad (12)$$

By taking into account the dependence of $n_{eff}$ and $\beta$ on the wavelength (which is called *dispersion*) and assuming a first order approximation for $\beta$ dispersion, i.e. $\beta(\lambda) \cong \beta_0 + \frac{\partial \beta}{\partial \lambda}(\lambda - \lambda_0)$, Eqn. (12) becomes:

$$\frac{\partial \beta}{\partial \lambda} \cong -\frac{k}{\lambda} n_g \qquad (13)$$

where it has been introduced the effective group index $n_g = n_{eff} - \lambda \frac{\partial n_{eff}}{\partial \lambda}$ which accounts for the dispersion.

From Equations (11) and (13), the FSR expression in terms of wavelength units has been derived:

$$\Delta \lambda \cong \frac{\lambda^2}{n_g L} \qquad (14)$$

In Fig. 3 the FSR is indicated by a double arrow and another figure is introduced, i.e. the Full Width at Half Maximum (FWHM).

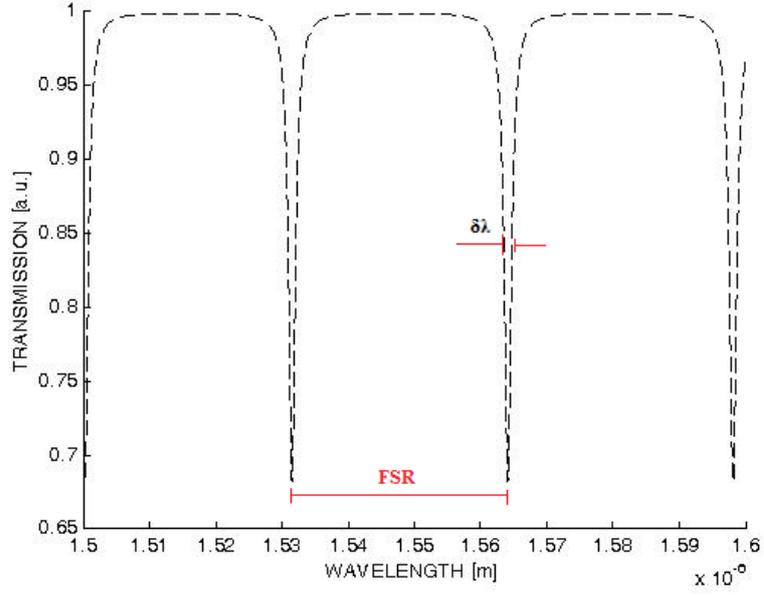

**Figure 3.** Free spectral range and full width at half maximum in the spectrum of a ring resonator.

The FWHM is defined as the 3dB bandwidth of the resonance line-width, $\delta\lambda$. To derive its expression we consider Equations (5) and (8) and find the detuning where

$$P_t(\theta) = P_{tmin} + \frac{P_{tmax} - P_{tmin}}{2} \qquad (15)$$

Without considering the phase contribution, $\varphi_t$, assuming $a \neq 1$ and using the real part of the series expansion of the Euler formula, $cos(\theta) \cong 1 - \frac{\theta^2}{2}$, for small $\theta$ the following expression is obtained:

$$\delta\theta \cong 2\sqrt{2}\ \frac{(a|\tau| - 1)}{\sqrt{1 + a^2|\tau|^2}} \qquad (16)$$

And:

$$\delta\lambda \cong \frac{\lambda^2}{\pi n_g L} \frac{\sqrt{2}\,(a|\tau|-1)}{\sqrt{1+a^2|\tau|^2}} \tag{17}$$

From a physical point of view, the FWHM accounts for the 50% attenuation of the optical signal propagating through the resonator.

Both the FSR and the FWHM are useful to evaluate the Finesse *F* of a resonator whose value is equal to the ratio of the free spectral range to the full width at half maximum:

$$\mathcal{F} = \frac{FSR}{FWHM} = \frac{\Delta\lambda}{\delta\lambda} = \frac{\pi\sqrt{1+a^2|\tau|^2}}{\sqrt{2}\,(a|\tau|-1)} \tag{18}$$

From a physical point of view, the finesse represents $2\pi$ times the number of round-trips made by the light in the ring.

For biosensing applications, this value should be higher than 5.

Another figure of merit widely employed to evaluate the performance of a resonant structure is the quality factor defined as:

$$Q = \frac{\lambda}{\delta\lambda} = \frac{n_g L}{\lambda}\mathcal{F} \tag{19}$$

For high Q values, this definition is equivalent to the one associated to the device energy storage, i.e. the Q factor is $2\pi$ times the ratio of the stored energy over the energy dissipated per oscillation cycle or, equivalently, the number of oscillation periods required for the stored energy to decay to 1/e ($\approx$ 37%) of its initial value.

A Q-factor of the order of $10^4$ is suggested for biosensing applications in order to ensure the achievement of a certain sensitivity and detection limit by the device.

Additionally, it is possible to define the figure of the extinction ratio that at the trough port reads as:

$$\mathrm{ER} = \frac{P_{tmax}}{P_{tmin}} \tag{20}$$

Its value, expressed in dB, should be higher than 15 for biosensing applications.

# APPENDIX 2: MICROSCOPY THEORY

According to the Collins Dictionary, a microscope is "an optical instrument that uses a lens or combination of lenses to produce a magnified image of a small, close object".

A primary microscopes classification relies on evaluating the method of interaction they use to measure an object, that could be the one of photons with an object as in optical microscopes.

Here we introduce the nomenclature employed in Chapter 6 starting with microscopes and focusing then on confocal microscopy.

As above defined, a conventional optical microscope is a lens or combination of lenses useful to correlate an object to its magnified image.

Both the object and its image are measured and described in terms of an *object space* and *image space* since they are three-dimensional. The *focus f* of a conventional optical microscope is the volume in the object space that the microscope clearly images.

In a conventional optical microscope, the first lens that light encounters after illuminating the object is referred to as the *objective* or *primary lens*.

Lens could introduce several defects during the imaging process, such as aberrations of different nature.

Aberrations can be classified as spherical and chromatic. When peripheral rays and axial rays have different focal points (i.e. in a spherical shaped lens), the *spherical aberration* occurs and it causes the image to appear hazy or blurred and slightly out of focus. This aspect is very important in terms of the resolution of the lens because it affects the coincident imaging of points along the optical axis and degrades the performance of the lens. *Chromatic aberration* could instead be axial or lateral. In axial aberration the blue light is refracted to the greatest extent followed by the green and red light. This phenomenon is commonly referred to as dispersion.

Lateral aberration instead relies on a chromatic difference of magnification: the blue image of a detail is slightly larger than the green image or the red image in white light, thus causing color ringing of specimen details at the outer regions of the field of view. If a converging lens is combined with a weaker diverging lens, it is possible to cancel the chromatic aberrations for certain wavelengths.

An additional defect referred to as *astigmatism* could occur when employing a lens. In this case, the off-axis image of a specimen point appears as a disc or blurred lines instead of a point.

Depending on the angle of the off-axis rays entering the lens, the line image may be oriented either tangentially or radially.

Even though it is possible to correct for spherical aberration, the light beams emitted from the object points off the axis may not converge on one point in the image field, leaving an asymmetric blur with a trail like a comet. This aberration is known as *coma aberration*.

If an *aplanatic lens* is employed, spherical aberration and coma could be corrected. Properly, spherical aberration produces a concentric defocus around all the bright elements in all parts of an image, and a blurring in extended and low contrast detail, preventing any part from coming into sharp focus. The aberration is uniform across the entire image area.

Despite spherical aberration which is related to an error in focusing light, coma is associated to an error of magnification. It indeed produces elongated, wedge shaped "tails" that extend radially from point images or luminance edges in off axis light and the magnification varies with distance from propagating axis.

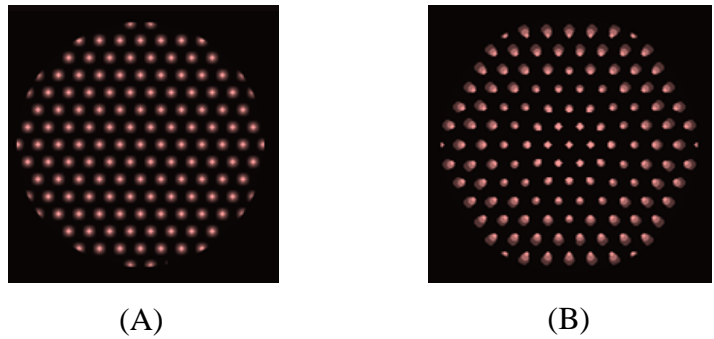

(A)          (B)

**Figure 1.** (A) Spherical aberration. (B) Coma.

In order to describe the above mentioned lens defects, an important microscope descriptor has been referred to, i.e. the *magnification M* or magnifying power of a microscope system. It is defined as the degree of enlargement of an image relative to the actual size of the object it represents. For example, a microscope that creates an image that is ten times as large as an object has a magnification of 10x.

An additional descriptor is the *light-gathering power* of an optical microscope which corresponds to the angular range and total amount of light coming from an object, that an optical microscope admits. A relative aperture commonly employed to measure the light-gathering power of a microscope is the numerical aperture NA, defined as n·$sin\theta$ where n is the refractive index of the object space and $\theta$ is angular semi-aperture (in the object space) which corresponds to the maximum angle that a microscope admits.

The objective NA defines also the image quality in terms of resolution which is one of the two components of image quality besides the contrast.

Resolution is dependent on the point-spread function (PSF) through the definition of the *resel*, i.e. the resolution element transverse to the optic axis.

The PSF is the intensity pattern illuminated or observed by a lens at its focal plane. It has the mathematical form of the Airy function for a circular aperture (see Fig. 2).

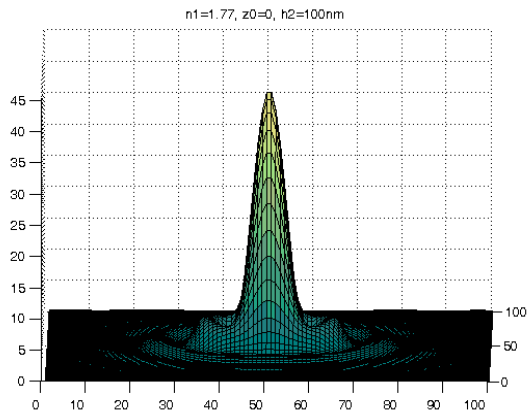

**Figure 2**. PSF

Properly the resel is defined as the resolution element due to a lens of NA = n·sinθ and coincides with the radius of the first dark fringe in the diffraction pattern, or half the diameter of the Airy disk as depicted in Fig. 3.

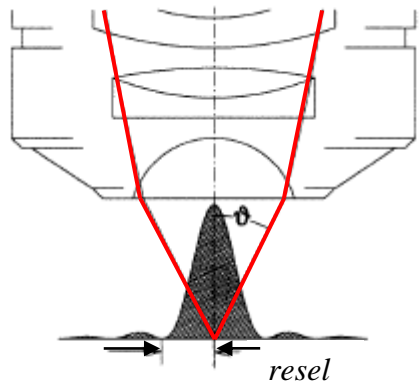

*resel*

**Figure 3.** Resel definition.

Resolution is then defined as

$$R = \frac{0.61\lambda}{NA} \qquad (1)$$

Where 0.61 is a geometrical term and λ is the illumination wavelength.

The PSF is a measure of resolution because two self-luminous points viewed by a microscope appear to be separate only if they are far enough, i.e. their PSFs are distinct [1].

In confocal microscopy, where a pinhole is placed before the detector to restrict the collected light, the axial resolution is increased since the pinhole itself spatially filters the fields in the image plane before they are sent to the detector. In this way, only the center part of the point-spread function reaches the detector [2].

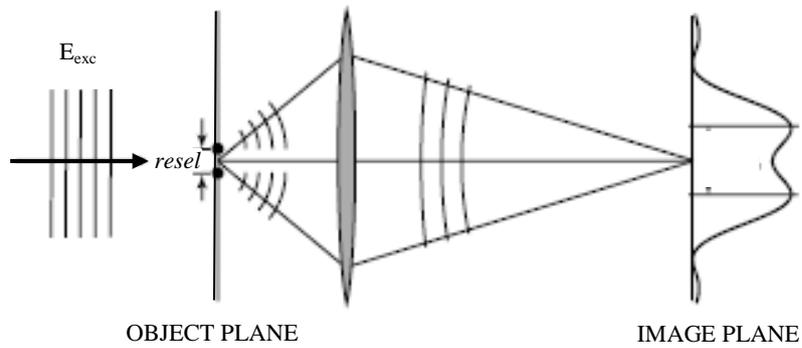

OBJECT PLANE                    IMAGE PLANE

References:
1. R.H Webb, "Confocal optical microscopy," Rep. Prog. Phys. 59, pp. 427–471, 1996.
2. L. Novotny and B. Hecht, "Principles of Nano Optics," Cambridge University Press: Cambridge, 2006.

*List of publications:*

**CHAPTERS IN INTERNATIONAL BOOKS**:

- C. Ciminelli, <u>C.M. Campanella</u>, M.N. Armenise, "Design, fabrication and characterization of a new hybrid resonator for biosensing applications", Lecture notes in Electrical Engineering, Springer, Vol. 91, **2011**, pp 229-233.
- C. Ciminelli, <u>C.M. Campanella</u>, M.N. Armenise, "Hybrid Ring-Resonator Optical Systems for Nanoparticle Detection and Biosensing Applications", Lecture notes in Electrical Engineering, Springer, Vol. 109, **2012**, pp 225-229.
- 

**PAPERS IN INTERNATIONAL CONFERENCE PROCEEDINGS:**

- C. Ciminelli, <u>C.M. Campanella</u>, M.N. Armenise, "Simulation and fabrication of a new photonic biosensor", Proceedings of the XII International Conference on Transparent Optical Networks ICTON 2010, Tu.P.16, Munich (Germany), 27 June - 1 July **2010**.
- C. Ciminelli, <u>C.M. Campanella</u>, M.N. Armenise, "Hybrid Optical Resonator for Nanostructured Virus Detection and Sizing", Proceedings of the International Symposium on Medical Measurements and Applications (MeMeA) 2011, Bari, Italy, 30-31 May **2011**.
- C. Ciminelli, <u>C.M. Campanella</u>, R. Pilolli, N. Cioffi, M.N. Armenise, "Optical Sensor for Nanoparticles", XIII International Conference on Transparent Optical Networks (ICTON) 2011, <u>Invited Paper</u>, Stockholm (Sweden), 27-30 June **2011**.
- C. Ciminelli, C.E. Campanella, F. Dell'Olio, <u>C.M. Campanella</u>, M.N. Armenise, "Multiple ring resonators in optical gyroscopes", XIV International Conference on Transparent Optical Networks (ICTON) 2012, <u>Invited Paper</u>, Coventry, England (UK), July 2 – 5, **2012**.

**PAPERS IN INTERNATIONAL JOURNALS:**

- C. Ciminelli, <u>C.M. Campanella</u>, F. Dell'Olio, C.E. Campanella, and M.N. Armenise, "Label-free optical resonant sensors for biochemical applications," (in press) Progress in Quantum Electronics **2013**.
- C. Ciminelli, C.E. Campanella, F. Dell'Olio, <u>C.M. Campanella</u>, and M.N. Armenise , 'Theoretical investigation on the scale factor of coupled waveguiding resonators under rotation conditions" submitted  for publication to Journal of the European Optical Society (JEOS).

**PAPERS IN NATIONAL CONFERENCE PROCEEDINGS:**

- C. Ciminelli, <u>C.M. Campanella</u>, M.N. Armenise, "New Hybrid Resonator for Biosensing Applications", Proceedings of the XV National Conference of the Italian Sensors and Microsystems Association (AISEM) 2010, Messina (Italy), 8-10 February **2010.**
- C. Ciminelli, <u>C.M. Campanella</u>, M.N. Armenise, "Sizing of a nanoparticle interacting with a hybrid photonic microresonator for biomedical applications", Proceedings of the XVI National Conference of the Italian Sensors and Microsystems Association (AISEM) 2011, Casaccia (Rome), Italy, 7-9 February **2011**.
- C. Ciminelli, <u>C.M. Campanella</u>, F. Dell'Olio, C.E. Campanella, M.N. Armenise, "Microrisonatore ottico per il dimensionamento di nanoparticelle ," Riunione del Gruppo nazionale di Elettronica, GE 2011, Trani (BAT), 6-8 Luglio **2011**.
- C. Ciminelli, F. Dell'Olio, C.E. Campanella, <u>C. Campanella</u>, M.N. Armenise, 'Giroscopi optoelettronici innovativi basati su risonatori passivi in tecnologia Silica-on-Silicon e in In GaAsP/InP', Riunione del Gruppo nazionale di Elettronica, GE 2011, Trani (BAT), 6-8 Luglio **2011**.
- C. Ciminelli, C.E. Campanella, <u>C.M. Campanella</u>, F. Dell'Olio, M.N. Armenise, 'Manipolazione della dispersione strutturale attraverso effetto

Franz Keldish in microrisonatori ottici in tecnologia InP', Riunione del Gruppo nazionale di Elettronica, GE 2011, Trani (BAT), 6-8 Luglio **2011**.

- C. Ciminelli, F. Dell'Olio, C.E. Campanella, <u>C. Campanella</u>, M.N. Armenise, 'Sensore ottico integrato per la misura del modulo e della fase di campi elettrici a microonde', Riunione del Gruppo nazionale di Elettronica, GE 2011, Trani (BAT), 6-8 Luglio **2011.**